\documentclass[11pt]{article}
\usepackage{geometry}                
\usepackage{setspace}
\usepackage{graphicx}
\usepackage{amssymb}
\usepackage{epstopdf}
\usepackage{placeins}
\usepackage[square,authoryear]{natbib}
\usepackage{lineno}
\usepackage[font=small,labelfont=bf]{caption}
\usepackage[titletoc,title]{appendix}
\usepackage{float}
\usepackage[caption = false]{subfig}
\newcommand{\ra}[1]{\renewcommand{\arraystretch}{#1}}

\mathcode`@="8000 
{\catcode`\@=\active\gdef@{\mkern1mu}}

\title{Crustal control of dissipative ocean tides\\in Enceladus and other icy moons}

\author{Mikael Beuthe\\
\it Royal Observatory of Belgium,\\
\it Avenue Circulaire 3, 1180 Brussels, Belgium\\
\it E-mail: mikael.beuthe@observatoire.be}      
\date{}      

\begin{document}
\maketitle

\begin{abstract}
Could tidal dissipation within Enceladus' subsurface ocean account for the observed heat flow?
Earthlike models of dynamical tides give no definitive answer because they neglect the influence of the crust.
I propose here the first model of dissipative tides in a subsurface ocean, by combining the Laplace Tidal Equations with the membrane approach.
For the first time, it is possible to compute tidal dissipation rates within the crust, ocean, and mantle in one go.
I show that oceanic dissipation is strongly reduced by the crustal constraint, and thus contributes little to Enceladus' present heat budget.
Tidal resonances could have played a role in a forming or freezing ocean less than $100\rm\,m$ deep. 
The model is general: it applies to all icy satellites with a thin crust and a shallow ocean.
Scaling rules relate the resonances and dissipation rate of a subsurface ocean to the ones of a surface ocean.
If the ocean has low viscosity, the westward obliquity tide does not move the crust.
Therefore, crustal dissipation due to dynamical obliquity tides can differ from the static prediction by up to a factor of two.
\end{abstract}


\vspace{\stretch{1}}

{\it \noindent Icarus, in press (www.elsevier.com/locate/icarus)}

\newpage

{\small
\tableofcontents

\listoffigures
\listoftables
}
\newpage

\section{Introduction}
\label{Introduction}

Ten years of Cassini flybys have provided incontrovertible evidence for an underground water reservoir at Enceladus' south pole: geysers of water vapor and ice crystals \citep{porco2006}; detection of ammonia \citep{waite2009} and salt-rich ice grains in the plume \citep{postberg2011}; compensation of the gravity signal \citep{iess2014}; hydrothermal activity \citep{hsu2015}.
Measurements of large librations now suggest that this liquid layer forms a global ocean and that the crust is not thicker than one-tenth of the surface radius \citep{thomas2016}.

Tides probably play a major role in Enceladus' geological activity, both as a source of the anomalous heat detected at the south pole \citep{howett2011} and as a trigger for the opening and closing of the tiger stripes from which geysers erupt \citep{hedman2013,nimmo2014}.
In icy satellites, the term `tides' usually refers to `solid tides', i.e.\  periodic crustal deformations due to the varying gravitational potential.
Of course, fluid tides simultaneously occur in the subsurface ocean.
Fluid tides differ from solid tides because dynamical effects due to fluid motion can be very large (if not, one speaks of static or equilibrium tide).
Oceanic dissipation, however, is usually neglected because obliquity tides, which dissipate much more energy in a deep ocean than eccentricity tides \citep{tyler2009,tyler2011}, are suppressed by the small obliquity allowed in the current Cassini state \citep{chen2011,baland2016}.
Nevertheless, \citet{matsuyama2014} and \citet{tyler2014} recently argued that resonant eccentricity tides could heat Enceladus up to the observed level even if the ocean is more than one kilometer thick.
A shallow global ocean has indeed a resonant response to tidal forcing for specific values of the forcing frequency or ocean depth.

Until now, dynamical ocean tides in icy satellites have been studied with the same Laplace Tidal Equations (LTE) that are used for Earth's tides.
That model implies a surface ocean, making it impossible to assess the impact of the crust on tides.
Examples of this approach are the studies of \citet{tyler2008,tyler2009,tyler2011,tyler2014}, \citet{chen2014}, and \citet{matsuyama2014}.
Crustal effects, however, are substantial in small and mid-size icy satellites.
Their magnitude can be estimated with a toy model in which the satellite does not rotate.
Since this model does not break spherical symmetry, tidal deformations are fully characterized by the dynamical Love numbers, which can be computed with analytical or numerical methods \citep{beuthe2015,kamata2015}.
I will not describe the results of the latter paper since they can easily be reproduced with the earlier formulas of \citet{beuthe2015}.
The main result is that the radial displacement (described by the Love number $h_2^T$) diverges if the ocean is very shallow because of a resonance due to a surface gravity mode, similar to the one predicted by Laplace more than 200 years ago \citep{lamb1932}.
A similar approach is taken by \citet{wunsch2016} who computes tides in a non-rotating `snowball' Earth assuming Cartesian LTE coupled to an elastic membrane.

In its simplest version, the non-rotating model involves three layers: an infinitely rigid mantle (or core), a homogeneous ocean of thickness (or depth) $D$, and an elastic crust of thickness $d$ and outer radius $R$, having the same density as the ocean.
If the crust is thin ($d/R\lesssim5-10\%$), the resonant ocean depth (or thickness) for tides of degree two is given by Eq.~(123) of \citet{beuthe2015}:
\begin{equation}
D_{res} = \frac{q_\omega R}{6} \, \frac{1}{ 1 - \xi_2 + Re( \Lambda^M_2) }  \, ,
\label{Dresthin}
\end{equation}
where $q_\omega$ is the dynamical parameter of Table~\ref{TableBulkOrbital}.
The degree-two density ratio $\xi_2=(3/5)(\rho/\rho_b)$ represents oceanic self-attraction ($\xi_2\approx0.373$ for Enceladus).
The membrane spring constant $\Lambda_2^M$ quantifies the resistance of the crust to deformation: $\Lambda_2^M\approx245(d/R)$ if Enceladus' crust is elastic (Eq.~(\ref{LambdaM})).
Thus, self-attraction increases the resonance depth of a surface ocean by 60\% ($1/(1-\xi_2)\approx1.6$), as found by \citet{matsuyama2014} for the largest tidal resonance.
On the other hand, the elastic crust strongly decreases the resonance depth unless crust thickness is less than, say, $100\rm\,m$.
If the crust is viscoelastic, the membrane spring constant is approximately $\Lambda_2^M\approx2(\bar\mu/\rho{}gR)(d/R)$, where $\bar\mu$ is the effective shear modulus of the crust ($\bar\mu$ is complex and smaller in absolute value than the elastic shear modulus).
Thus, the magnitude of crustal effects sensitively depends on the reduced shear modulus $(\bar\mu/\rho{}gR)$ and on the relative crust thickness $(d/R)$.
Large moons such as Europa and Titan have higher surface gravities and larger radii, so that crustal effects are smaller on these bodies than on Enceladus.

The non-rotating model gives an idea of what to expect, but does not provide accurate predictions because tides cannot be dissociated from rotation.
For Enceladus as for other synchronously rotating satellites, the rotation frequency is equal to the forcing frequency of the dominant eccentricity or obliquity tides.
The effect of synchronous rotation is twofold: (1) the unique resonance at $D=D_{res}$ is split into an infinite series of surface gravity modes, and (2) the ocean has a specific response, called the westward obliquity tide, which is proportional to the obliquity.
While the westward obliquity tide is probably irrelevant to Enceladus (due to its small obliquity), it could be important for other icy satellites.
I will thus examine whether it still occurs under an elastic crust.

In this paper, I explain how to write and solve LTE for a subsurface ocean.
Interestingly, the crust shifts the resonances in the same way in the rotating and non-rotating models.
More generally, I will show that the energy spectrum and the dissipation rate scale with the crust in a very simple manner.
In the modified LTE, the crust is modelled as a viscoelastic membrane and the mantle as a deformable body;
dynamical effects are neglected in the crust and mantle.
For the first time, it is possible to compute the dissipation rate simultaneously in the crust and ocean; mantle dissipation is also computable, but it is negligible.
The new formalism is sufficiently general to be applicable to any synchronously rotating satellite having a spherically symmetric structure, a thin crust,  and a shallow ocean.

\begin{table}[h]\centering
\ra{1.3}
\small
\caption[Orbital and bulk parameters of Enceladus]{Orbital and bulk parameters of Enceladus.}
\begin{tabular}{@{}llllll@{}}
\hline
Parameter &  Symbol & Definition & Value & Unit
\\
\hline
Mean eccentricity${}^a$ & $e$ & - & $0.0047$ & - \\
Obliquity${}^b$ & $I$ & - & $4\times10^{-4}$ & \mbox{degree} \\
Rotation rate${}^c$    & $\Omega$ & - & $5.307\times10^{-5}$ & $\rm\,s^{-1}$ \\
Surface radius${}^d$  & $R$ & - & $252.1\times10^3$  & m \\
GM${}^e$ & $GM$ & - & $7.2104\times10^9$ & $\rm m^3 \, s^{-2}$ \\
Bulk density${}^f$ & $\rho_b$ & $GM/(GV)$ & $1610$ & $\rm kg \, m^{-3}$ \\
Surface gravity & $g$ & $GM/R^2$ & $0.113$ & $\rm m \, s^{-2}$ \\
Dynamical parameter${}^g$ & $q_\omega$ & $\omega^2R^3/GM$ & $6.26\times10^{-3}$ & -
\vspace{0.5mm} \\
\hline
\multicolumn{4}{l}{\scriptsize ${}^a$ \citet{porco2006}.}
\vspace{-1.5mm}\\
\multicolumn{4}{l}{\scriptsize ${}^b$ Theoretical upper bound \citep{baland2016}.}
\vspace{-1.5mm}\\
\multicolumn{4}{l}{\scriptsize ${}^c$ JPL satellite ephemerides (http://ssd.jpl.nasa.gov/).}
\vspace{-1.5mm}\\
\multicolumn{4}{l}{\scriptsize ${}^d$ Radius of the sphere of equivalent volume \citep{thomas2010}.}
\vspace{-1.5mm}\\
\multicolumn{4}{l}{\scriptsize ${}^e$ Table~S3 in \citet{iess2014}.}
\vspace{-1.5mm}\\
\multicolumn{4}{l}{\scriptsize ${}^f$ $G=6.674\times10^{-11}\rm\,m^3kg^{-1}s^{-2}$; $V$ is the spherical volume.}
\vspace{-1.5mm}\\
\multicolumn{4}{l}{\scriptsize ${}^g$ $\omega$ is the tidal forcing frequency; $\omega=\Omega$ if synchronous rotation.}
\end{tabular}
\label{TableBulkOrbital}
\end{table}

\begin{table}[h]\centering
\ra{1.3}
\small
\caption[Parameters of interior model]{Parameters of interior model}
\begin{tabular}{@{}lllll@{}}
\hline
\vspace{0.3mm}
Parameter &  Symbol & Definition & Value & Unit \\
\hline
Ocean depth (or thickness) & $D$ & - & $1-10^5$ & m \\
Density of ice and ocean${}^a$ & $\rho$ & - & 1000 & $\rm kg/m^3$ \\
Shear modulus of elastic mantle${}^a$ & $\mu_m$ & - & $4\times10^{10}$ & Pa \\
Shear modulus of ice${}^a$ (elastic value) & $\mu_E$ & -& $3.5\times10^9$ & Pa \\
Poisson's ratio of ice${}^a$ & $\nu_E$ &  - & 0.33 & - \\
Ocean-to-bulk density ratio & $\xi_1$ & $\rho/\rho_b$ & 0.621 & - \\
Reduced shear modulus of ice & $\hat\mu_E$ & $\mu_E/(\rho{}gR)$ & 122.86 & -
\vspace{0.5mm} \\
\hline
\multicolumn{4}{l}{\scriptsize ${}^a$ Values taken from \citet{beuthe2014}.}
\end{tabular}
\label{TableParamInterior}
\end{table}%

\section{Laplace Tidal Equations (LTE)}
\label{LaplaceTidalEquations}

The Laplace Tidal Equations (LTE) are the approximate equations of motion for a shallow ocean of uniform depth on a rotating spherical body.
Underlying assumptions are reviewed by \citet{hendershott1981} (part of the MIT open course `Evolution of Physical Oceanography' available at http://ocw.mit.edu) while the LTE are derived from Navier-Stokes equations in \citet{hendershott2004}.
I assume in this paper that the ocean is shallow and homogeneous.
One should keep in mind, however, that the LTE also hold for a deep ocean if it is stratified in density.
In that case, the ocean depth is replaced in the LTE by a smaller `equivalent depth' whose size depends on the density profile \citep{hendershott1981,hendershott2004,tyler2011}.
The associated resonant depths do not correspond to the real depth of the ocean, but should rather be seen as dynamical parameters which can only be related to the geometric depth if the density stratification is known.
On the other hand, the hydrothermal activity inferred within Enceladus \citep{hsu2015} may cause turbulent convection and destroy density stratification \citep{thomson2001,goodman2004,goodman2012,soderlund2014}.
Another implicit assumption is that the top and bottom boundaries of the ocean are spherical in absence of tides.
In other words, the topography of the seafloor and crust-ocean interface does not fundamentally alter the flow.

\subsection{General form of the LTE}

The LTE for a surface ocean including self-attraction and solid-body deformation were first written by \citet{hendershott1972}:
\begin{eqnarray}
\frac{\partial}{\partial t} \mathbf{u} + 2 \mathbf{\Omega \times u} &=&
- \frac{1}{\rho} \mathbf{\nabla} p + {\cal F}(\mathbf{u}) \, ,
 \nonumber \\
\frac{\partial}{\partial t} \eta + D \, \mathbf{\nabla \cdot u} &=& 0 \, .
 \label{LTE}
\end{eqnarray}
The variables characterizing fluid motion are the depth-averaged horizontal velocity vector $\mathbf{u}$ and the radial tide $\eta$.
The latter is defined as the difference between the radial displacements of the top and bottom of the ocean:
\begin{equation}
\eta=\eta^{top}-\eta^{bot} \, .
\label{zeta}
\end{equation}
The operators $\nabla$ and $\nabla\cdot$ denote the surface gradient and divergence.
The parameter $D$ is the uniform depth of the ocean if the tidal perturbation is turned off.
If there is a crust, `ocean depth' denotes the thickness of the ocean layer.
The parameter $\mathbf{\Omega} =  \Omega \cos\theta\,\mathbf{\hat r}$ is the radial component of the rotation vector.
In the right-hand side of the LTE, $-\mathbf{\nabla}p/\rho$ is the LTE forcing term while ${\cal F}(\mathbf{u})$ is the dissipative stress.
The parameter $\rho$ is the density of the crust and ocean.

The LTE forcing term is the surface gradient of a pressure-like potential $p$ defined by
\begin{equation}
p = \rho g \left( \eta^{top} - \Gamma/g \right) + q \, ,
 \label{pressure}
\end{equation}
where $\Gamma$ is the sum of all tide-producing potentials (or total perturbing potential) while $q$ is the pressure at the top of the ocean, for example due to the atmospheric tide ($q$ is positive inwards).
For a surface ocean, $q$ is usually set to zero because atmospheric tides on Earth have a negligible effect on ocean tides \citep{marchuk1984}.
The term ($\eta^{top} - \Gamma/g$) represents the elevation of the water level above the tidally perturbed geoid $\Gamma/g$.
When the fluid acceleration and the pressure $q$ are both negligible, the water level adjusts itself to the geoid and the LTE forcing term vanishes: this is the equilibrium tide (see Section~\ref{SurfaceOcean}).

In this paper, the dissipative stress is proportional to the velocity so that the LTE can be solved with linear methods:
\begin{equation}
{\cal F}(\mathbf{u}) = - \alpha \mathbf{u} + \nu \nabla^2 \mathbf{u} \, ,
\label{DissipTerm}
\end{equation}
where $\alpha$ is the linear (top and bottom) drag coefficient and $\nu$ is the Navier-Stokes viscosity (not to be confused with Poisson's ratio $\nu_E$).
The vector Laplacian $\nabla^2 \mathbf{u}$ is restricted to its tangential components \citep{arfken2013}.
More realistically, top and bottom drag in shallow water should be modelled as a quadratic term (e.g.\ \citet{jayne2001,egbert2001}):
\begin{equation}
{\cal F}(\mathbf{u}) = - \frac{c_D}{D} \, |\mathbf{u}|\mathbf{u} \, ,
\label{nonlineardrag}
\end{equation}
where $c_D$ is an empirical coefficient usually set to $0.002-0.003$ for oceans on Earth (\citet{sohl1995} discuss other values which may be relevant to Titan).
This value can be doubled if drag also occurs at the top of the ocean \citep{tyler2014}.
LTE with nonlinear dissipation should be solved numerically on a grid, as done by \citet{sears1995} and \citet{chen2014}.
Besides the quadratic term, recent Earth tidal models include a linear term describing internal wave generation over rough topography in the deep ocean \citep{jayne2001,egbert2001}; this is a subject of ongoing research \citep{green2013}.
The magnitude of this process depends on the amplitude and wavelength of topography and on the density stratification within the ocean.
On Earth, `deep ocean' means a depth of a few kilometers, thus a ratio $D/R\approx10^{-3}$.
On Enceladus, the same ratio corresponds to a depth of about $250\rm\,m$ so that the subsurface ocean is deep in that sense.
In short, we don't know which dissipation mechanism is most relevant to Enceladus and what could be its magnitude.

\subsection{Tidal potential}
\label{TidalPotential}

Although it arises from various sources, the total perturbing potential $\Gamma$ ultimately depends on the external tidal potential $U^T$: $\Gamma=0$ if $U^T=0$.
Suppose that there is only one tidal forcing frequency $\omega$, chosen to be positive, which may a priori differ from the rotation rate $\Omega$.
The external tidal potential can be written as
\begin{equation}
U^T(t,\theta,\varphi) = \frac{1}{2} \sum_{n=2}^\infty \, U_n^T(\theta,\varphi) \, e^{i\omega t} \,\, + \,\, c.c. \, ,
\label{UT}
\end{equation}
where \textit{c.c.} denotes the complex conjugate.
$U_n^T(\theta,\varphi)$ is an eigenfunction of the Laplacian of harmonic degree $n$ and can thus be expanded in spherical harmonics $Y_n^m(\theta,\varphi)$ of degree $n$ and order $m$ (Appendix~\ref{AppendixSH}).
As dynamical tides depend on the direction of the tidal wave, $U_n^T(\theta,\varphi)$ is further decomposed into westward (W) and eastward (E) components:
\begin{equation}
U_n^T(\theta,\varphi) = \sum_{m=0,1,2} \Big( U_{nW}^m \, Y_n^m(\theta,\varphi) + U_{nE}^m \, Y_n^{m*}(\theta,\varphi) \Big)  \, .
\label{UnT}
\end{equation}
If $m=0$, this decomposition is just a matter of convenience because a zonal tidal wave is a standing wave: $U_{nW}^0=U_{nE}^0$ for all $n$.

Contrary to \citet{chen2014} and \citet{matsuyama2014}, I do not define the eastward component as the coefficient of $Y_n^{m}e^{-i\Omega t}$, but rather as the coefficient of $(Y_n^{m}e^{-i\Omega t})^*$.
Thus synchronous rotation implies $\omega=\Omega$ whatever the direction of the tide.
My choice is justified by a complication due to crustal viscoelasticity: viscoelastic corrections are the same for westward and eastward components if the Fourier transform is identical for the two components.
Choosing the convention of \citet{chen2014} and \citet{matsuyama2014} would force me to complex-conjugate the viscoelastic terms associated with eastward components.

The potential for eccentricity tides (eccentricity $e$) and obliquity tides (obliquity $I$) is given in \citet{tyler2011}, \citet{beuthe2013} (no Condon-Shortley phase), and \citet{chen2014} (potential of opposite sign).
As eccentricity and obliquity tides are usually treated separately, I arbitrarily fix the phase between the eccentricity and obliquity components as in \citet{tyler2011}.
The westward and eastward components are given in Table~\ref{TableU}.
I do not follow \citet{tyler2011} and \citet{chen2014} in calling the component $U_{2X}^0$ the `radial' tide because variations in the semi-major axis also generate tidal components of order two.
In this paper, the term `radial tide' refers to the radial displacement $\eta$ (Eq.~(\ref{zeta})).

\begin{table}[ht]\centering
\ra{1.3}
\small
\caption[Westward and eastward components of the tidal potential]
{Westward and eastward components of the tidal potential of degree two for eccentricity ($e$) and obliquity ($I$) tides if the body is in synchronous rotation.
If the spherical harmonics are normalized (Eq.~(\ref{Nnm})), $(N_2^0,N_2^1,N_2^2)=(\sqrt{5/4\pi},\sqrt{5/24\pi},\sqrt{5/96\pi})$.
If not, $N_n^m=1$.
}
\begin{tabular}{@{}ccc@{}}
\hline
 &  Westward ($X=W$) & Eastward ($X=E$) \\
\hline
$U_{2X}^0$ & $(\Omega R)^2 e \, (-3/4) /N_2^0$ &  $(\Omega R)^2 e \, (-3/4)/N_2^0$
\vspace{1mm}\\
$U_{2X}^1$ & $(\Omega R)^2 \sin I \, (-1/2)/N_2^1$ &  $(\Omega R)^2 \sin I \, (-1/2)/N_2^1$
\vspace{1mm}\\
$U_{2X}^2$ & $(\Omega R)^2 e  \, (-1/8)/N_2^2$ &  $(\Omega R)^2 e \, (7/8)/N_2^2$
\vspace{1mm}\\
\hline
\end{tabular}
\label{TableU}
\end{table}%

In the static limit, the whole-body dissipation rate due to a tidal potential of degree $n$ is proportional to $U_n^{sq}$, the surface average of the squared norm of $U_n^T$ (see Eqs.~(35) and (41) of \citet{beuthe2013}).
This term also appears in the whole-body dissipation rate including dynamical ocean tides (Eq.~(\ref{E1E2E3})) and is equal to
\begin{eqnarray}
U_n^{sq} &=& \frac{1}{4\pi} \int_S |U_n^T(\theta,\varphi)|^2 \, \sin\theta \, d\theta \, d\varphi
\nonumber \\
&=& \frac{1}{4\pi} \sum_{m=0}^2 \left( 1+\delta_{m0} \right) \sum_{X=E,W} \left| U_{nX}^m \right|^2 .
\label{Unsq}
\end{eqnarray}
Using Table~\ref{TableU}, one can check that $U_2^{sq}=(\Omega R)^4(21/5)e^2$ for eccentricity tides, while $U_2^{sq}=(\Omega R)^4(3/5)\sin^2 I$ for obliquity tides.

\subsection{Surface ocean and rigid mantle}
\label{SurfaceOcean}

Consider an infinitely rigid mantle which does not deform under tidal loading ($\eta^{bot}=0$).
Let $\Gamma_n$, $U_n^T$, and $\eta_n$ be the components of degree $n$ in the spherical harmonic expansions of $\Gamma$, $U^T$, and $\eta$, respectively (as in Eq.~(\ref{UT})).
These components are related by
\begin{equation}
\Gamma_n = U_n^T + g \, \xi_n \, \eta_n \, ,
\label{GammaRigid}
\end{equation}
where the degree-$n$ density ratio is given by
\begin{equation}
\xi_n = \frac{3}{2n+1} \, \frac{\rho}{\rho_b} \, .
\label{xin}
\end{equation}
The second term of Eq.~(\ref{GammaRigid}) is the gravitational contribution due to the change in ocean depth \citep[][Eq.~(2.1.25)]{kaula1968}, also called \textit{self-attraction}.
Self-attraction is a significant correction which was already considered by \citet{hough1898} and \citet{lamb1932}.
It accounts for nearly all the difference between the results of \citet{matsuyama2014} and those of \citet{tyler2011} and \citet{chen2014}.

Combining Eqs.~(\ref{zeta}), (\ref{pressure}) and (\ref{GammaRigid}), I can write the degree-$n$ component of the LTE forcing term for a surface ocean above a rigid mantle as
\begin{equation}
p_n /\rho = \left( 1 -\xi_n \right) g \, \eta_n - U_n^T \, .
\label{pnsurf}
\end{equation}
The \textit{equilibrium tide} is defined as the ocean tide in the limit of negligible fluid acceleration and dissipation, which is equivalent to setting $p_n=0$.
The equilibrium tide is usually written in terms of a nondimensional admittance $Z_n$,
\begin{equation}
\eta_n = Z_n \, \frac{U_n^T}{g} \, ,
\label{Zn}
\end{equation}
where $Z_n=1/(1-\xi_n)$ for the simple model considered here.

\subsection{Subsurface ocean and rigid mantle}
\label{SubsurfaceOcean}

\subsubsection{Enceladus as a membrane world}
\label{MembraneWorld}

On icy satellites, global oceans are always covered by an icy shell which exerts inward (resp.\ outward) pressure on the ocean where there is a tidal bulge (resp.\ depression). 
The equations of motion of the fluid should thus be modified so as to include this force.
The minimum change consists in modelling the crust as a massless membrane of zero thickness but finite rigidity.
The membrane or thin shell approximation is valid for shells whose thickness is less than five to ten percents of the surface radius.
Librations of large amplitude imply that Enceladus' crust is 14 to $26\rm\,km$ thick \citep{thomas2016,vanhoolst2016}.
It is thus reasonable to model Enceladus' crust as a thin shell, although the approximation will not be as good as for Europa and Titan.
The massless membrane approach has been thoroughly validated for tidal deformations (including tectonics) and tidal dissipation \citep{beuthe2014}.
In particular, it predicts the tidal Love numbers $(k_2^T,h_2^T)$ of Europa and Titan with an error less than one percent if the crust thickness is less than five percents of the surface radius.
This approach has the great advantage that it takes into account the dependence on depth of crustal viscoelasticity, which markedly differs between conductive and convective crusts.
Crustal rheology appears through effective viscoelastic parameters which have been rigorously defined in the framework of thick shell theory taken in the membrane limit \citep{beuthe2015}.
In the same paper, the membrane approach has been extended to models with crust-ocean density contrasts.
Membrane formulas are implemented in the Mathematica notebook \textit{MembraneWorlds.nb} with numerous examples (available from the author or at http://library.wolfram.com, subject class Science/Geology and Geophysics).

\begin{figure}
   \centering
   \includegraphics[width=7.3cm]{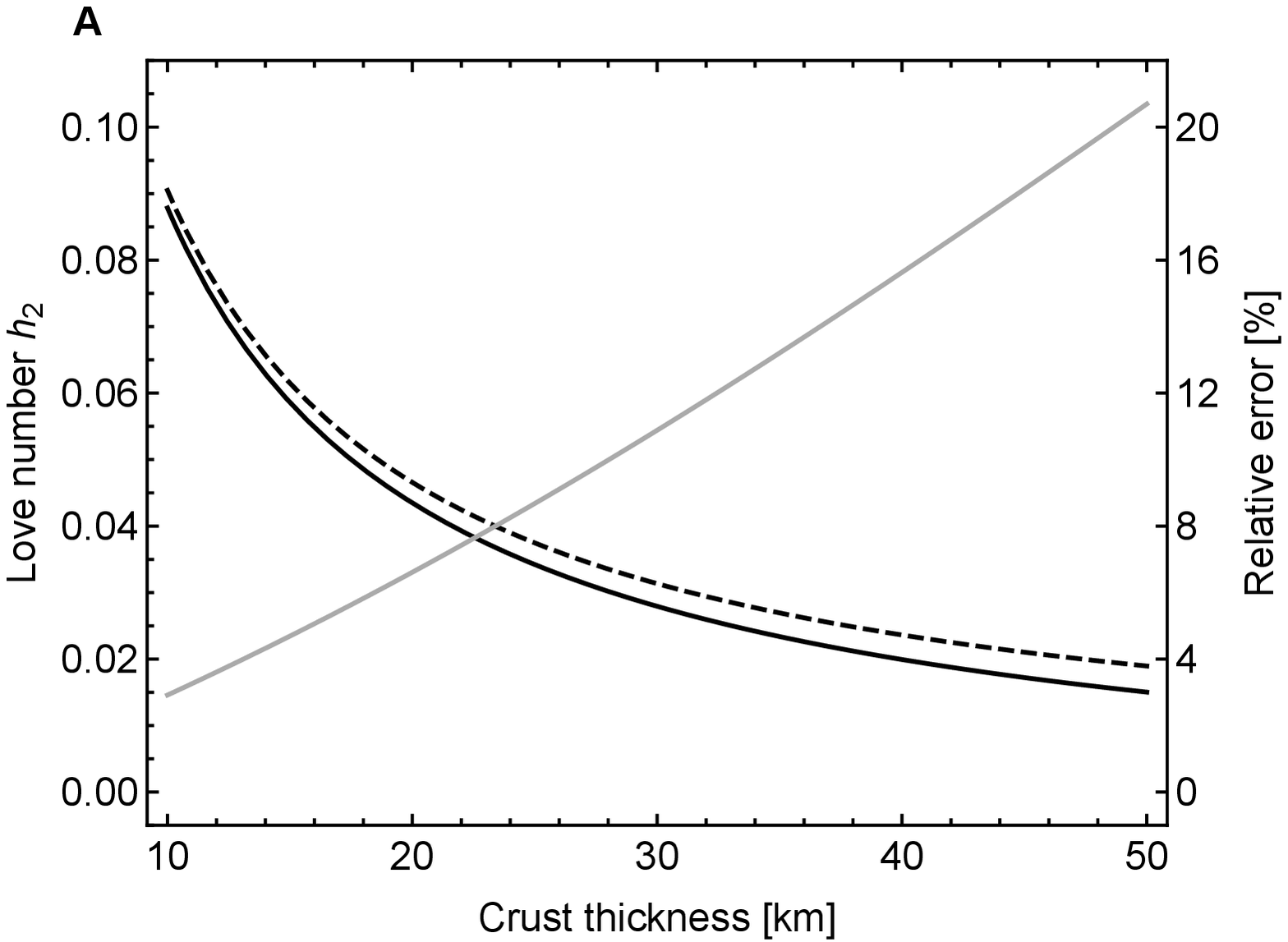}
    \includegraphics[width=7.3cm]{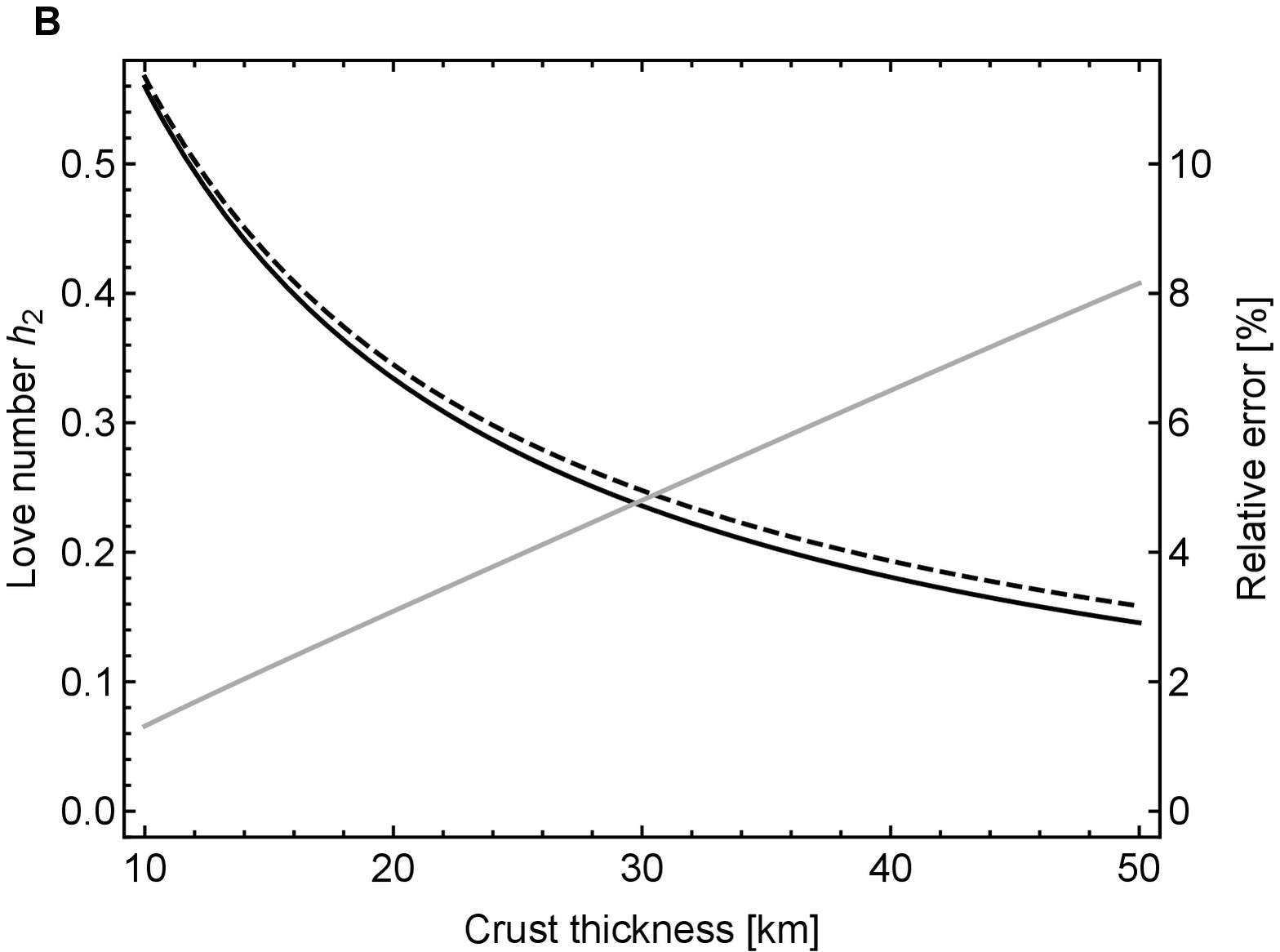}
   \caption[Membrane versus thick shell predictions of equilibrium tide]
   {Membrane versus thick shell predictions of the equilibrium tide: (A) $h_2$ if the lithosphere and crust coincide, (B) $h_2$ if the lithosphere makes up one tenth of the crust thickness.
   Dashed (resp.\ solid) black curves show membrane (resp.\  thick shell) predictions.
   Gray curves show the relative error between the two.
     See Section~\ref{MembraneWorld} for details.
}
   \label{FigLove}
\end{figure}

The reduced shear modulus of Enceladus is much larger than the one of Europa and Titan, so that elasticity plays a much greater role than self-gravitation in tidal deformations (see Section~\ref{Introduction}).
In that case, does the membrane approximation generate a much larger error for Enceladus than for Europa and Titan?
I will start by estimating the effect of finite crust thickness on static tides before considering its impact on dynamical tides in the non-rotating model.
Fig.~\ref{FigLove} shows a comparison of membrane and thick shell predictions for the radial displacement of the surface, as measured by the Love number $h_2$ (computed with the static propagator matrix method \citep{sabadini2004}).
The satellite is modelled as a three-layer body made of an infinitely rigid mantle, an ocean, and an incompressible elastic crust with shear modulus equal to $3.5\rm\,GPa$ (Table~\ref{TableParamInterior}).
The total thickness of the $\rm{}H_2{}O$ layer (ocean plus crust) is $60\rm\,km$.
Membrane estimates are computed with crust and ocean densities equal to $1020\rm\,kg/m^3$ (Eq.~(\ref{Dresthin})).
Thick shell predictions are computed with crust and ocean densities equal to $930\rm\,kg/m^3$ and $1020\rm\,kg/m^3$, respectively.
In panel~A, the crust is conductive and fully elastic so that the lithosphere makes up the whole crust.
The error due to the membrane approximation is less than ten percents if the crust is thinner than $28\rm\,km$.
 In panel~B, the crust is viscoelastic, with an elastic top layer (stagnant lid or lithosphere) making up the top tenth of the crust, while the bottom layer is convective with a shear modulus equal to one hundredth of its elastic value.
The error due to the membrane approximation is much smaller than in the fully elastic case, being less than five percents if the crust is thinner than $31\rm\,km$.
The accuracy of the membrane approach thus depends on crustal rheology.
An error of ten percents is perfectly acceptable given that most parameters of the interior model are poorly known.
 
\begin{figure}
   \centering
   \includegraphics[width=7.5cm]{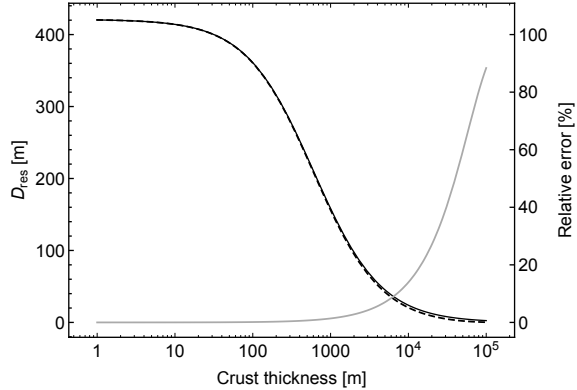}
   \caption[Membrane versus thick shell predictions of resonant depths]
   {Membrane versus thick shell predictions of resonant depths (non-rotating model).
   Solid and dashed black curves show the membrane and thick shell predictions, respectively.
   The gray curve shows the relative error between the two.
     See Section~\ref{MembraneWorld} for details.
}
   \label{FigResNR}
\end{figure}

Consider now dynamical tides in the three-layer non-rotating model described in Section~\ref{Introduction}.
 Fig.~\ref{FigResNR} shows the resonant depth as a function of the crust thickness. 
 The solid curve shows the membrane prediction (Eq.~(\ref{Dresthin})) while the dashed curve shows the thick shell prediction made with the dynamical homogeneous crust model (see Appendix~E of \citet{beuthe2015} or \textit{MembraneWorlds.nb}).
Crust and ocean densities are equal to $1020\rm\,kg/m^3$.
 The error due to the membrane approximation is less than ten percents if the crust is thinner than $7\rm\,km$.
Beyond this threshold, the resonant depth becomes so small that it is physically irrelevant anyway.

\subsubsection{LTE-membrane approach}
\label{LTEmembrane}

Suppose that the crust is thin ($d/R\lesssim0.1$ where $d$ is the crust thickness and $R$ the surface radius) and that the crust and the ocean have the same density $\rho$.
As before, the mantle is infinitely rigid.
With these assumptions, the effect of the crust can be very simply included in the LTE by modifying the pressure at the top of the ocean ($q$ term in Eq.~(\ref{pressure})).
The crust-ocean interface is modelled as a solid-liquid boundary.
The tangential stresses exerted by the ocean on the crust are set to zero because their effect on crustal deformations is negligible.

By Newton's third law, the pressure exerted by the crust on the ocean (positive inwards) is equal to the bottom load exerted by the ocean on the crust (positive outwards).
Let $q_n$ and $\eta^{top}_{n}$ be the components of degree $n$ in the spherical harmonic expansions of $q$ and $\eta^{top}$, respectively (as in Eq.~(\ref{UT})).
In the membrane approximation, the load is related to the radial deformation of the crust by Hooke's law (Eq.~(\ref{hookeMAppendix})),
\begin{equation}
q_n = \rho g \, \Lambda^M_n \, \eta^{top}_{n} \, .
\label{hookeM}
\end{equation}
The \textit{membrane spring constant} $\Lambda^M_n$ is a nondimensional parameter characterizing the extensional response of the crust defined by Eq.~(\ref{LambdaM}).
It is proportional to the effective shear modulus $\bar\mu$ and to the crust thickness $d$; it also depends on the effective Poisson's ratio $\bar\nu$ but this does not cause much variation in $\Lambda^M_n$.
If the crust is viscoelastic with depth-dependent rheology, $\bar\mu$ is smaller than the elastic shear modulus $\mu_E$ by a factor depending on the viscosity of ice and on the rheology model (see Fig.~2 of \citet{beuthe2014} and Fig.~3 of \citet{beuthe2015}).
Note that the effective viscoelastic parameters ($\bar\mu$, $\bar\nu$) do not vary with depth: they are defined for the whole crust so that we can solve the viscoelastic equations as if crustal viscoelasticity were uniform \citep{beuthe2014,beuthe2015}.

The membrane approximation is valid if the deformation is of long wavelength (see below).
While ocean tides are predominantly of long wavelength, they also include short-wavelength components.
It is thus a priori safer to consider the short-wavelength response of the crust, though we will see that it is negligible if the ocean is deeper than a few meters.
If membrane and bending responses are both included, Hooke's law takes the form of
\begin{equation}
q_n = \rho g \, \Lambda_n \, \eta^{top}_{n} \,,
\label{hookeMB}
\end{equation}
where the \textit{thin-shell spring constant} $\Lambda_n$ is defined by
\begin{equation}
\Lambda_n =  \Lambda^M_n + \Lambda^B_n \, .
\label{LambdaMB}
\end{equation}
The \textit{bending spring constant} $\Lambda^{B}_n$ is a nondimensional parameter characterizing the bending response of the crust (Eq.~(\ref{LambdaB})).
Bending effects are significant if $\Lambda^B_n\gtrsim\Lambda^M_n$, which occurs if
\begin{equation}
n\gtrsim\kappa\sqrt{R/d} \, ,
\label{BendingThreshold}
\end{equation}
where $\kappa=(12(1-\nu_E^2))^{1/4}\approx1.8$.
For example, bending becomes significant for $n\gtrsim18$ if $d=2.5\rm\,km$.
In Section~\ref{KineticEnergy}, I will show that the dominant components of the ocean response are of low harmonic degree, so that the crust mainly deforms as a membrane.
In this paper, I will neglect $\Lambda^B_n$ when the crust is viscoelastic with depth-dependent rheology.

The LTE forcing term in an ocean sandwiched between a rigid mantle and a flexible crust follows from Eqs.~(\ref{pressure}), (\ref{pnsurf}) and (\ref{hookeMB}):
\begin{equation}
p_n/\rho = \left( 1 -\xi_n + \Lambda_n \right) g \, \eta_n - U^T_n \, .
\end{equation}
The equilibrium tide admittance (see Eq.~(\ref{Zn})) is here
\begin{equation}
Z_n = \frac{1}{1 -\xi_n + \Lambda_n} \, .
\label{Znrigid}
\end{equation}
If $n=2$, this expression is identical to the radial Love number of the rigid mantle model (Eq.~(\ref{h2r})) because the ocean tide is equal to the radial displacement of the crust if the mantle is infinitely rigid.

\subsection{Subsurface ocean and nonrigid mantle}
\label{Nonrigidmantle}

\subsubsection{Tidal, load, and pressure Love numbers}
\label{LoveNumbers}

If the mantle is not infinitely rigid, one should compute the displacement of the mantle-ocean boundary $(\eta^{bot})$ as well as the gravitational perturbation associated with it.
For a surface ocean, \citet{hendershott1972} parameterizes these quantities with the \textit{tidal Love numbers} $(k^T_n, h^T_n)$ and the \textit{load Love numbers} $(k^L_n, h^L_n)$ of the solid body without ocean (see also \citet{hendershott1981}).
The tidal forcing consists of the tidal potential $U^T_n$ while the mass-load forcing results from the surface density $\sigma_n=\rho\eta_n$, which can be replaced by a mass-load potential $U_n^L$ (see Appendix~\ref{AppendixLoveTL}).
For a subsurface ocean, one also needs the \textit{pressure Love numbers} of the solid body \citep{molodensky1977,guo2004}.
These numbers describe the response of the body to the surface pressure exerted by the membrane (Eq.~(\ref{hookeMB})), which can be replaced by a pressure potential $U_n^P$.
In Appendix~\ref{AppendixLoveP}, I show that pressure Love numbers are linear combinations of tidal and load Love numbers: $k^P_n=-h^T_n$ and $h^P_n=h^L_n-h^T_n$.

In terms of forcing potentials, the total perturbing potential reads
\begin{equation}
\Gamma_n
= \left( 1 + k^T_n \right) U^T_n + \left( 1 + k^L_n \right)  U^L_n + k^P_n \, U^P_n \, .
\end{equation}
The last term differs from the first two because pressure has no direct gravitational effect.
Eqs.~(\ref{UL}) and (\ref{UP}) yield $U^L_n=g\xi_n\eta_n$ and $U^P_n=g\xi_n\Lambda_n\eta^{top}_n$ where $\xi_n$ is given by Eq.~(\ref{xin}).
The total perturbing potential can thus be written as
\begin{equation}
\Gamma_n
= \left( 1 + k^T_n \right) U^T_n + g  \, \xi_n \left( \left( 1 + k^L_n \right)  \eta_n + k^P_n \, \Lambda_n \, \eta^{top}_n \right) \, .
\label{Gamman}
\end{equation}

Similarly, the degree-$n$ component of the displacement of the mantle-ocean boundary reads
\begin{equation}
\eta^{bot}_n =  \frac{1}{g} \left( h^T_n \, U^T_n +  h^L_n \, U^L_n + h^P_n \, U^P_n \right) \, ,
\end{equation}
which can be expressed as
\begin{equation}
\eta^{bot}_n =  h^T_n \, U^T_n/g +  \xi_n \left( h^L_n \, \eta_n + h^P_n \, \Lambda_n \, \eta^{top}_n \right) \, .
\label{zetabot0}
\end{equation}
Combining this expression with Eq.~(\ref{zeta}), I write the displacements of the top and bottom of the ocean in terms of the external tidal potential $U_n^T$ and the radial tide $\eta_n$:
\begin{eqnarray}
\eta^{top}_n &=&  \frac{h^T_n \, U^T_n/g + \left( 1 + \xi_n \, h^L_n \right) \eta_n}{1-\xi_n \, h^P_n \, \Lambda_n} \, ,
\label{zetatop} \\
\eta^{bot}_n &=& \frac{h^T_n \, U^T_n/g + \xi_n \left( h^L_n + h^P_n \, \Lambda_n \right) \eta_n}{1-\xi_n \, h^P_n \, \Lambda_n} \, .
\label{zetabot}
\end{eqnarray}

\subsubsection{LTE forcing term}
\label{LTEforcing}

\begin{figure}
   \centering
   \includegraphics[width=7.5cm]{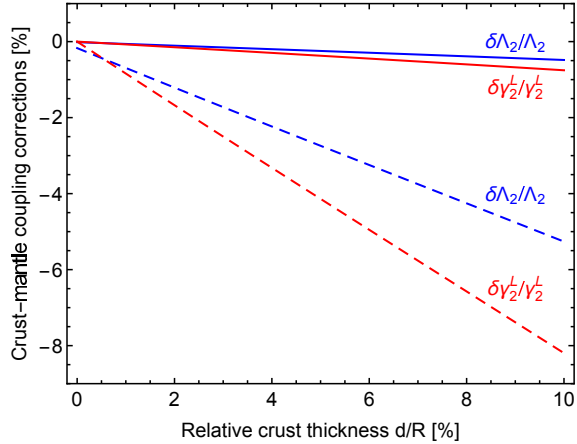}
   \caption[Crust-mantle coupling corrections to the LTE forcing term]{Crust-mantle coupling corrections to the LTE forcing term: $\delta\Lambda_2/\Lambda_2$ and $\delta\gamma^T_2/\gamma^T_2$ as defined by Eq.~(\ref{delgammaT}).
   The mantle is approximated as an incompressible homogenous sphere which has either the rigidity of silicates ($\mu_m=40\rm\,GPa$, solid lines), or the rigidity of ice ($\mu_m=3.5\rm\,GPa$, dashed lines).
}
   \label{FigCMcorr}
\end{figure}

The LTE forcing term (Eq.~(\ref{pressure})) can now be expressed in terms of $U^T_n$ and $\eta_n$:
\begin{equation}
p_n/\rho = \beta_n \, g \, \eta_n - \upsilon_n \, U^T_n \, ,
\label{pressureflexi}
\end{equation}
where the coefficients multiplying the radial tide and the potential are defined as
\begin{eqnarray}
\beta_n &=& 1 - \xi_n \gamma^L_n + \Lambda_n + \delta\Lambda_n \, ,
\nonumber \\
\upsilon_n &=& \gamma^T_n + \delta\gamma^T_n \, .
\label{betan}
\end{eqnarray}
The symbols $(\gamma^T_n,\gamma^L_n)$ denote the diminishing (or tilt) factors of classical geodesy (see Eq.~(\ref{gammaJ})).
If the mantle is infinitely rigid, $\gamma^T_n=\gamma^L_n=1$.
The corrections due to crust-mantle coupling are negative numbers (see Fig.~\ref{FigCMcorr}) given by
\begin{eqnarray}
\delta\Lambda_n &=& - \left( 1 - \frac{ \left(1+\xi_n \, h^L_n\right)^2 }{1+\xi_n \left( h^T_n - h^L_n \right) \Lambda_n} \right) \Lambda_n \, ,
\nonumber \\
\delta\gamma^T_n &=&  - \left(  \frac{ 1+ \xi_n h^L_n }{1+\xi_n \left( h^T_n - h^L_n \right) \Lambda_n} \, h^T_n   \right) \Lambda_n \, ,
\label{delgammaT}
\end{eqnarray}
where I used relations between Love numbers (Eqs.~(\ref{saitomolo})-(\ref{kLgammaT}) and (\ref{hPkP})).
These corrections vanish if the crust is fluid-like ($\Lambda_n=0$) or if the mantle is infinitely rigid ($h^T_n=h^L_n=0$).
The equilibrium tide admittance (see Eq.~(\ref{Zn})) is here
\begin{equation}
Z_n = \frac{\upsilon_n}{\beta_n} \, .
\label{ZnNonrigid}
\end{equation}

As a first consistency check, I will compute the equilibrium tide of degree two in a simple but nontrivial example.
The core-mantle system is approximated as an incompressible homogenous sphere, with Love numbers given by Eqs.~(\ref{LovekH})-(\ref{hnTincomp}).
Substituting these formulas into Eq.~(\ref{gammaJ}) and Eqs.~(\ref{betan})-(\ref{ZnNonrigid}), I get
\begin{equation}
Z_2 = \frac{ 57\hat\mu_m + 5 \left(5\xi_2-3\right) \Lambda_2 }{ 2 \left( 3 - 5\xi_2 \right) + 57 \left( 1 - \xi_2 + \Lambda_2 \right) \hat\mu_m + \left( 6 + 5\xi_2 \left(1-5\xi_2\right)  \right) \Lambda_2} \, .
\label{Zn3layers}
\end{equation}
This expression agrees with the result of the static propagator matrix method in the limit of a thin crust and shallow ocean (Eq.~(\ref{Zn3layersPMM})).
It also agrees with Eq.~(\ref{Znrigid}) in the rigid mantle limit ($\hat\mu_m\rightarrow\infty$).

Let us now check the limit of a surface ocean.
If crustal rigidity vanishes, the LTE forcing term (Eq.~(\ref{pressureflexi})) reduces to the well-known formula for a surface ocean,
\begin{equation}
p_n/\rho = \left( 1 - \xi_n \, \gamma^L_n  \right) g \, \eta_n - \gamma^T_n \, U^T_n \, .
\label{pnNonrigid}
\end{equation}
This forcing term is used in numerical models of ocean tides on Earth \citep{hendershott1981,egbert2001} and on icy satellites \citep{matsuyama2014}.
The corresponding equilibrium tide admittance, $Z_n=\gamma^T_n/(1 - \xi_n\gamma^L_n)$, agrees with \citet{dahlen1976} and \citet{agnew1978}.

Fig.~\ref{FigCMcorr} shows the magnitude of the crust-mantle coupling corrections as a function of crust thickness.
The mantle is approximated as an incompressible homogenous sphere.
For a silicate-rich mantle ($\mu_M=40\rm\,GPa$, solid curves), the relative corrections are less than 1\% (if $d/R<10\%$), whereas they are ten times larger in the unlikely case of a very soft mantle ($\mu_M=3.5\rm\,GPa$, dashed curves).
Therefore, it is a very good approximation to treat Enceladus' mantle as an infinitely rigid layer.

\section{Solving LTE}

Following \citet{love1913} and many others, I write the LTE in terms of scalar potentials (Helmholtz decomposition) before expanding them in spherical harmonics.
The method of solution is well-known so that we can dispense with detailed derivations.
I quote the resulting formulas, however: partly because of the new crustal effects, and partly because of the new convention chosen for eastward and westward waves (Section~\ref{TidalPotential}).
The standard reference for the normal modes analysis is \citet{longuet1968} with typos detailed by \citet{tyler2011} but our problem also includes the tidal forcing and dissipation terms required in ocean tidal models (e.g.\ \citet{tyler2011}).

\subsection{Scalar equations}

On the surface of a sphere, the fluid velocity can always be expressed in terms of two scalar potentials:
\begin{equation}
{\bf u} = \nabla \Phi + \nabla \times (\Psi {\rm\bf \hat r}) \, .
\label{helmholtz}
\end{equation}
$\Phi$ is sometimes called the \textit{consoidal potential} while $\Psi$ is the \textit{toroidal potential} or \textit{stream function}.
In spherical coordinates (colatitude $\theta$, eastern longitude $\varphi$), the velocity components are $u_\theta=(\partial_\theta \Phi+\partial_\varphi\Psi/\sin\theta)/R$ and $u_\varphi=(\partial_\varphi\Phi/\sin\theta-\partial_\theta\Psi)/R$.
After standard manipulations, the LTE (Eq.~(\ref{LTE})) become three scalar equations:
\begin{eqnarray}
{\cal L}_0 \, \Phi + 2\Omega \, {\cal L}_1 \, \Psi 
&=& - \frac{1}{\rho} \Delta p \, ,
\label{LTEdiv} \\
{\cal L}_0 \Psi - 2\Omega \, {\cal L}_1 \, \Phi &=& 0 \, ,
\label{LTEvort} \\
\frac{\partial}{\partial t} \eta + \frac{D}{R^2} \, \Delta \Phi &=& 0 \, ,
\label{LTEz}
\end{eqnarray}
where $\Delta$ is the spherical Laplacian (Eq.~(\ref{defLaplacian})).
In spherical coordinates, the operators ${\cal L}_0$ and ${\cal L}_1$ are given by
\begin{eqnarray}
{\cal L}_0 &=& \left( \frac{\partial}{\partial t} + \alpha - \frac{\nu}{R^2} \, \Delta \right) \Delta + 2\Omega \, \frac{\partial}{\partial \varphi} \,  ,
\nonumber \\
{\cal L}_1 &=& \cos\theta \, \Delta - \sin\theta \, \frac{\partial}{\partial\theta} \, .
\end{eqnarray}
${\cal L}_0$ does not change the degree of a spherical harmonic, whereas  ${\cal L}_1$ changes it by one unit.

\subsection{Spherical harmonic expansion}

Now that the LTE are in a scalar form, the velocity potentials can be Fourier-transformed and expanded in spherical harmonics of degree $n$ and order $m$.
Projection on complex-conjugated spherical harmonics then yields an infinite number of coupled linear equations.
The harmonic order $m$ takes a definite value (always positive); the general solution follows by superposition.
The spherical harmonic expansion of the consoidal velocity potential reads
\begin{equation}
\Phi = \frac{1}{2} \sum_{n=m}^{\infty} \left( \tilde \Phi_{nW}^m \, Y_n^m(\theta,\varphi) + \tilde \Phi_{nE}^m \, Y_n^{m*}(\theta,\varphi) \right) e^{i\omega t} \,\, + \,\, c.c. \, ,
\label{FT}
\end{equation}
where \textit{c.c.} denotes the complex conjugate.
For synchronous rotation, $\omega=\Omega$ whatever the direction of the tide.
$\Psi$ and $\eta$ are similarly expanded in $\tilde \Psi_{nX}^m$ and $\tilde \eta_{nX}^m$.

In order to obtain real equations in the inviscid limit (and simplify the notation at the same time), let us define (e.g.\ \citet{kasahara1976})
\begin{eqnarray}
\tilde \Phi_{nX}^m &=& i \, \Phi_{nX}^m \, ,
\nonumber \\
\left( \tilde \Psi_{nX}^m, \tilde \eta_{nX}^m \right) &=& \Big( \Psi_{nX}^m, \eta_{nX}^m \Big) \, .
\label{Phidef}
\end{eqnarray}
The LTE for vertical equilibrium (Eq.~(\ref{LTEz})) yields an equation for the radial tide:
\begin{equation}
\eta_{nX}^m = \frac{D}{\omega R^2} \, n(n+1) \, \Phi_{nX}^m \, .
\label{zetanm}
\end{equation}
Combining this equation with Eq.~(\ref{pressureflexi}), I eliminate the radial tide $\eta$ in the tangential LTE (Eqs.~(\ref{LTEdiv})-(\ref{LTEvort})).
This procedure yields an infinite system of equations for the spherical harmonic components of the velocity potentials:
\begin{eqnarray}
q_{n-1}^m \, \Psi_{n-1,X}^m + K_{nX}^m \, \Phi_{nX}^m + p_{n+1}^m \, \Psi_{n+1,X}^m &=& \frac{\upsilon_n}{2\Omega} \, U_{nX}^{m} \, ,
\nonumber \\
q_{n-1}^m \, \Phi_{n-1,X}^m + L_{nX}^m \, \Psi_{nX}^m + p_{n+1}^m \,  \Phi_{n+1,X}^m &=& 0 \, ,
\label{LTEsh}
\end{eqnarray}
which hold for $n\geq{}m$, it being understood that $\Psi_{m-1,X}^m=\Phi_{m-1,X}^m=0$ (recall that $m$ is fixed).
Besides, $n\geq{}1$ if $m=0$ (constant potentials do not contribute to Eq.~(\ref{helmholtz})).

The diagonal coefficients of Eq.~(\ref{LTEsh}) read
\begin{eqnarray}
L_{nX}^m &=& - \lambda \pm \frac{m}{n(n+1)} + \frac{i}{4Q_n} \, ,
\label{defL} \\
K_{nX}^m &=& L_{nX}^m + \frac{n(n+1)\beta_n}{\lambda \, \epsilon_L} \, .
\label{defK}
\end{eqnarray}
In Eq.~(\ref{defL}), the sign `plus' holds for westward tides ($X=W$) and the sign `minus' for eastward tides ($X=E$).
The nondimensional tidal quality factors $Q_n$ are defined by
\begin{equation}
\frac{1}{Q_n} = \frac{2}{\Omega} \left( \alpha + n(n+1) \, \frac{\nu}{R^2} \right) .
\label{qualityfactor}
\end{equation}
If $\nu=0$, all $Q_n$ are equal to the effective tidal quality factor $Q=\Omega{}E_{kin}/\dot{E}$ of \citet{tyler2011,tyler2014} and \citet{chen2014}.
If $\alpha=0$, $Q_1$ is the same as the factor $Q_{obl,W}$ of \citet{chen2014}.
Tides are weakly (resp.\ strongly) damped by ocean viscosity if $Q_n\gg1$ (resp.\ $Q_n\approx1$).

In Eq.~(\ref{defK}), $\beta_n$ includes the effects of self-attraction, mantle deformat
ion, and crust rigidity (see Eq.~(\ref{betan})).
$\lambda$ and $\epsilon_L$ are the nondimensional frequency and Lamb parameter, respectively:
\begin{eqnarray}
\lambda &=& \frac{\omega}{2\Omega} = \frac{1}{2} \, ,
\label{lambda} \\
\epsilon_L &=& \frac{4\Omega^2 R^2}{gD} = 4 q_\omega \frac{R}{D} \, .
\label{LambParam}
\end{eqnarray}
In each line, the second equality holds for synchronous rotation ($q_\omega$ is defined in Table~\ref{TableBulkOrbital}).
For Enceladus, $\epsilon_L\approx1$ if $D\approx6.4\rm\,km$.

The nondiagonal coefficients $(p_n,q_n)$ of Eq.~(\ref{LTEsh}) depend on the spherical harmonic normalization (see Eq.~(\ref{Nnm})).
If the basis is normalized (e.g.\ \citet{chen2014}), then
\begin{equation}
\left( p_n^m , \, q_n^m \right) = \left( \frac{n+1}{n} \, C_n^m  \, , \frac{n}{n+1} \, C_{n+1}^m  \right) ,
\label{pnqnN}
\end{equation}
where
\begin{equation}
C_n^m = \sqrt{ \frac{n^2-m^2}{4 n^2 -1} } \, .
\end{equation}
If the basis is not normalized (e.g.\ \citet{longuet1968}), then
\begin{equation}
\left( p_n^m , \, q_n^m \right) = \left( \frac{n+1}{n} \frac{n+m}{2n+1} \, (1-\delta_{mn}) \, , \frac{n}{n+1} \frac{n-m+1}{2n+1} \right) ,
\label{pnqn}
\end{equation}
where I inserted the factor $(1-\delta_{mn})$ so that $p_n^n=0$ as in the normalized case (this modification does not affect Eq.~(\ref{LTEsh})).
Both definitions (Eqs.~(\ref{pnqnN}) and (\ref{pnqn})) thus satisfy
\begin{equation}
p_m^m = q_{m-1}^{m} = 0 \, .
\label{property}
\end{equation}

\subsection{Matrix solution}
\label{MatrixSolution}

In practice, the system formed by Eq.~(\ref{LTEsh}) is truncated at degree $N$ and written in matrix form.
If the matrices are nonsingular, the solution is regular and is found by inverting the linear system.
In general, $\Phi$ and $\Psi$ are superpositions of spherical harmonics of all degrees $n\geq{}m$, although a selection rule eliminates half the components for a forcing of given degree and order.
Indeed, Eq.~(\ref{LTEsh}) is made of two independent systems \citep{hough1898,longuet1968,kasahara1976}, each for one half of the variables:
\begin{enumerate}
\item symmetric case: the variables are $(\Phi_{mX}^m,\Psi_{m+1,X}^m,\Phi_{m+2,X}^m,\Psi_{m+3,X}^m,...)$, from which $\Phi_{0X}^0$ is dropped if $m=0$.
Motion is mirrored with respect to the equatorial plane, and there is no motion across the equator.
\item antisymmetric case: the variables are $(\Psi_{mX}^m,\Phi_{m+1,X}^m,\Psi_{m+2,X}^m,\Phi_{m+3,X}^m,...)$, from which $\Psi_{0X}^0$ is dropped if $m=0$.
Motion at the equator is normal to the equator.
\end{enumerate}
The matrices associated with the two cases can be found in \citet{longuet1968} (see his Eqs.~(3.23)-(3.24)).
The components of the external tidal potential with even $n-m$ (eccentricity tides) and odd $n-m$ (obliquity tides) are sources for the symmetric and antisymmetric systems, respectively.
If the external tidal potential is of degree two, only three possibilities remain: symmetric if $m=0$ or $2$, antisymmetric if $m=1$.
I gather the three cases into the following system:
\begin{equation}
\left(
\begin{array}{ccccc}
L_{1X}^m & p_2^m & 0 & 0 & ... \\
q_1^m & K_{2X}^m & p_3^m & 0 & ... \\
 0 & q_2^m & L_{3X}^m & p_4^m & ... \\
 ... & ... & ... & ... & ...
\end{array}
\right)
\left(
\begin{array}{c}
\Psi_{1X}^m \\
\Phi_{2X}^m \\
\Psi_{3X}^m \\
 ... 
\end{array}
\right)
=
\frac{\upsilon_2}{2\Omega}
\left(
\begin{array}{c}
0 \\
U_{2X}^{m} \\
0 \\
 ... 
\end{array}
\right)
.
\label{matrix}
\end{equation}
If $m=2$, Eq.~(\ref{property}) ensures that the unphysical component $\Psi_{1X}^2$ vanishes because $p_2^2=q_1^2=0$.
The components that do not appear in Eq.~(\ref{matrix}) vanish: $\Phi_{0X}^0=0$, $\Phi_{nX}^m=0$ if $n$ is odd, and $\Psi_{nX}^m=0$ if $n$ is even.

\subsection{Slow-rotation and deep-ocean limits}
\label{Limits}

In the \textit{slow-rotation limit}, tides are artificially decoupled from rotation: $\Omega\ll\omega$.
In that limit, Eq.~(\ref{LTEsh}) tends to a diagonal system,
\begin{eqnarray}
\left( 2\Omega \, K_{nX}^0  \right) \Phi_{nX}^m &=& \upsilon_n \, U_{nX}^{m} \, ,
\nonumber \\
\Psi_{nX}^m &=& 0 \, ,
\label{LTEslowrot}
\end{eqnarray}
which yields the following solution for the radial tide:
\begin{equation}
\eta_{nX}^m = \frac{\upsilon_n}{\beta_n + \left(1-i/(4Q_n\lambda)\right) \Lambda_\omega} \,  \frac{U_{nX}^m}{g} \, ,
\label{slowrotlim}
\end{equation}
where $\Lambda_\omega= -( q_\omega{}R)/(n(n+1)D)$.
If the mantle is infinitely rigid, the factor multiplying $U_{nX}^m/g$ coincides with the radial Love number of the non-rotating model (Eq.~(122) of \citet{beuthe2015}) except for the new oceanic dissipation term $-i/(4Q_n\lambda)$.
The degree-two solution is resonant if
\begin{equation}
D = \frac{q_\omega R}{6} \, \frac{1}{Re(\beta_2)} \, ,
\label{Dslowrot}
\end{equation}
in agreement with the prediction of the non-rotating model (Eq.~(\ref{Dresthin})).

In the \textit{deep-ocean limit}, the ocean is deep enough so that dynamical effects become negligible: the term proportional to $\beta_n$ dominates in Eq.~(\ref{defK}), so that Eq.~(\ref{LTEsh}) becomes similar to Eq.~(\ref{LTEslowrot}) but with the supplementary condition that $\lambda\approx0$.
The solution for the radial tide is thus Eq.~(\ref{slowrotlim}) in which $\Lambda_\omega=0$, i.e.\  the equilibrium tide given by Eq.~(\ref{ZnNonrigid}):
\begin{equation}
\eta_{nX}^m \approx Z_n \, \frac{U_{nX}^m}{g} \, .
\label{deepocean}
\end{equation}
The westward obliquity tide must be treated separately (Section~\ref{WestwardObliquityTide}).

\section{Resonant and nonresonant tidal waves}

In this section, I look more closely at LTE solutions that stand out by their magnitude: the nonresonant westward obliquity tide, and the resonant solutions for all the other tidal waves.
The resonances can be investigated either as peaks in the energy spectrum, or as solutions of an eigenvalue problem.

\subsection{Westward obliquity tide (WOT)}
\label{WestwardObliquityTide}

Suppose that the body is in synchronous rotation.
The westward obliquity tide, hereafter denoted WOT or indexed as \textit{wot}, differs from other tidal waves.
The reason is that the inviscid LTE (Eqs.~(\ref{LTEdiv})-(\ref{LTEz})) have a closed-form solution \citep{tyler2009}:
\begin{eqnarray}
\Phi_{tor} &=& 0 \, ,
\nonumber \\
\Psi_{tor} &=& \frac{3}{2} \, \Omega R^2 \, \upsilon_2 \, \sin I \, \sin\theta \, \cos \left( \varphi + \Omega t \right) ,
\label{toroidalsol}
\end{eqnarray}
where the subscript $tor$ stands for `toroidal'.
In terms of spherical harmonics, the only nonzero component is $\Psi_{1W}^1$.
The particular form of this solution originates in the identity ${\cal L}_0\Psi_{tor}=0$ or, equivalently, in $Re(L_{1W}^1)=0$ if $\lambda=1/2$, $m=1$ and $n=2$ (see Eq.~(\ref{defL})).

This solution has the following characteristics:
\begin{itemize}
\item
It is purely toroidal ($\Phi=0$). This means that the radial tide vanishes: $\eta=0$.
\item
It is not a resonance. A resonance is a free oscillation mode whose periodic excitation by the tidal potential causes a divergent response.
By contrast, the toroidal solution does not exist without tidal forcing and does not diverge.
\item
It does not depend on ocean depth.
\end{itemize}

If the fluid is viscous, other components of $\Psi$ as well as of $\Phi$ become nonzero, though $\Psi_{1W}^1$ remains the dominant component.
Following \citet{chen2014}, I find an approximate viscous WOT solution by solving the system (\ref{matrix}) truncated to degree two:
\begin{eqnarray}
\Psi_{1W}^1 &=& \frac{\sqrt{5}}{1 - i \, (5K_2^1)/(3Q_1) } \, \frac{\upsilon_2}{\Omega} \, U_{2W}^1 \, ,
\nonumber \\
\Phi_{2W}^1 &=& - \frac{\sqrt{5}}{6} \frac{i}{Q_1} \, \Psi_{1W}^1 \, ,
\label{toroidalcomp}
\end{eqnarray}
where
\begin{equation}
K_2^1 = - \frac{1}{3} + \frac{12\beta_2}{\epsilon_L} + \frac{i}{4 Q_2} \, .
\label{defK21}
\end{equation}
If viscosity is low ($Q_1,Q_2\gg1$), the solution is dominated by the toroidal potential and $K_2^1\approx12\beta_2/\epsilon_L$.
In that case, the viscous WOT solution is approximately given in the spatial domain  by
\begin{equation}
\Psi_{wot} \approx \frac{1}{1 - i \, (20 \beta_2)/(\epsilon_L Q_1) } \, \Psi_{tor} \,  .
\label{toroidalsolviscous}
\end{equation}
Obliquity tides now depend on ocean depth through the Lamb parameter $\epsilon_L$: the magnitude of $\Psi_{wot}$ is nearly constant if $\epsilon_L\gtrsim20|\beta_2|/Q_1$ but decreases as $1/D$ if $\epsilon_L\lesssim20|\beta_2|/Q_1$.

For a surface ocean without self-attraction ($\beta_2=1$) and with low Navier-Stokes viscosity, the solution (\ref{toroidalcomp}) is equivalent to Eqs.~(41), (42), (46), and (47) of \citet{chen2014}.
Their threshold $\epsilon_L{}Re\approx80$ (where $Re=4Q_1$ is the Reynolds number) is a special case of the threshold $\epsilon_L\approx20|\beta_2|/Q_1$ if $\beta_2=1$.

In Section~\ref{Limits}, I showed that the radial displacement is given by the equilibrium tide in the deep-ocean limit.
There is however no such limit for the inviscid WOT solution because $(\Phi_{tor},\Psi_{tor})$ do not depend on ocean depth.
Is this conclusion changed by viscosity?
Plugging Eq.~(\ref{toroidalcomp}) into Eq.~(\ref{zetanm}), I can write
\begin{equation}
\eta_{2W}^1 \approx \frac{1}{1 + i \, (\epsilon_L Q_1)/(20 \beta_2) } \, Z_2 \, \frac{U_{2W}^1}{g} \, .
\label{deepoceanW}
\end{equation}
Thus the radial WOT tends, in the deep-ocean limit, to the equilibrium tide (Eq.~(\ref{deepocean})) if the viscosity is high enough: $(\epsilon_L Q_1)/|20\beta_2|\ll1$.

\subsection{Energy spectrum}
\label{KineticEnergy}

The LTE matrix (Eq.~(\ref{matrix})) is singular at specific frequencies.
Physically, these singularities correspond to free oscillations of the ocean (normal modes), which are of two main types: surface gravity waves and inertial (Rossby) modes \citep{longuet1968}.
Our tidal forcing problem is slightly different because the frequency is fixed (it is equal to the forcing frequency $\Omega$) while the ocean depth is unknown.
The frequency of Rossby modes is in the range $[0,2m\Omega/(n(n+1))]$ which is strictly smaller than $\Omega$.
Thus we only need to bother about surface gravity waves.
Strictly speaking, they should be called `surface gravity waves modified by rotation', but I will stick in this paper to the shorter name `surface gravity waves'.

The average kinetic energy stored in ocean tides gives a global picture of the tidal response, in which resonances appear as diverging energy peaks for specific values of the ocean depth.
Computing the kinetic energy serves other ends.
On the one hand, it is a good preliminary to dissipation: the dissipation rate for linear top and bottom drag is indeed proportional to the kinetic energy.
On the other, it gives an estimate of the average flow speed, which is needed when modeling nonlinear dissipation (Section~\ref{OceanicDissipation}).

The kinetic energy of the fluid integrated over the ocean volume $V_O$ and averaged over the orbital period $T$ reads
\begin{equation}
E_{kin} = \frac{1}{T} \int_T dt \int_{V_O} dV \left(  \frac{\rho}{2} \, \mathbf{u} \cdot \mathbf{u} \right) .
\end{equation}
Following \citet{chen2014}, I evaluate the kinetic energy in terms of spherical harmonic coefficients,
\begin{equation}
E_{kin} =  \sum_{m=0}^2 \sum_{n=m}^N \sum_{X=E,W} E_{nX}^m \, ,
\label{Ekin}
\end{equation}
where $E_{nX}^m$ is the kinetic energy for the component of degree $n$, order $m$, and direction $X$:
\begin{equation}
E_{nX}^m = \frac{\rho D}{4} \, n(n+1) \left(1+\delta_{m0} \right) \left( | \Phi_{nX}^m |^2 + | \Psi_{nX}^m |^2 \right) .
\label{EnmX}
\end{equation}
This formula is equivalent to Eq.~(24) of \citet{chen2014} taking into account the following differences.
First, they do not sum over both directions when $m=0$.
Second, their SH coefficients for $m>0$ are half the size of mine because they define their Fourier transform without factor $1/2$ (see Eq.~(\ref{FT})).

\begin{figure}
   \centering
   \includegraphics[width=7.3cm]{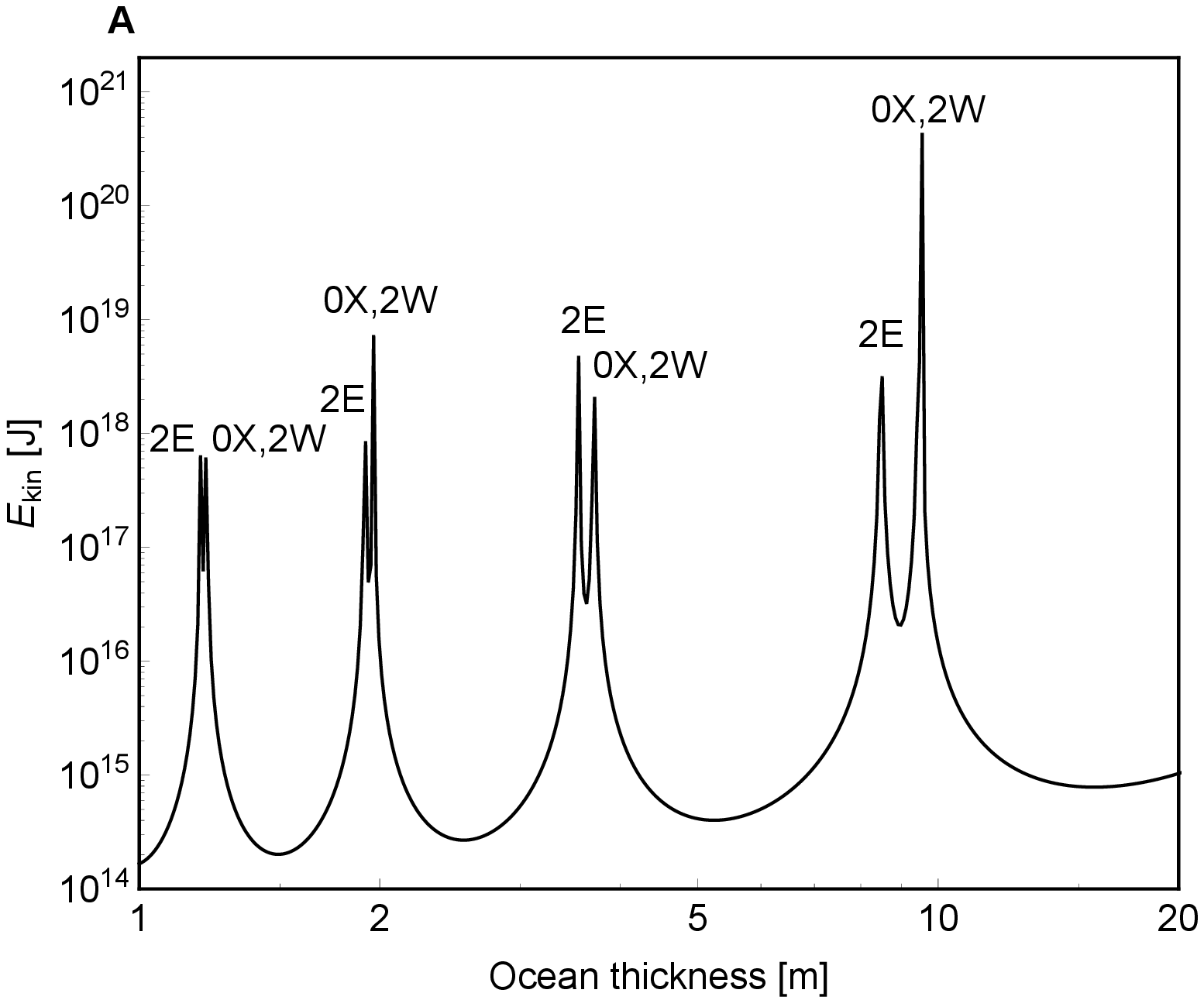}
    \includegraphics[width=7.3cm]{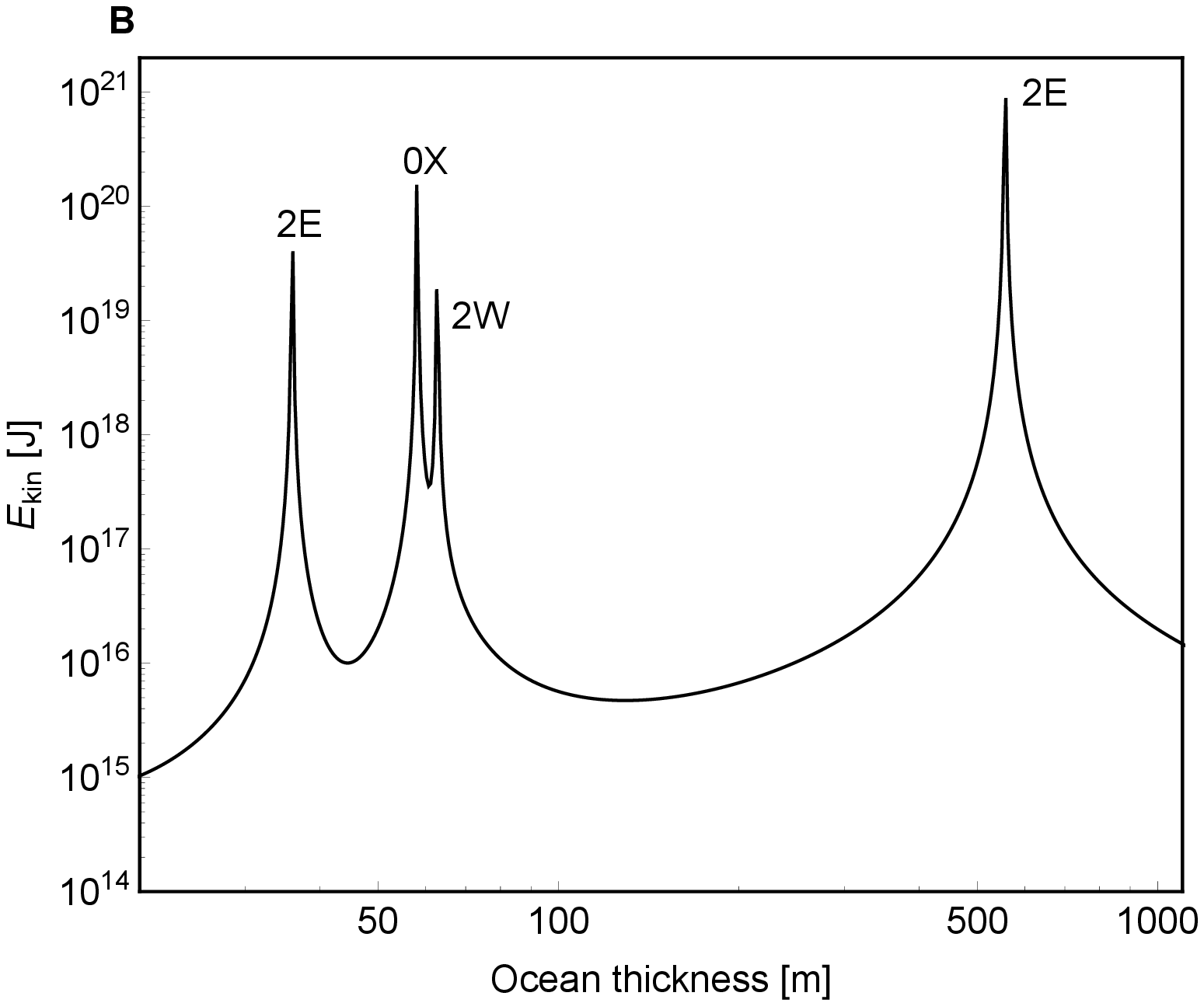}
   \caption[Average kinetic energy of eccentricity tides in a surface ocean]
   {Average kinetic energy of eccentricity tides in a surface ocean: (A) ocean depth between $1\rm\,m$ and $20\rm\,m$, (B) ocean depth between $20\rm\,m$ and $1000\rm\,m$.
      The ocean is inviscid.
   Resonances are labeled with $mX$ where $m$ is the order and $X$ is the direction (the resonances $0X$ and $2W$ are indistinguishable except the two largest ones).
}
   \label{FigEnergyEcc}
\end{figure}

\begin{figure}
   \centering
   \includegraphics[width=7.3cm]{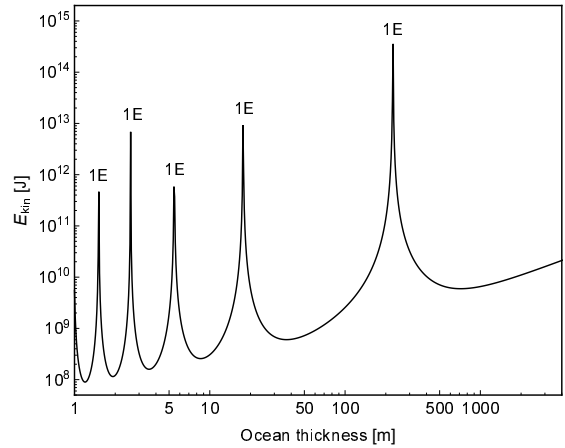}
   \caption[Average kinetic energy of obliquity tides in a surface ocean]
   {Average kinetic energy of obliquity tides in a surface ocean.
   The ocean is inviscid.
   All resonances are associated with eastward obliquity tides ($m=1$).
   The background trend is due to the toroidal solution for the westward obliquity tide.
}
   \label{FigEnergyObli}
\end{figure}

Fig.~\ref{FigEnergyEcc} shows the average kinetic energy stored in an inviscid surface ocean for eccentricity tides (crustal effects are addressed in Section~\ref{SpecificScaling}).
In this figure, each peak has been associated with a tidal wave of given order and direction by plotting separately the energy components $E_{nX}^m$ before adding them.
The largest resonances are well separated into modes $0X$, $2W$, and $2E$ (Fig.~\ref{FigEnergyEcc}B).
As the ocean gets shallower, resonances occur in tight groups and the modes $(0X,2W)$ become indistinguishable.
Fig.~\ref{FigEnergyObli} is similar to Fig.~\ref{FigEnergyEcc} but for obliquity tides.
In this case, the surface gravity modes are only due to eastward tides, whereas the westward tide does not depend on the ocean depth (Eq.~(\ref{toroidalsol})).
The background trend results from the linear dependence of the energy on the ocean depth $D$ (Eq.~(\ref{EnmX})).

\subsection{Resonant ocean depths}

It is not necessary to compute the full energy spectrum in order to identify resonances.
Resonant depths are the values of $D$ for which the LTE matrix (Eq.~(\ref{matrix})) is singular.
In order to obtain a standard eigenvalue problem (with eigenvalue $1/\epsilon_L$), I eliminate $\Psi_{nX}^m$ between the two equations in Eq.~(\ref{LTEsh}) in which the tidal forcing is set to zero:
\begin{equation}
- \frac{\lambda}{n(n+1)\beta_n} \Big( {q'}_{n-2,X}^m \, \Phi_{n-2,X}^m + {L'}_{nX}^m \, \Phi_{nX}^m + {p'}_{n+2,X}^m \, \Phi_{n+2,X}^m \Big) = \frac{1}{\epsilon_L} \, \Phi_{nX}^m \, .
\label{LTEeigen}
\end{equation}
These equations hold for $n\geq{}m$, it being understood that $\Phi_{m-2,X}^m=\Phi_{m-1,X}^m=0$.
The primed factors are defined by
\begin{eqnarray}
{L'}_{nX}^m &=& L_{nX}^m - \frac{p_{n+1}^m \, q_n^m}{L_{n+1,X}^m} - \frac{p_n^m \, q_{n-1}^m}{L_{n-1,X}^m} \, ,
\nonumber \\
\left( {p'}_{nX}^m , \, {q'}_{nX}^m \right) &=& \left( - \frac{p_{n-1}^m \, p_n^m}{L_{n-1,X}^m} \, , \, - \frac{q_n^m \, q_{n+1}^m}{L_{n+1,X}^m} \right) .
\end{eqnarray}
If written in terms of the ocean tide $\eta$ instead of $\Phi$, these equations are equivalent to Eq.~(43) of \citet{hough1898} for an inviscid surface ocean above a rigid mantle ($\alpha=\nu=0$ and $\beta_n=1-\xi_n$).

Similarly to Eq.~(\ref{LTEsh}), Eq.~(\ref{LTEeigen}) falls into independent systems: symmetric (if $n-m$ is even), and antisymmetric (if $n-m$ is odd).
After truncating each system to $n\leq{}N$ and writing it in matrix form, it becomes easy to compute the eigenvalues with mathematical software (I used both Mathematica and the LAPACK library in FORTRAN).
The eigenvalues for the symmetric (resp.\ antisymmetric) case are relevant to eccentricity (resp.\ obliquity) tides.
Since the eigenvalue problem is independent of the choice of the basis, the factors $(p_n^m,q_n^m)$ can be defined with either Eq.~(\ref{pnqnN}) or Eq.~(\ref{pnqn}).
The positive eigenvalues are denoted, in order of decreasing magnitude,
\begin{equation}
D^{(1)}_{mX} , \, D^{(2)}_{mX}, ...
\label{orderingindex}
\end{equation}
For the westward obliquity tide in the inviscid limit, eigenvalues must be computed by approaching the synchronous frequency from above (since $L_{1W}^1=0$).
However, these eigenvalues do not cause resonances, even in presence of viscosity, because of the existence of the toroidal solution specific to the westward obliquity tide (Eq.~(\ref{toroidalsol})).

\section{Crustal effects: inviscid-elastic response}
\label{InviscidElastic}

Before tackling dissipation, let me analyze the impact of the crust on the tides of an ocean with zero viscosity, sandwiched between an elastic mantle and an elastic crust: the \textit{inviscid-elastic response}.
I propose to look at the degree-two response, the positions of resonances, the energy spectrum, and the energy concentration.
To begin with, the degree-two response provides a connection with the Love number approach of the non-rotating model.
Next, I show that positions of resonances are related to the surface ocean solutions by an approximate scaling rule, which also reproduces the energy spectrum over a large range.
Finally, computing the energy concentration in the harmonic domain allows me to check the membrane assumption.

\subsection{Degree-two response}
\label{DegreeTwoResponse}

In the non-rotating model, the radial response of the body to tidal forcing of degree two is described by one Love number ($h_2^T$), which is degenerate in harmonic order and does not depend on the tidal direction (east or west).
Once rotation is taken into account, the response of the fluid differs according to $m$ and to the tidal direction (except if $m=0$).
Moreover, tidal forcing of degree two generates tides of other degrees, so that an infinite set of Love numbers becomes necessary.
The response is however negligible at most harmonic degrees.
With the exception of the westward obliquity tide, the dominant response, far from resonances, to a forcing of degree two and order $m$ is given by $\Phi_{2X}^m$ and the nearest-coupled toroidal components, which are $\Psi_{1X}^m$ (if $m<2$) and $\Psi_{3X}^m$ (see Fig.~1 in \citet{chen2014}).

The connection with the Love number approach becomes obvious if one works with the radial tide $\eta$ instead of the velocity potential $\Phi$.
Thus, let us analyze the dominant radial response, component by component, by defining parameters $z_{2X}^m$ analogous to the admittance $Z_n$ (Eq.~(\ref{Zn})):
\begin{equation}
\eta_{2X}^m = z_{2X}^m \, \frac{U_{2X}^m}{g} \, .
\label{defz}
\end{equation}
In the limit of equilibrium tide, all $z_{2X}^m$ (except $z_{2W}^1$) tend to the equilibrium tide admittance $Z_2$ (Eq.~(\ref{Zn})), which is equal to the radial Love number at the surface of the satellite if the mantle is infinitely rigid.
The above definition is not useful for the westward obliquity tide in the inviscid limit because fluid motion can occur without surface deformation (Eq.~(\ref{toroidalsol})).

\begin{figure}
   \centering
   \includegraphics[width=16cm]{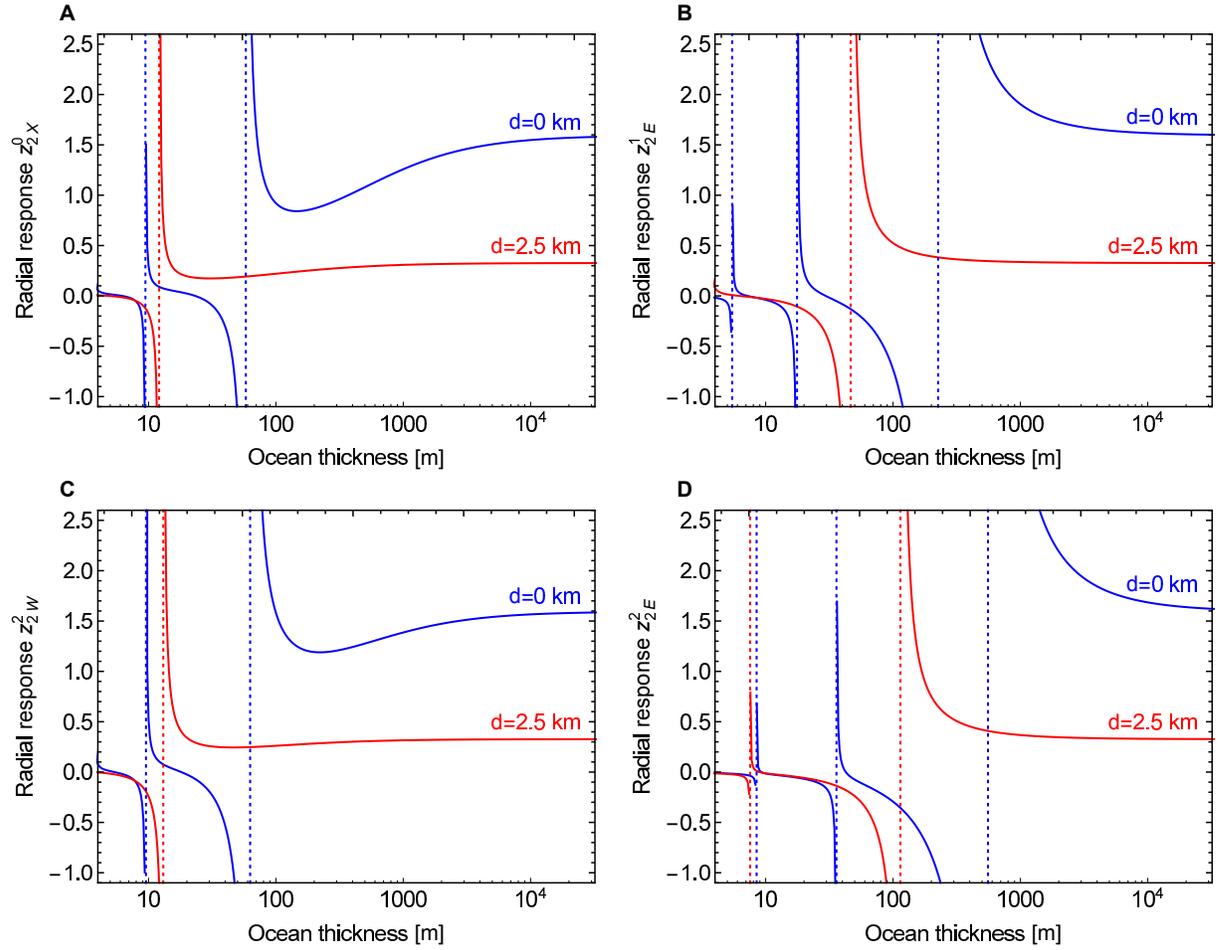}
   \caption[Degree-two radial response to tidal forcing of given order and direction]{Degree-two radial response to tidal forcing of given order and direction: (A) $z_{2X}^0$ with $X=E$ or $W$, (B) $z_{2E}^1$, (C) $z_{2W}^2$, (D) $z_{2E}^2$.
   In each panel, two cases are considered: surface ocean ($d=0\rm\,km$) or subsurface ocean under a thin elastic crust ($d=2.5\rm\,km\approx{}R/100$).
   Vertical dotted lines show the positions of the resonances.
   The ocean is inviscid.
}
   \label{Figh2}
\end{figure}

Fig.~\ref{Figh2} shows the degree-two radial response, component by component.
Two cases are presented: surface ocean or subsurface ocean with a $2.5\rm\,km$-thick crust.
If the ocean is thicker than a few kilometers, the response tends to the equilibrium admittance (Eq.~(\ref{deepocean})):
$Z_2=1.60$ if $d=0\rm\,km$ and $Z_2=0.32$ if $d=2.5\rm\,km$.
The crust actually reduces the equilibrium admittance by the factor $(1-\xi_2)/\beta_2\approx1/4.88$ ($\upsilon_2$ is nearly constant).
Besides decreasing the radial response, the crust shifts the largest resonance $D^{(1)}_{mX}$ to the left by the same factor $4.88$ whatever the order $m$ and the direction $X$.
This factor quantifying the amplitude reduction and the resonance shift coincides with the reduction factor of the slow-rotation limit (Eq.~(\ref{Dslowrot})), and thus agrees with the predictions of the non-rotating model (analytical \citep{beuthe2015} and numerical \citep{kamata2015}), although the positions of the resonances are different.

The responses can be grouped into westward-type (panels A and C)  and eastward-type (panels B and D), although the $m=0$ case is not properly directional.
Resonances occur at much smaller ocean depths in the former category than in the latter.
Besides, the westward-type response shows a dip just after the largest resonance which is not seen in the eastward-type response.

\subsection{Universal scaling rule}
\label{UniversalScaling}

In Section~\ref{DegreeTwoResponse}, we saw that the crust reduces the magnitude of the radial tide in the deep-ocean limit by the factor $(1-\xi_2)/\beta_2$.
It also shifts by the same factor the largest resonance to a smaller ocean depth.
Let us try to understand why this is so.

\begin{figure}
   \centering
   \includegraphics[width=7.5cm]{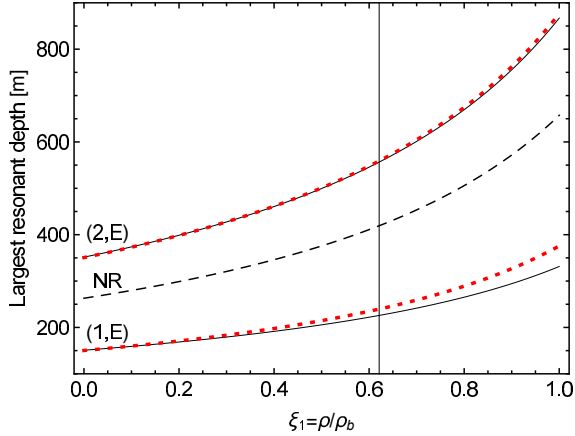}
   \caption[Largest resonant depths for a surface ocean]{Largest resonant depths for a surface ocean as a function of the ocean-to-bulk density ratio.
   Solid curves show the largest resonant ocean depth $D^{(1)}_{mX}$ for the eastward tidal waves (labeled by $(m,X)$).
   Dotted curves result from the approximation (\ref{DresApprox1}).
   The dashed curve is the prediction of the non-rotating (NR) model (Eq.~(\ref{Dresthin})).
   The vertical axis indicates the bulk-to-ocean density ratio of Table~\ref{TableParamInterior}.
   The ocean is inviscid and the mantle is infinitely rigid.
}
   \label{FigEigenvalSurface}
\end{figure}

The largest eigenvalue for all tidal waves does not depend much on degrees higher than two: truncating the system (\ref{LTEeigen}) to $N=2$ yields reasonable estimates of the two largest resonant depths ($D^{(1)}_{1E}$ and $D^{(1)}_{2E}$).
Increasing the size of the eigensystem leads to the determination of smaller eigenvalues, but does not change much the largest ones.
Thus,  as a first approximation, the factors $\beta_n$ can be replaced by $\beta_2$ in the LTE eigenvalue equation:
\begin{equation}
- \frac{\lambda}{n(n+1)} \, \Big( ... \Big) = \frac{1}{\epsilon'_L} \, \Phi_{nX}^m \, ,
\label{LTEeigenApprox1}
\end{equation}
where the quantity within brackets is the same as in Eq.~(\ref{LTEeigen}) and $\epsilon'_L$ is a rescaled Lamb factor defined by $\epsilon'_L=\epsilon_L/\beta_2$.
Eq.~(\ref{LTEeigenApprox1}) corresponds to the LTE eigenvalue problem for a surface ocean without self-attraction.
As the left-hand side of this equation does not depend on the ocean depth, neither does the eigenvalue $1/\epsilon'_L$.
In this approximation, the factor $\epsilon_L$ is thus proportional to $\beta_2$, meaning that $D^{(1)}_{mE}$ is inversely proportional to $\beta_2$.
Thus $D^{(1)}_{mE}$ is given by a scaling rule similar to the one of the non-rotating model (Eq.~(\ref{Dresthin})):
\begin{equation}
D^{(1)}_{mE} \approx f \, \frac{q_\omega R}{6} \, \frac{1}{\beta_2} \, ,
\label{DresApprox1}
\end{equation}
where $f$ is a nondimensional factor specific to each tidal wave but otherwise independent of the physical parameters of the satellite: $f\approx4/7$ if $m=1$ and $f\approx4/3$ if $m=2$.
This rough approximation is universal, in the sense that it yields the largest resonance ocean depths for all synchronously rotating satellites (see Table~3 of \citet{matsuyama2014} for the case of a surface ocean).
Fig.~\ref{FigEigenvalSurface} shows the two largest resonant depths for a surface ocean as a function of the ocean-to-bulk density ratio.
For Enceladus, the error is less than $1\%$ and close to $6\%$ for eccentricity and obliquity tides, respectively.
For satellites whose mean density is close to water density ($\xi_1\approx1$), the error reaches 1.5\% and 14\% for eccentricity and obliquity tides, respectively.

\subsection{Scaling rule for Enceladus}
\label{SpecificScaling}

\begin{figure}
   \centering
   \includegraphics[width=7.5cm]{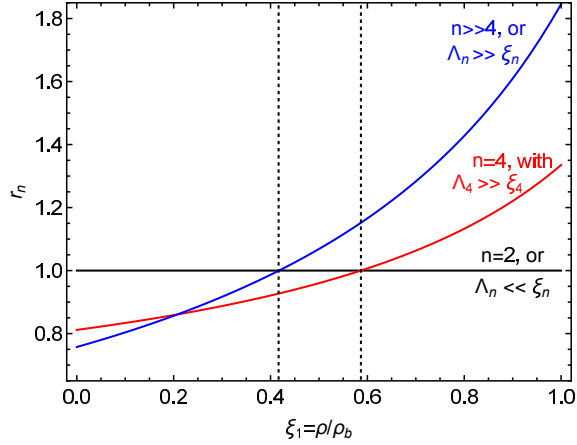}
   \caption[Deviation from scaling rule]{Deviation from scaling rule: factor $r_n$ as a function of the ocean-to-bulk density ratio $\xi_1$ (Eq.~(\ref{rn})).
   $r_n$ tends to the horizontal line $r_2=1$ if the crust is so thin that its effect is negligible: $\Lambda_n\ll\xi_n$.
   The two increasing curves show $r_n$ if either $n=4$ or $n\gg4$ in the limit of a rigid crust.
   The vertical dotted lines indicate the corresponding $\xi_1$-values for which either $r_4$ (rightmost line) or $r_n$ with $n\gg4$ (leftmost line) does not depend on crust thickness.
   The ocean is inviscid and the crust is elastic.
}
   \label{FigEigenApprox}
\end{figure}

Approximating all $\beta_n$ by $\beta_2$ in Eq.~(\ref{LTEeigenApprox1}) is straightforward but brutal, and is not very accurate for eigenvalues other than the largest one.
Instead, let us rewrite the LTE eigenvalue system (\ref{LTEeigen}) so that self-attraction explicitly appears:
\begin{equation}
- \frac{\lambda}{n(n+1)} \, \frac{r_n}{1-\xi_n} \, \Big( ... \Big) =  \frac{1}{\epsilon''_L} \, \Phi_{nX}^m \, ,
\label{LTEeigenApprox2}
\end{equation}
where the quantity within brackets is the same as in Eq.~(\ref{LTEeigen}) and $\epsilon''_L$ is a rescaled Lamb factor defined by $\epsilon''_L=\epsilon_L(1-\xi_2)/\beta_2$.
The factor $r_n$ in the left-hand side of Eq.~(\ref{LTEeigenApprox2}) is given by
\begin{equation}
r_n = \frac{\beta_2}{\beta_n} \, \frac{1-\xi_n}{1-\xi_2} \, ,
\label{rn}
\end{equation}
and verifies $r_2=1$.
If $r_n\approx1$ for all $n$, Eq.~(\ref{LTEeigenApprox2}) is identical to the LTE eigenvalue problem for a surface ocean including self-attraction.
Fig.~\ref{FigEigenApprox} shows that $r_4=1$ if $\xi_1\approx0.59$, a value which is very close to Enceladus' ocean-to-bulk density ratio for a pure water ocean ($\xi_1=0.62$).
Furthermore, $r_4$ remains within 20\% of unity in most of the density range.
That $r_2$ and $r_4$ are close to unity guarantees that $r_n\approx1$ is a good approximation for at least the few largest eigenvalues.
If $n\gg4$, $r_n$ remains within 20\% of unity for $0.10<\xi_1<0.63$.

Following the above, we consider Eq.~(\ref{LTEeigenApprox2}) under the approximation $r_n\approx1$.
As the eigenvalues $1/\epsilon_L''$ do not depend on ocean depth, the corresponding resonant depths are inversely proportional to $\beta_2/(1-\xi_2)$.
Thus the resonant depths are given by a scaling rule similar to Eq.~(\ref{DresApprox1}):
\begin{equation}
D^{(i)}_{mX} \approx f^{(i)}_{mX} \, \frac{q_\omega R}{6} \, \frac{1}{\beta_2} \, .
\label{DresApprox2}
\end{equation}
The nondimensional factors $f^{(i)}_{mX}$ are adjusted so as to yield the correct resonant depths $D^{(1)}_{mX}$ for a surface ocean including self-attraction.
Thus they slightly vary with the ocean-to-bulk density ratio.
Table~\ref{TableEigenApprox} gives the numerical values of these factors for the first three series of resonant depths ($i=1,2,3$), assuming that $\xi_1=0.621$ for Enceladus.
Observe that $f^{(1)}_{2E}=1.328$ differs by less than 1\% from the universal factor $f=4/3$ appearing in Eq.~(\ref{DresApprox1}) (no self-attraction).
For obliquity tides, the difference reaches 6\%.
Though most accurate if the ocean-to-bulk density ratio is close to $\xi_1\approx0.6$, the above scaling rule can also be applied to synchronously rotating satellites with other density ratios, from Europa ($\xi_1\approx1/3$) to Tethys ($\xi_1\approx1$).

\begin{table}[ht]\centering
\ra{1.3}
\small
\caption[Factors $f^{(i)}_{mX}$ parameterizing the resonant depths]
{Factors $f^{(i)}_{mX}$ parameterizing the resonant depths in Eq.~(\ref{DresApprox2}), to $(5-i)$ significant figures, assuming that the ocean-to-bulk density ratio is $\xi_1=0.621$.}
\begin{tabular}{@{}ccccc@{}}
\hline
$i$ & \multicolumn{4}{c}{$(m,X)$}
\\
 &  $(0,W)$ & $(1,E)$ &  $(2,W)$ &  $(2,E)$ \\
\hline
1  &  0.1386 &  0.5389 & 0.1498 & 1.328
\vspace{1mm}\\
2 &  0.0225 & 0.0421 & 0.0228 & 0.0858
\vspace{1mm}\\
3 &  0.0088 & 0.013 & 0.0089 & 0.020
\\
\hline
\end{tabular}
\label{TableEigenApprox}
\end{table}%

\begin{figure}
   \centering
   \includegraphics[width=7.3cm]{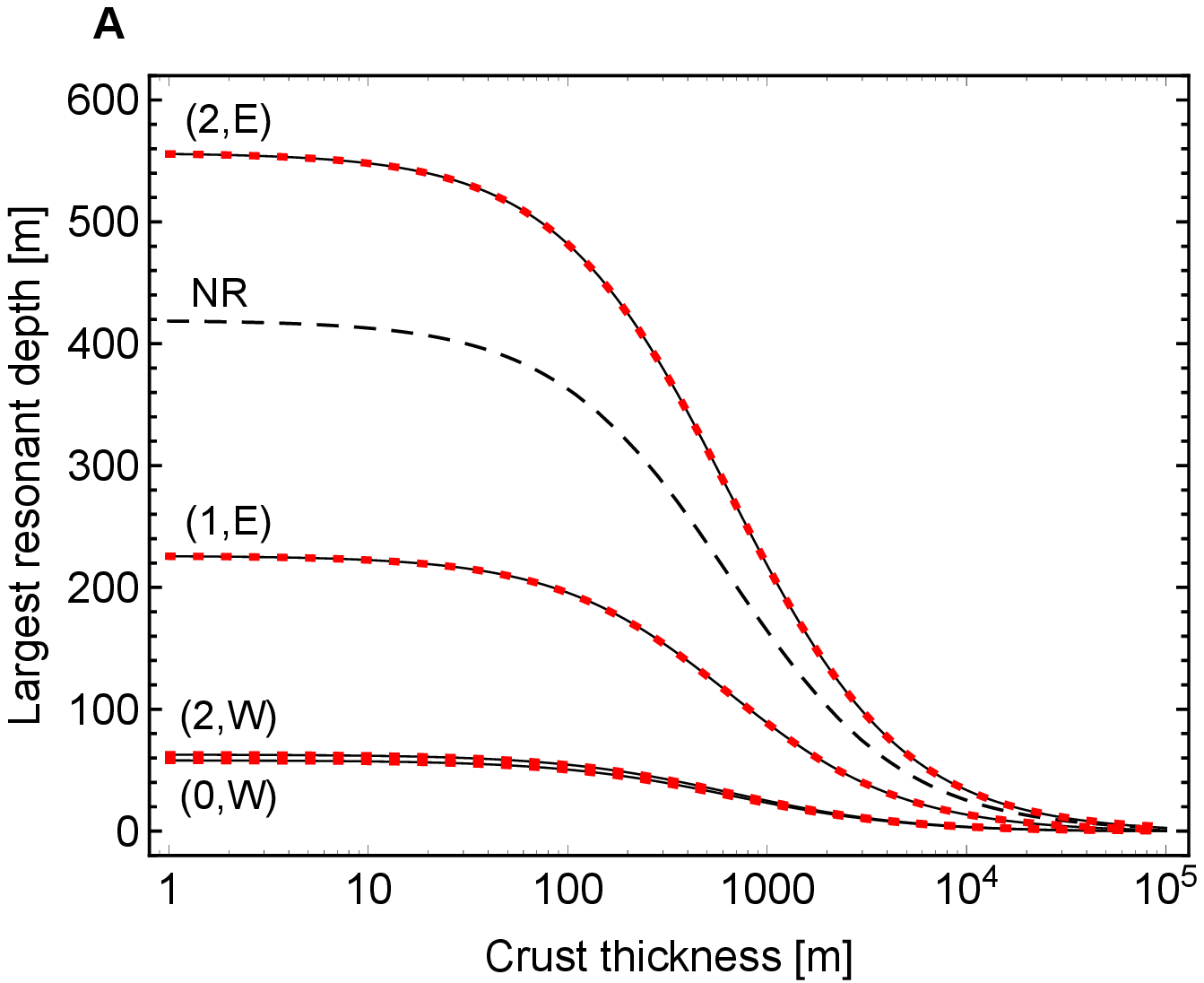}
      \includegraphics[width=7.3cm]{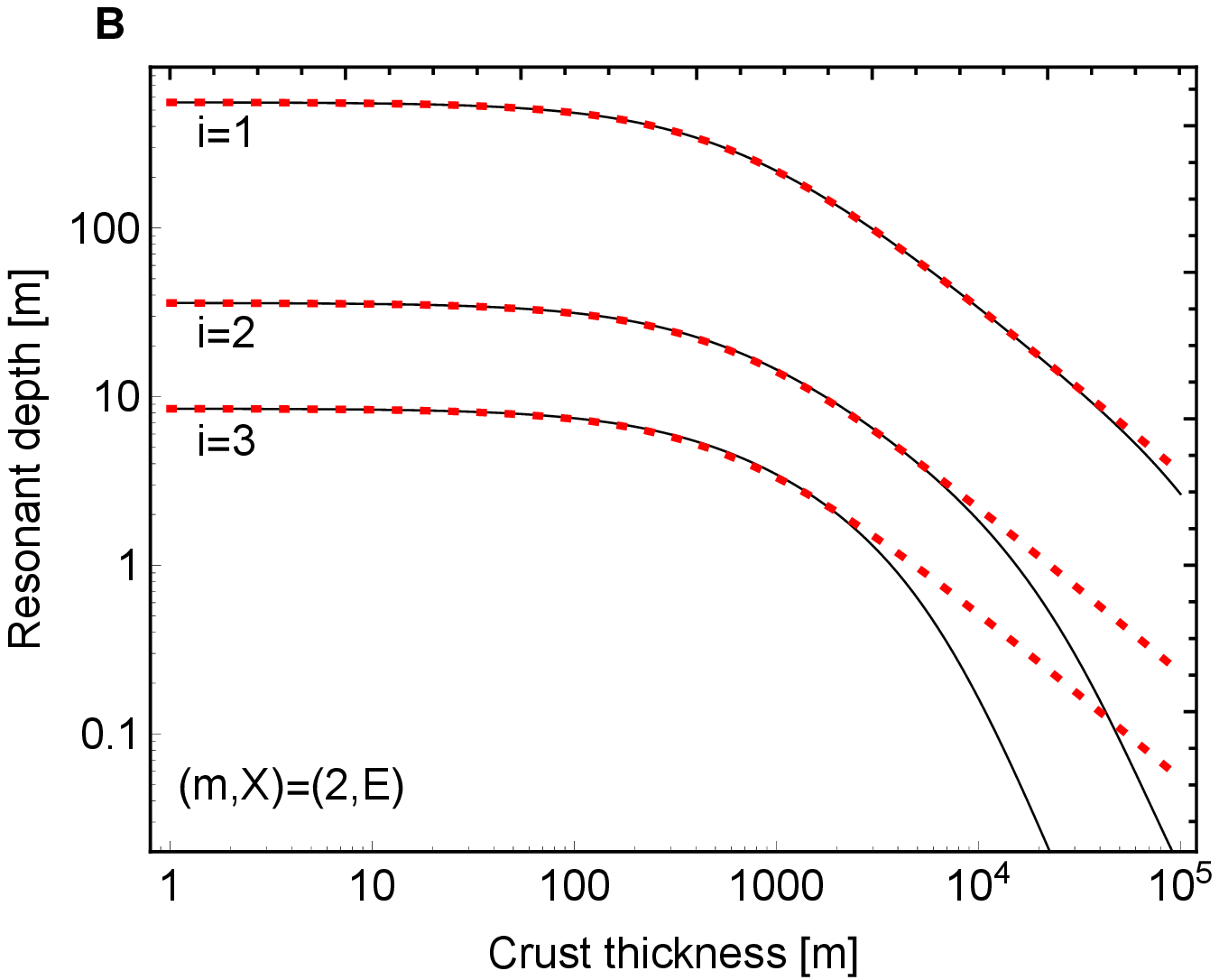}
   \caption[Resonant ocean depth as a function of crust thickness]{Resonant ocean depth as a function of crust thickness: (A) largest resonant depth $D^{(1)}_{mX}$ for each mode, (B) three largest resonant depths for the $m=2$ eastward mode.
   Solid curves are the results of the full LTE; dotted curves are estimates from the scaling rule (Eq.~(\ref{DresApprox2})).
   In panel~A, curves are labeled by the tidal waves indices $(m,X)$;
   the dashed curve is the prediction of the non-rotating (NR) model (Eq.~(\ref{Dresthin})).
   In panel~B, curves are labeled by the ordering index $i$ as in (\ref{orderingindex}).
   The ocean is inviscid and the crust is elastic.
}
   \label{FigEigenValCrust}
\end{figure}

Fig.~\ref{FigEigenValCrust}A shows how crust thickness affects $D^{(1)}_{mX}$ for each tidal wave $(m,X)$.
Solid curves are computed by solving the system (\ref{LTEeigen}) truncated to $N=100$, whereas dashed curves show the estimates given by Eq.~(\ref{DresApprox2}).
If $d\lesssim100\rm\,m$, $D^{(1)}_{mX}$ is close to the value for a surface ocean, whereas if $d\gtrsim100\rm\,m$, $D^{(1)}_{mX}$ quickly decreases to very small values (one order of magnitude drop between $d\approx100\rm\,m$ and $d\approx5\rm\,km$).
Fig.~\ref{FigEigenValCrust}B shows the three largest resonant depths for the $m=2$ eastward component.
The estimate from the approximate scaling rule deviates from the exact result when the resonant depth decreases to less than a few meters.

\begin{figure}
   \centering
     \includegraphics[width=7.3cm]{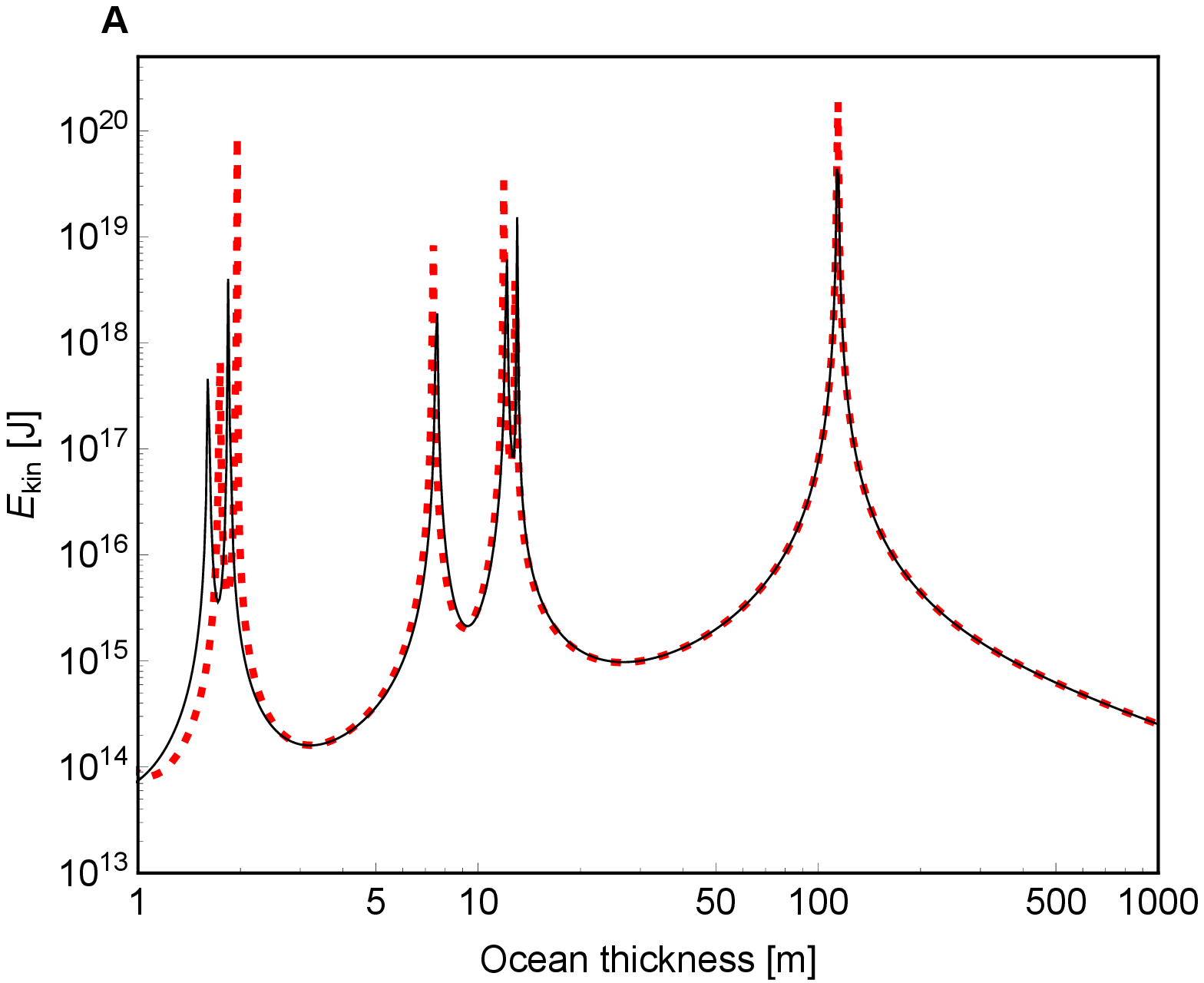}
    \includegraphics[width=7.3cm]{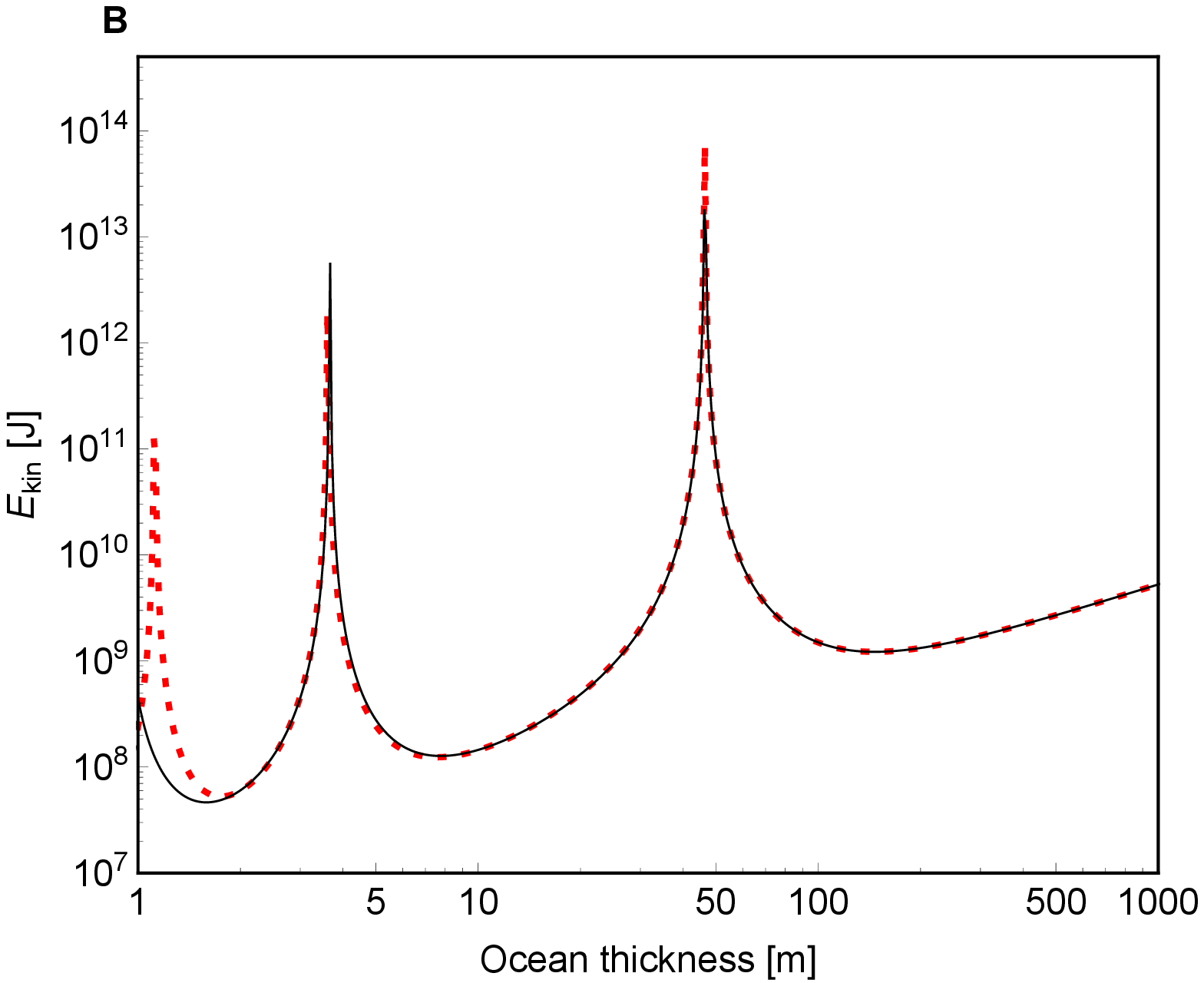}
   \caption[Average kinetic energy of subsurface ocean tides]{
   Average kinetic energy of subsurface ocean tides: (A) eccentricity tides, (B) obliquity tides.
   The ocean is inviscid.
   The crust is elastic and 2.5~km thick.
   In each panel, the solid curve is the solution of subsurface LTE, while the dotted curve shows the surface ocean solution rescaled (on $x$ and $y$ axes) by the factor $(1-\xi_2)/\beta_2\approx1/4.88$.
   Panels~A and B should be compared to Figs.~\ref{FigEnergyEcc} and \ref{FigEnergyObli} showing the energy spectra of eccentricity and obliquity tides in a surface ocean.
   }
   \label{EkinSubsurface}
\end{figure}

The same type of reasoning can be applied to the full LTE system (Eq.~(\ref{LTEsh})): the velocity potentials $(\Phi,\Psi)$ for a subsurface ocean are approximately given by the potentials for a surface ocean, but shifted to smaller ocean depths by the factor $(1-\xi_2)/\beta_2$.
According to Eq.~(\ref{EnmX}), the energy spectrum is shifted in the same way; it is also reduced in magnitude by the same factor because the energy is proportional to ocean depth.
Fig.~\ref{EkinSubsurface} shows that the energy spectrum for $d=2.5\rm\,km$ is very well approximated by the surface ocean solution (including self-attraction) shifted and reduced in magnitude by the factor $(1-\xi_2)/\beta_2\approx1/4.88$.

\subsection{Energy concentration}
\label{EnergyConcentration}

In order to check that the crust deforms as a membrane (Section~\ref{SubsurfaceOcean}), we must determine the harmonic degree range in which most of the energy is concentrated.
Far from the resonances, most of the energy resides in $\Psi_{1X}^m$, $\Phi_{2X}^m$, and $\Psi_{3X}^m$ (see Section~\ref{DegreeTwoResponse}).
Let us quantify the energy concentration in degrees $n\leq{}N'$ for a given mode $(m,X)$ by
\begin{equation}
C_{mX}(N') = \Big( \sum_{n=m}^{N'} E_{nX}^m \Big) /  \Big( \sum_{n=m}^N E_{nX}^m \Big) \, ,
\label{CXm}
\end{equation}
where $E_{nX}^m$ is defined by Eq.~(\ref{EnmX}).

At smaller ocean depths, where resonances occur, the energy is distributed in a wider degree range.
For a given mode, the degree threshold is defined as the degree $N'$ such that the energy concentration is higher than some threshold, say 0.99:
\begin{equation}
C_{mX}(N') > 0.99 \, .
\label{DegreeThreshold}
\end{equation}

\begin{figure}
   \centering
   \includegraphics[width=7.3cm]{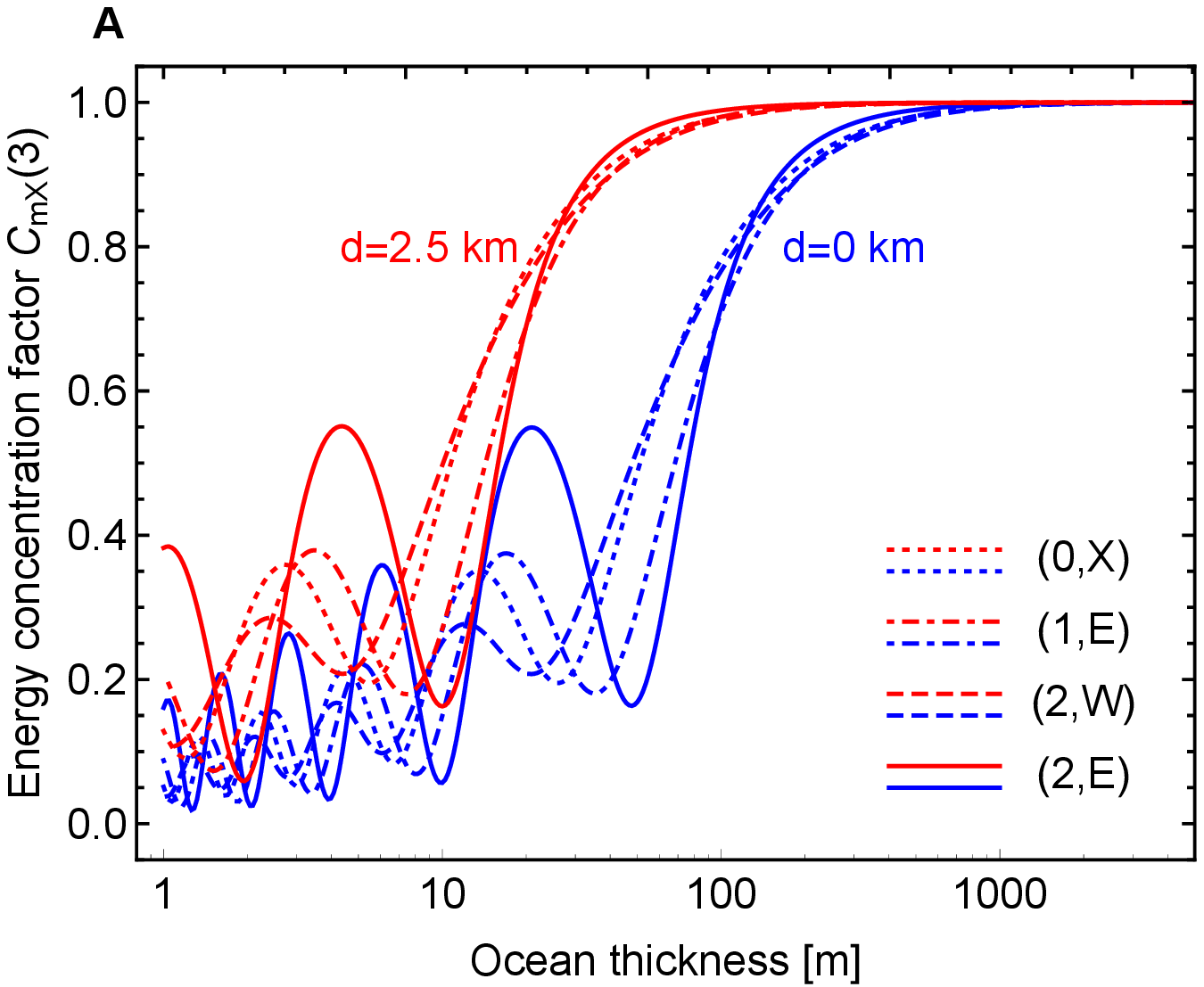}
   \includegraphics[width=7.3cm]{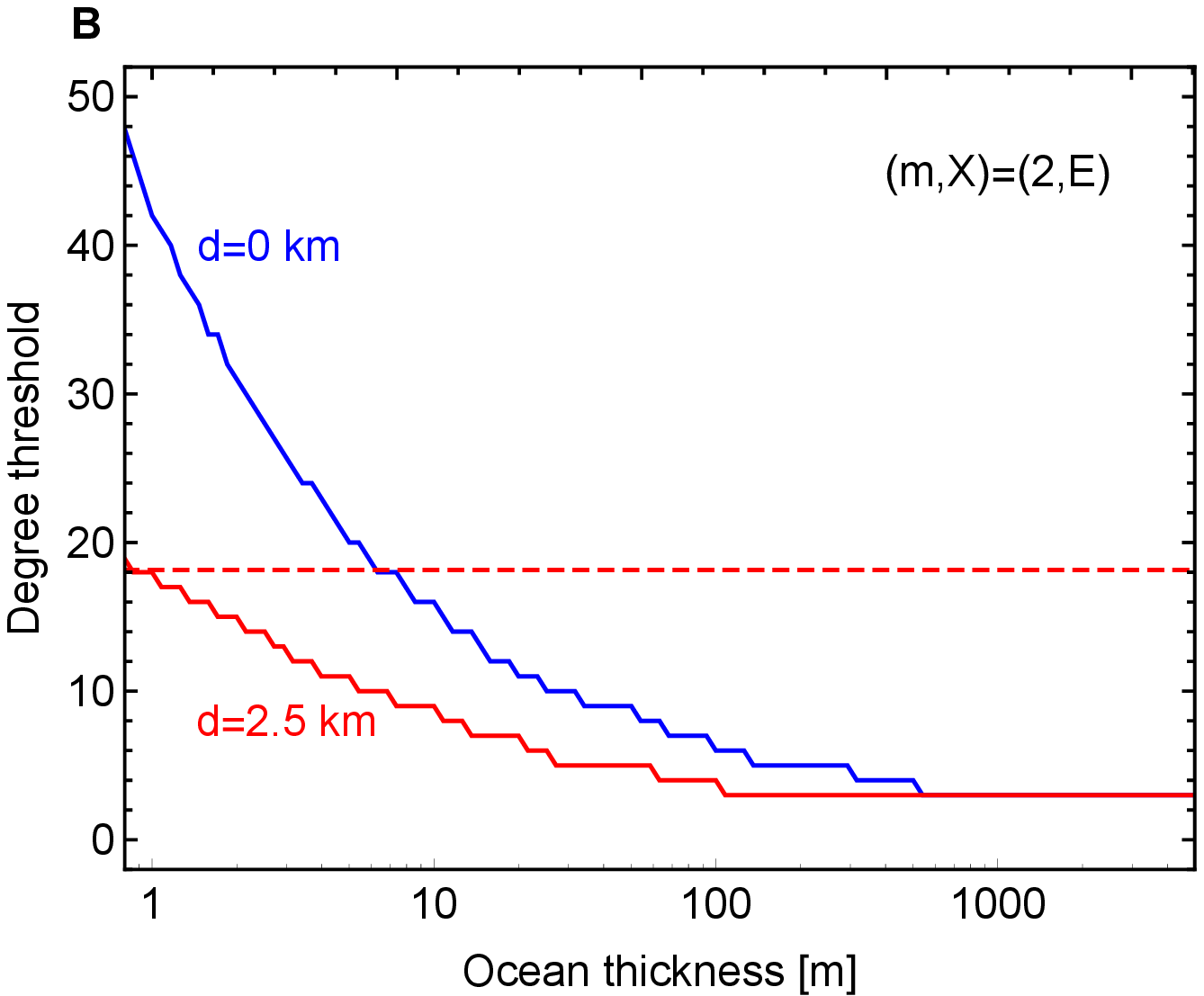}
   \caption[Energy concentration]{Energy concentration:
   (A) energy concentration factor in the first three harmonic degrees (Eq.~(\ref{CXm}) with $N'=3$), (B) degree threshold (Eq.~(\ref{DegreeThreshold})).
   In each panel, two cases are considered: surface ocean ($d=0\rm\,km$) or subsurface ocean under a thin elastic crust ($d=2.5\rm\,km\approx{}R/100$).
   In panel~A, the different types of dashing correspond to the different tidal waves $(m,X)$.
   In panel~B, the step functions show the degree threshold $N'$ for the $m=2$ eastward tidal wave.
   The horizontal line is the membrane-bending threshold if $d=2.5\rm\,km$ (Eq.~(\ref{BendingThreshold})).
}
   \label{FigConc}
\end{figure}

For a surface ocean, Fig.~\ref{FigConc}A shows that the kinetic energy of Enceladus' ocean tides is indeed concentrated in harmonic degrees $n\leq3$ if the ocean is deeper than about 100 meters.
Adding a crust shifts these curves, as well as the resonances, to smaller ocean depths by a factor $(1-\xi_2)/\beta_2$.
Fig.~\ref{FigConc}B shows the degree threshold in function of the ocean depth.
Bending effects become significant when the degree threshold is equal to the bending threshold (Eq.~(\ref{BendingThreshold})).
If $d=2.5\rm\,km$, this occurs for an ocean depth of about one meter.
If the crust thickness increases, the bending threshold decreases but the degree threshold is shifted to smaller ocean depths.
Therefore bending effects are always negligible if the ocean is deeper than a few meters.

\section{Dissipation}
\label{Dissipation}

Tides dissipate energy in the ocean, crust, and mantle.
I start by giving the formulas for the dissipation rate in each layer and in the whole body in terms of the velocity potentials $(\Phi,\Psi)$.
Next, I analyze oceanic dissipation for eccentricity and obliquity tides if the crust is elastic.
Finally, I study simultaneous dissipation in the ocean and crust.

\subsection{How-to}
\label{DissipationLayer}

\subsubsection{Ocean}
\label{DissipationO}

The global dissipation rate in the ocean, averaged over an orbital period, is equal to the power developed by the dissipative stress ${\cal F}(\mathbf{u})$ (Eq.~(\ref{DissipTerm})):
\begin{equation}
\dot E_{O} = - \frac{1}{T} \int_T dt \int_{V_O} dV \Big(  \rho \, {\cal F}(\mathbf{u}) \cdot \mathbf{u} \Big) .
\label{EdissO1}
\end{equation}
For linear top and bottom drag (${\cal F}(\mathbf{u})=-\alpha\mathbf{u}$), the dissipation rate is equal to the kinetic energy (Eq.~(\ref{Ekin})) multiplied by $2\alpha$ \citep{tyler2011}.
For Navier-Stokes viscosity, the vector Laplacian in Eq.~(\ref{DissipTerm}) introduces a supplementary factor of $n(n+1)$ \citep{chen2014}.
The result can be written as follows ($Q_n$ is the tidal quality factor, see Eq.~(\ref{qualityfactor})):
\begin{equation}
\dot E_{O} = \frac{\rho D \Omega}{4} \sum_{m=0}^2 \left( 1+\delta_{m0} \right) \sum_{n=m}^N \frac{n(n+1)}{Q_n} \sum_{X=E,W} \Big( \, | \Phi_{nX}^m |^2 + | \Psi_{nX}^m |^2 \Big) \, .
\label{EdissO2}
\end{equation}

\subsubsection{Crust}
\label{DissipationC}

The global dissipation rate in the crust, averaged over an orbital period, can be computed with two equivalent methods: micro or macro \citep{beuthe2014}.
In the micro approach, it results from the stress times the strain rate integrated over the volume of the shell.
In the macro approach, it is equal to the average work done by the bottom load on the crust integrated over the surface (the tidal potential does no direct work on the shell because it is massless in the membrane approach).
Choosing the latter method, I can write the global dissipation rate in the crust as
\begin{eqnarray}
\dot E_{C} &=& \frac{1}{T} \int_T dt \int_S dS \left( q \, \dot \eta^{top} \right)
\nonumber \\
&=& \frac{\Omega}{2} \sum_n \int_S dS \, Im \left( q_n \, {\eta_n^{top}}^* \right) ,
\label{EdissC1}
\end{eqnarray}
where $q$ is the bottom load and $\dot\eta^{top}$ is the radial velocity of the crust.
Substituting Eq.~(\ref{hookeM}) into Eq.~(\ref{EdissC1}) and expanding into E-W spherical harmonic components yields
\begin{equation}
\dot E_{C} = \frac{\rho{}gR^2\Omega}{2} \sum_{m=0}^2 \left(1+\delta_{m0} \right) \sum_{n=m}^N Im(\Lambda_n) \sum_{X=E,W} | (\eta^{top})_{nX}^{m} |^2 \, .
\label{EdissC2}
\end{equation}
In the general case of a nonrigid mantle, $\eta^{top}$ is related to $\eta$ by Eq.~(\ref{zetatop}), which is in turn related to $\Phi$ by Eq.~(\ref{zetanm}).
If the mantle is rigid, $\eta^{top}=\eta$ so that the crustal dissipation rate reads
\begin{equation}
\dot E_{C}^{rigid} = \frac{2\rho D \Omega}{\epsilon_L} \sum_{m=0}^2 \left(1+\delta_{m0} \right) \sum_{n=m}^N n^2(n+1)^2 \, Im(\Lambda_n) \sum_{X=E,W} | \Phi_{nX}^{m} |^2 \, .
\label{EdissC3}
\end{equation}
In this paper, depth-dependent rheology is not included in the bending component of $\Lambda_n$ (see Appendix~\ref{AppendixMembrane}), so that I always assume that $\Lambda_n=\Lambda^M_n$ when computing crustal dissipation.
This approximation is good if the ocean is deeper than a few meters (see Section~\ref{EnergyConcentration}).

\subsubsection{Mantle}
\label{DissipationM}

Suppose that the mantle is incompressible and of uniform density.
The mantle can have a depth-dependent rheology, or include a liquid core with the same density.
In that case, I show in Appendix~\ref{AppendixMantleDissipation} that the global dissipation rate in the mantle is given by
\begin{equation}
\dot E_M = \Omega R \sum_{m=0}^2 \left( 1+\delta_{m0} \right) \sum_{n=m}^N \frac{n-1}{2n+1} \, Im \left(\mu_m \, S_n \right) \sum_{X=E,W} \left| (\eta^{bot})_{nX}^m \right|^2 ,
\label{EM}
\end{equation}
where $\mu_m$ is the shear modulus of the top of the mantle and $S_n$ is a function depending on the structure of the core-mantle system.
If the mantle is homogeneous and there is no core, $S_n=(2n^2+4n+3)/n$ (Eq.~(\ref{Sn})).
The displacement of the mantle-ocean boundary $\eta^{bot}$ is related to the ocean tide $\eta$ by Eq.~(\ref{zetabot}), which is in turn related to $\Phi$ by Eq.~(\ref{zetanm}).
If the mantle is rigid, $\eta^{bot}=0$ so that $\dot E_M=0$.

\subsubsection{Whole body}
\label{TotalDissipation}

The formula for the dissipation rate in the ocean (Eq.~(\ref{EdissO2})) depends on unknown parameters, such as the ocean depth and the viscosity.
For similar reasons, the dissipation rate in Earth's oceans is determined, in an indirect way, from the global planetary dissipation rate.
This quantity is proportional to the phase shift between the induced and external tidal potentials which is observable because it affects satellite orbits \citep{egbert2001}.
For icy satellites, such measurements are not yet available; the heat output of Enceladus' south polar terrain has been measured \citep{howett2011}, but it is not clear how it relates to present-day tidal dissipation.
That being said, the global planetary dissipation rate is an observable quantity which serves as a global constraint on the various contributions.

In practice, one often constructs an energy balance equation with the LTE as building blocks (e.g.\ \citet{hendershott1972,egbert2001,tyler2011}).
It is difficult, however, to give a physical meaning to the various terms when ocean, crust and mantle interact.
I will rather follow the approach of \citet{zschau1978} and \citet{platzman1984} who compute the global dissipation rate from the work done by the external tidal potential.
The only difference is that \citet{platzman1984} treats the ocean and the atmosphere as two mass loads, whereas I model the ocean as a mass load and the crust as a pressure load (Section~\ref{Nonrigidmantle}).
As before, I approximate the external tidal potential by its degree-two component.
The dissipation rate integrated over the whole satellite and averaged over an orbital period is given by Eq.~(7) of \citet{platzman1984}:
\begin{equation}
\dot E = \frac{\Omega}{2} \, \frac{5}{4\pi{}GR} \, \int_S dS \, Im \left( U_2^T \, {U_2'}^* \right) ,
\label{Etot1}
\end{equation}
where $U_2'$ is the induced (or secondary) potential due to the deformation of the body,
\begin{equation}
U_2' = \Gamma_2 - U_2^T \, .
\end{equation}
In Appendix~\ref{AppendixTotalDissipation}, I transform this formula into an explicit function of the velocity potential $\Phi$ for the ocean tide (Eq.~(\ref{Ediss3})).
I consider here two special cases: elastic mantle/elastic crust, or rigid mantle/viscoelastic crust.
In these two cases, Eq.~(\ref{Ediss3}) reduces to
\begin{equation}
\dot E = -3\rho D \left( \gamma_2^T + \delta\gamma_2^T \right) \sum_{m=0}^2 \left( 1+\delta_{m0} \right) \sum_{X=E,W} Im \left( \Phi_{2X}^m \right) U_{2X}^{m} \, ,
\label{Ediss4}
\end{equation}
where I assumed that $U_{2X}^{m}$ is real (as in Table~\ref{TableU}).
If the mantle is infinitely rigid, $\gamma_2^T=1$ and $\delta\gamma_2^T=0$, in which case Eq.~(\ref{Ediss4}) is equivalent to the formula for global dissipation given by \citet{chen2014} for a surface ocean (see their Eq.~(28); $\Phi$ differs by a factor $i$ and the tidal potential is of opposite sign; other differences in normalization are noted after my Eq.~(\ref{EnmX})).

\subsection{Oceanic dissipation}
\label{OceanicDissipation}

\subsubsection{All tides except WOT}

For simplicity, suppose that oceanic dissipation results from linear top and bottom drag.
In that case, the quality factor $Q_n$ does not depend on $n$ so that the oceanic dissipation rate (Eq.~(\ref{EdissO2})) is proportional to the average kinetic energy stored in the ocean (Eq.~(\ref{Ekin})).
If ocean viscosity is low, the kinetic energy is close to what is shown in Fig.~\ref{EkinSubsurface}, and the dissipation rate should also be similar, except for a change in magnitude by the factor $2\alpha=\Omega/Q_n$.

\begin{figure}
   \centering
   \includegraphics[width=7.3cm]{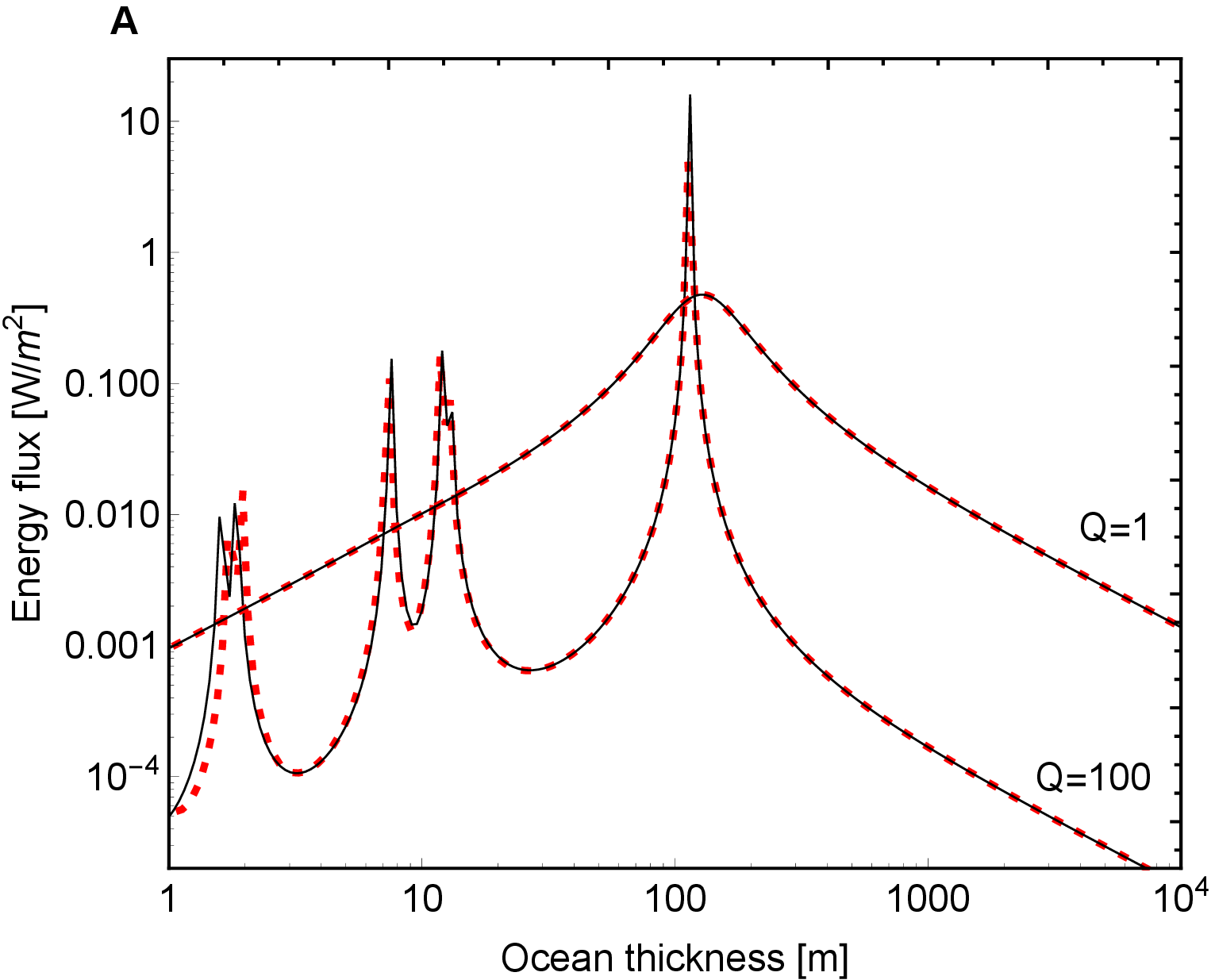}
   \includegraphics[width=7.3cm]{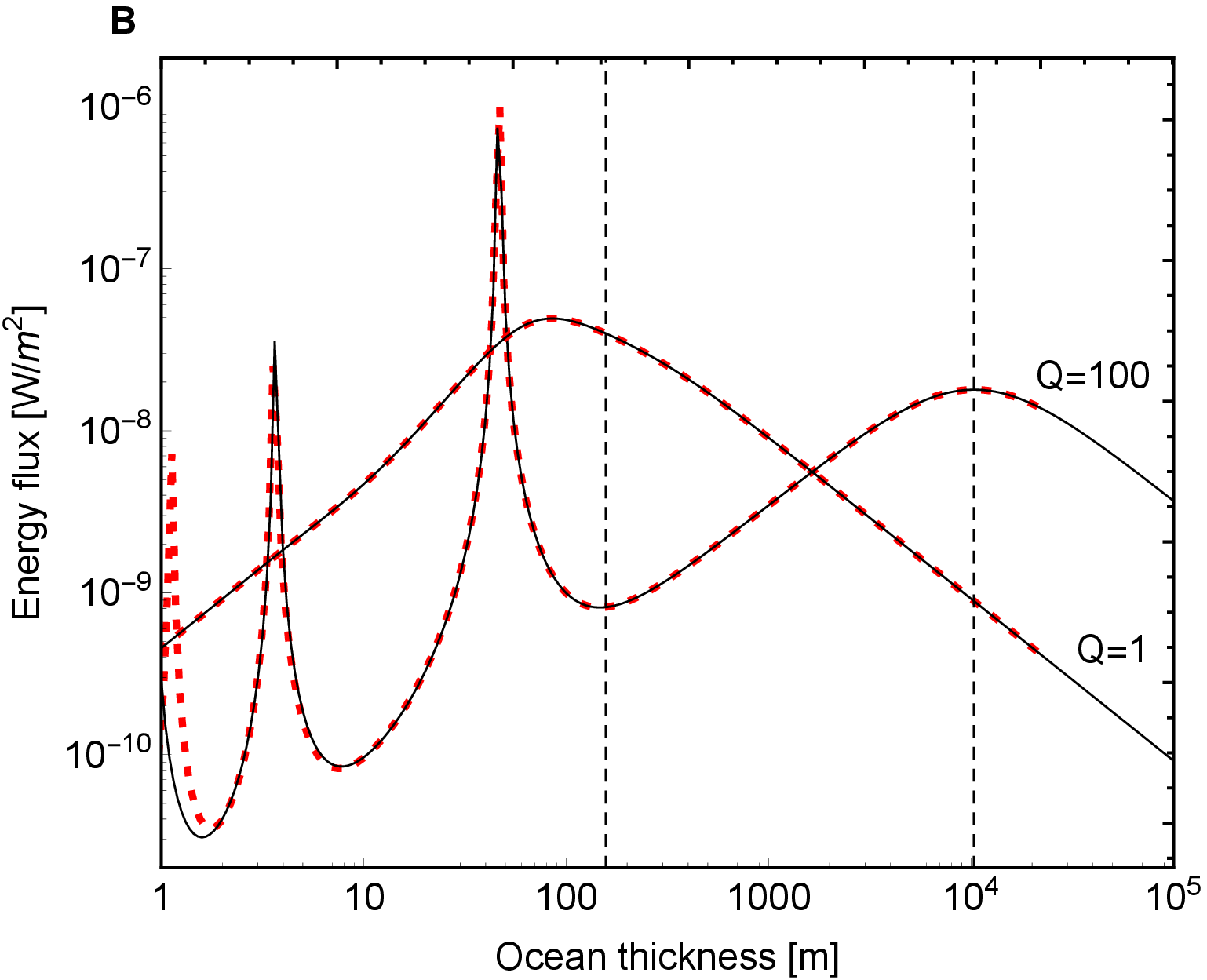}
   \caption[Oceanic dissipation if elastic crust]{
   Oceanic dissipation if elastic crust: (A) eccentricity tides, (B) obliquity tides.
   The crust is 2.5 km thick.
   Dissipation results from linear top and bottom drag with either $Q_n=100$ or $Q_n=1$.
   The surface energy flux is computed by solving the LTE for a subsurface ocean (solid curves).
   The dotted curves show the surface ocean solutions rescaled on both axes by the factor $(1-\xi_2)/\beta_2\approx1/4.88$.
   In panel~B, the vertical dashed lines indicate where dissipation is maximum for the westward obliquity tide (Eq.~(\ref{Dwot})).
}
   \label{FigDissElastic}
\end{figure}

Consider first eccentricity tides.
Fig.~\ref{FigDissElastic}A shows the surface energy flux, as a function of ocean depth, due to eccentricity tides in a subsurface ocean.
In this example, the crust is elastic and 2.5~km thick.
As expected, the curve for low viscosity ($Q_n=100$) is similar in shape to the kinetic energy shown in Fig.~\ref{EkinSubsurface}A.
If viscosity is high ($Q_n=1$), resonances are damped and the energy flux has a unique peak close to the largest resonance.
In both cases, the surface energy flux is well approximated by rescaling on both axes the surface ocean solution by the factor $(1-\xi_2)/\beta_2\approx1/4.88$.

Consider now obliquity tides.
Fig.~\ref{FigDissElastic}B shows the surface energy flux, as a function of ocean depth, due to obliquity tides in a subsurface ocean.
As above, the crust is elastic and 2.5~km thick.
If viscosity is low and the ocean is not too deep, the surface energy flux is similar in shape to the kinetic energy for an inviscid ocean (Fig.~\ref{EkinSubsurface}B).
As the ocean gets deeper, the westward tide solution is damped by ocean viscosity so that the energy flux reaches a peak before decreasing to zero (see Section~\ref{disswot}).
Similarly to eccentricity tides, the surface energy flux for obliquity tides is well approximated by rescaling on both axes the surface ocean solution by the factor $(1-\xi_2)/\beta_2\approx1/4.88$.

What happens if the crust is viscoelastic?
As it should be expected, the resonances peaks are damped: if viscoelasticity increases, the different peaks become more diffuse until they form only one broad peak (see example in Section~\ref{DissExamples}).
The dissipation rate, however, cannot be scaled from the surface ocean solution as before since viscoelasticity changes the spectrum shape.
If ocean viscosity is low, the positions of the resonances follow a scaling rule similar to Eq.~(\ref{DresApprox2}):
\begin{equation}
D^{(i)}_{mX} \approx f^{(i)}_{mX} \, \frac{q_\omega R}{6} \, \frac{1}{Re(\beta_2)} \, ,
\label{DresApprox3}
\end{equation}
where the coefficients $f^{(i)}_{mX}$ are given by Table~\ref{TableEigenApprox}.
If the ocean is deeper than the largest resonance (either for eccentricity or eastward obliquity tides), it is a good approximation to solve the LTE truncated to $n=3$, as done by \citet{chen2014}.
If ocean viscosity is low, oceanic dissipation tends to
\begin{eqnarray}
\dot E_{O,ecc} &\approx& \frac{1}{|\beta_2|^2} \left( \frac{57\pi}{28} \, \alpha + \frac{519\pi}{35} \, \frac{\nu}{R^2} \right) \frac{\rho\Omega^6 R^8}{g^2 D} \, e^2 \, ,
\label{EOeccdeeplim} \\
\dot E_{O,1E} &\approx& \frac{1}{|\beta_2|^2} \left( \frac{2397\pi}{13720} \, \alpha + \frac{9237\pi}{6860} \, \frac{\nu}{R^2} \right) \frac{\rho\Omega^6 R^8}{g^2 D} \, \sin^2 I \, .
\label{EO1Edeeplim}
\end{eqnarray}
These formulas agree with Table~4 of \citet{chen2014} if $\alpha=0$ (Navier-Stokes dissipation) and $\beta_2=1$ (surface ocean without self-attraction); they are also illustrated in Fig.~\ref{FigDissEccVisco}.
Analog expressions can be found for the kinetic energy (the terms within brackets changes); again, they agree with Table~4 of \citet{chen2014} if $\alpha=0$ and $\beta_2=1$.
Asymptotic formulas for the kinetic energy are useful when estimating the effective viscosity for nonlinear dissipation (see Section~\ref{nonlindiss}).

\subsubsection{WOT dissipation}
\label{disswot}

The westward obliquity tide is at the origin of the extremely broad peaks in Fig.~\ref{FigDissElastic}B.
What are the position and the magnitude of these peaks?
For more generality, let us do it for a viscoelastic crust, though it does not hurt to assume a nearly rigid mantle ($\upsilon_2\approx1$).
Substituting the approximate solution (\ref{toroidalcomp}) into Eq.~(\ref{EdissO2}), I can write the oceanic WOT dissipation rate as
\begin{eqnarray}
\dot E_{O,wot} &=& \frac{\rho D \Omega}{2} \, \frac{a}{Q_1} | \Psi_{1W}^1 |^2 \, ,
\label{EdotOwot}
\end{eqnarray}
where $a=1+5/(12Q_1Q_2)$ and $\Psi_{1W}^1$ is known from Eq.~(\ref{toroidalcomp}).
In the shallow- and deep-ocean limits, the WOT dissipation rate tends to
\begin{eqnarray}
\dot E_{O,wot} &\approx& 3\pi \, \frac{a}{a^2+b^2} \, \frac{1}{Q_1} \,  \rho D \, \Omega^3 R^4 \, \sin^2 I
\hspace{5mm} \mbox{(shallow)} \, ,
\label{EWOTshallowlim} \\
\dot E_{O,wot} &\approx& \frac{1}{|\beta_2|^2} \, \frac{3\pi}{25} \, (a \, Q_1) \,  \frac{\rho \, \Omega^7 R^8}{g^2 D} \, \sin^2 I
\hspace{5mm} \mbox{(deep)} \, ,
\label{EWOTdeeplim}
\end{eqnarray}
where $a$ is as above and $b=5/(9Q_1)$.
If ocean viscosity is low ($Q_1\gg1$), $a\approx1$ and $b\approx0$.
The threshold between the two regimes corresponds to $\epsilon_L\approx20|\beta_2|/Q_1$ (see Section~\ref{WestwardObliquityTide}).
If $\beta_2=1$ and ocean viscosity is low, Eqs.~(\ref{EdotOwot})-(\ref{EWOTdeeplim}) agree with Table~4 of \citet{chen2014}.
These limits are illustrated in Fig.~\ref{FigDissObliVisco}.

$\dot E_{O,wot}$ is maximum for
\begin{equation}
D_{wot} \approx \sqrt{a^2+b^2} \, \, \frac{q_\omega R}{5} \, \frac{Q_1}{|\beta_2|} \, .
\label{Dwot}
\end{equation}
Compared to the largest eastward obliquity resonance (Eq.~(\ref{DresApprox2})), the WOT peak typically occurs for a deeper ocean:
\begin{equation}
\frac{D_{wot}}{D_{1E}^{(1)}} \approx 2.24 \, \sqrt{a^2+b^2} \,\, Q_1 \, .
\end{equation}
If ocean viscosity is low, the maximum of $\dot E_{O,wot}$ is given by
\begin{equation}
\mbox{Max}\left( \dot E_{O,wot} \right) \approx \frac{9}{40} \, \xi_1 \, \frac{(\Omega R)^5}{G} \, \frac{\sin^2 I}{|\beta_2|+Im(\Lambda_2)} \, ,
\label{Ewotmax}
\end{equation}
which depends neither on ocean depth nor on ocean viscosity.
In Fig.~\ref{FigDissElastic}B, the WOT peak for $Q=1$ is higher than the one for $Q=100$ because the former combines the WOT peak with the (very damped) largest resonance.
For the same reason, the peak for $Q=1$ occurs at a smaller ocean depth than predicted by Eq.~(\ref{Dwot}).

\subsubsection{Nonlinear dissipation}
\label{nonlindiss}

Locating the WOT peak as in Eqs.~(\ref{Dwot}) to (\ref{Ewotmax}) is all right for linear dissipation (top and bottom drag or Navier-Stokes), but is not correct if dissipation is nonlinear.
On the basis of numerical tests, \citet{chen2014} proposed that the nonlinear drag (Eq.~(\ref{nonlineardrag})) is equivalent, in a time-averaged sense, to the Navier-Stokes formulation:
\begin{equation}
<-(c_D/D)|\mathbf{u}|\mathbf{u}> \, = \,  <\nu_{\mbox{\footnotesize\textit{eff}}}\nabla^2\mathbf{u}>  \, ,
\end{equation}
where $\nu_{\mbox{\footnotesize\textit{eff}}}$ is an effective Navier-Stokes viscosity.
For the westward obliquity tide, this equivalence yields $\nu_{\mbox{\footnotesize\textit{eff}},1W}\approx(c_D/2D)R^2\bar u$.
The average velocity $\bar u$ is estimated from the kinetic energy: $\bar u\approx\sqrt{E_{kin}/(2\pi{}R^2\rho{}D)}\approx|\Psi_{1W}^1|/\sqrt{4\pi{}R^2}$.
If the ocean viscosity is low, the effective viscosity is given by
\begin{equation}
\nu_{\mbox{\footnotesize\textit{eff}},1W} \approx \sqrt{ \frac{3}{8} } \, \frac{c_D}{D} \, \frac{\Omega R^3 \sin I}{|1-20 i \, \beta_2 \, \nu_{\mbox{\footnotesize\textit{eff}}} \, (g D)/(\Omega^3 R^4) | } \, ,
\label{nueff}
\end{equation}
which agrees with Eq.~(63) of \citet{chen2014} if $\beta_2=1$ (surface ocean without self-attraction).
Eq.~(\ref{nueff}) can be solved for $\nu_{\mbox{\footnotesize\textit{eff}},1W}$.
Substituting the result into Eq.~(\ref{Ewotmax}), one obtains a formula for quadratic WOT dissipation in a subsurface ocean.
Using this formula for a surface ocean ($\beta_2=1$), \citet{nimmo2015} recently argued that Triton's geological activity is driven by obliquity tides.
It would be interesting to check whether their conclusions still hold for a subsurface ocean.

\subsection{Crustal dissipation}
\label{CrustalDissipation}

\subsubsection{Magnitude}

Unless it is purely elastic, the crust dissipates energy when deforming under tidal stresses.
The relative magnitude of crustal dissipation (compared to oceanic dissipation) depends on several factors:
\begin{itemize}
\item the ocean depth. In Eqs.~(\ref{EdissO2}) and (\ref{EdissC3}), the prefactors of the squared velocity potentials are linear and quadratic in $D$, respectively.
\item the ratios of $Im(\Lambda_n)$ to the quality factors $Q_n$.
\item the relative magnitudes of consoidal ($\Phi$) and toroidal ($\Psi$) fluid motions: crustal dissipation only depends on the former, whereas oceanic dissipation depends on both.
\end{itemize}
The westward obliquity tide is a particularly interesting case.
For simplicity, suppose that the mantle is infinitely rigid.
Substituting the approximate solution (\ref{toroidalcomp}) into Eq.~(\ref{EdissC3}), I can write the crustal dissipation rate as
\begin{eqnarray}
\dot E_{C,wot} &=&  10 \, \frac{\rho D \Omega}{\epsilon_L} \, \frac{Im (\Lambda_2)}{(Q_1)^2} \, | \Psi_{1W}^1 |^2 \, ,
\end{eqnarray}
while the oceanic dissipation rate is given by Eq.~(\ref{EdotOwot}).
As a consistency check, one can verify that $\dot E_{O,wot}+\dot E_{C,wot}$ is equal to the dissipation rate in the whole body (Eq.~(\ref{Ediss4})).
If viscosity is low ($Q_1,Q_2\gg1$), the ratio of crustal to oceanic dissipation, whatever the ocean depth, is given by
\begin{equation}
\frac{\dot E_{C}}{\dot E_{O}}\bigg|_{wot} \approx \frac{5}{q_\omega} \, \frac{D}{R} \, \frac{Im(\Lambda_2)}{Q_1} \, .
\end{equation}
In the inviscid limit ($Q_1\rightarrow\infty$), crustal dissipation vanishes for the westward obliquity tide because the radial tide tends to zero.
This property is unique to the westward obliquity tide.
For other tidal waves, crustal dissipation does not vanish if the ocean is inviscid.

\subsubsection{Deep-ocean limit}

Dynamical effects become negligible in the deep-ocean limit (Section~\ref{Limits}).
Crustal dissipation should thus tend in that limit to the classical formula derived for static tides.
For simplicity, assume that the mantle is infinitely rigid (crust and ocean have the same density as before).
In that case, the radial ($H_2$) and gravitational ($K_2$) degree-two tidal Love numbers of the whole body are simply related by $K_2=(3\xi_1/5)H_2$ (see Eq.~(42) of \citet{beuthe2014}).
If the mantle is infinitely rigid, the Love number $H_2$ and the admittance $Z_2=1/(1-\xi_2+\Lambda_2)$ are equal (see Eq.~(\ref{Znrigid})).
Putting these two relations together, I obtain
\begin{equation}
Im(K_2) = -\frac{3}{5} \, \xi_1 \, |Z_2|^2 \, Im(\Lambda_2) \, ,
\label{ImK2}
\end{equation}
which is a special case of Eq.~(101) of \citet{beuthe2014}.

\begin{figure}
\begin{minipage}{\textwidth}
   \centering
   \includegraphics[width=7.3cm]{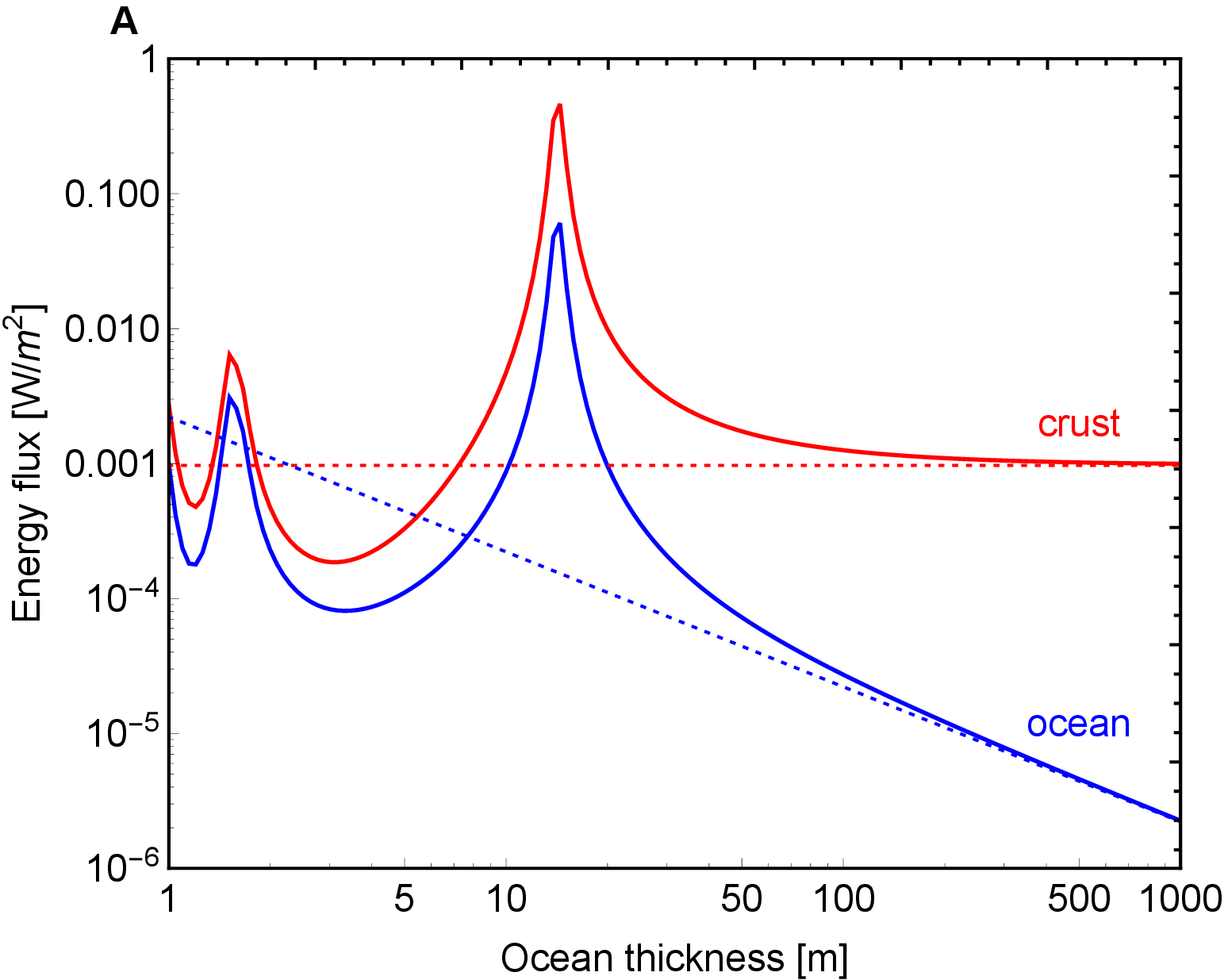}
   \includegraphics[width=7.3cm]{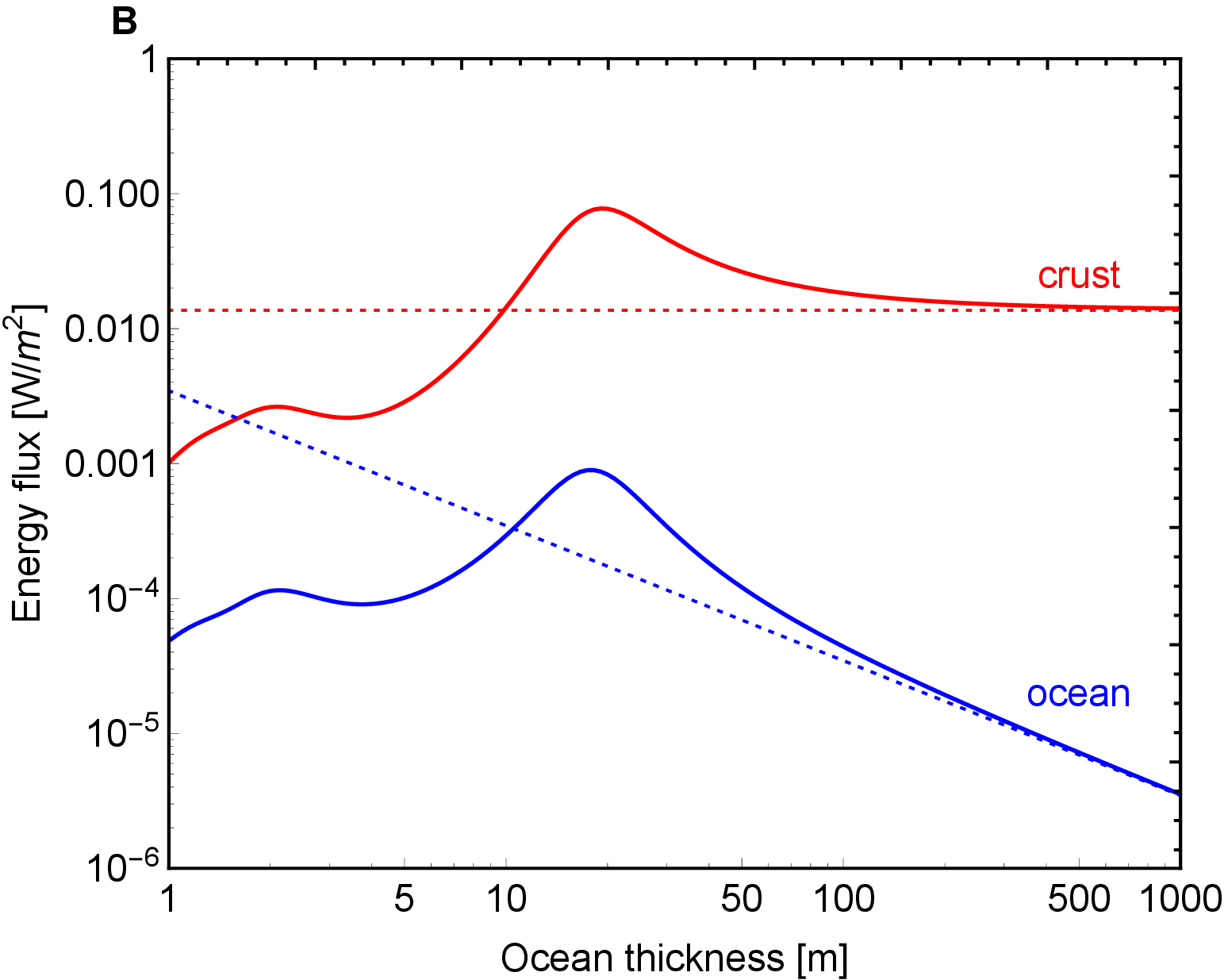}
   \caption[Dissipation in the crust and ocean due to eccentricity tides]{
    Dissipation in the crust and ocean due to eccentricity tides: (A) conductive crust, (B) convective crust.
    The contributions of crustal and oceanic dissipation to the surface energy flux are shown as separate curves.
    In each panel, the horizontal dotted line shows the deep-ocean limit for crustal dissipation (Eq.~(\ref{ECecc2})), whereas
    the oblique dotted line shows the deep-ocean asymptotic limit for oceanic dissipation (Eq.~(\ref{EOeccdeeplim})).
    See Section~\ref{DissExamples} for details.
}
   \label{FigDissEccVisco}
\vspace{5mm}
   \includegraphics[width=7.3cm]{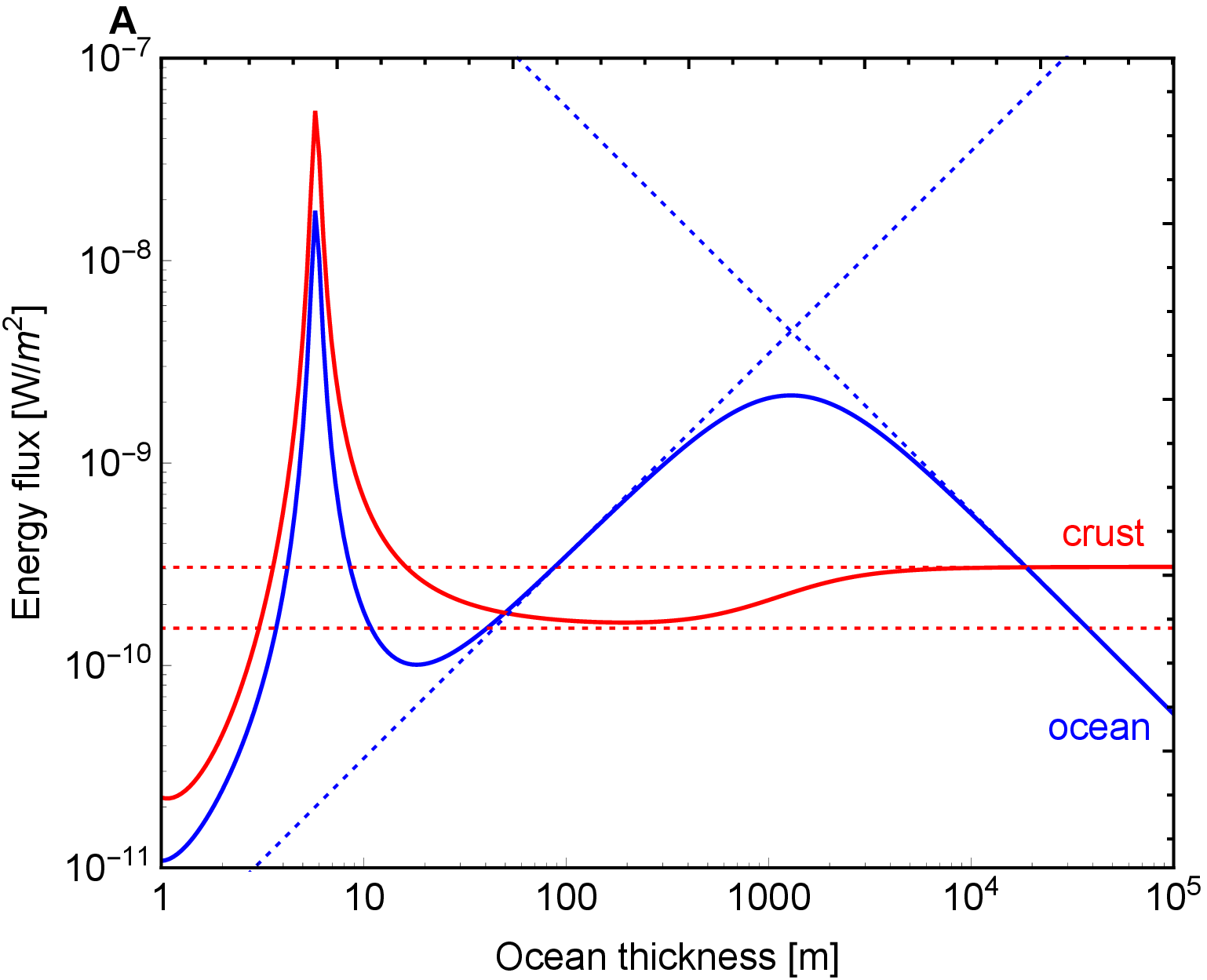}
   \includegraphics[width=7.3cm]{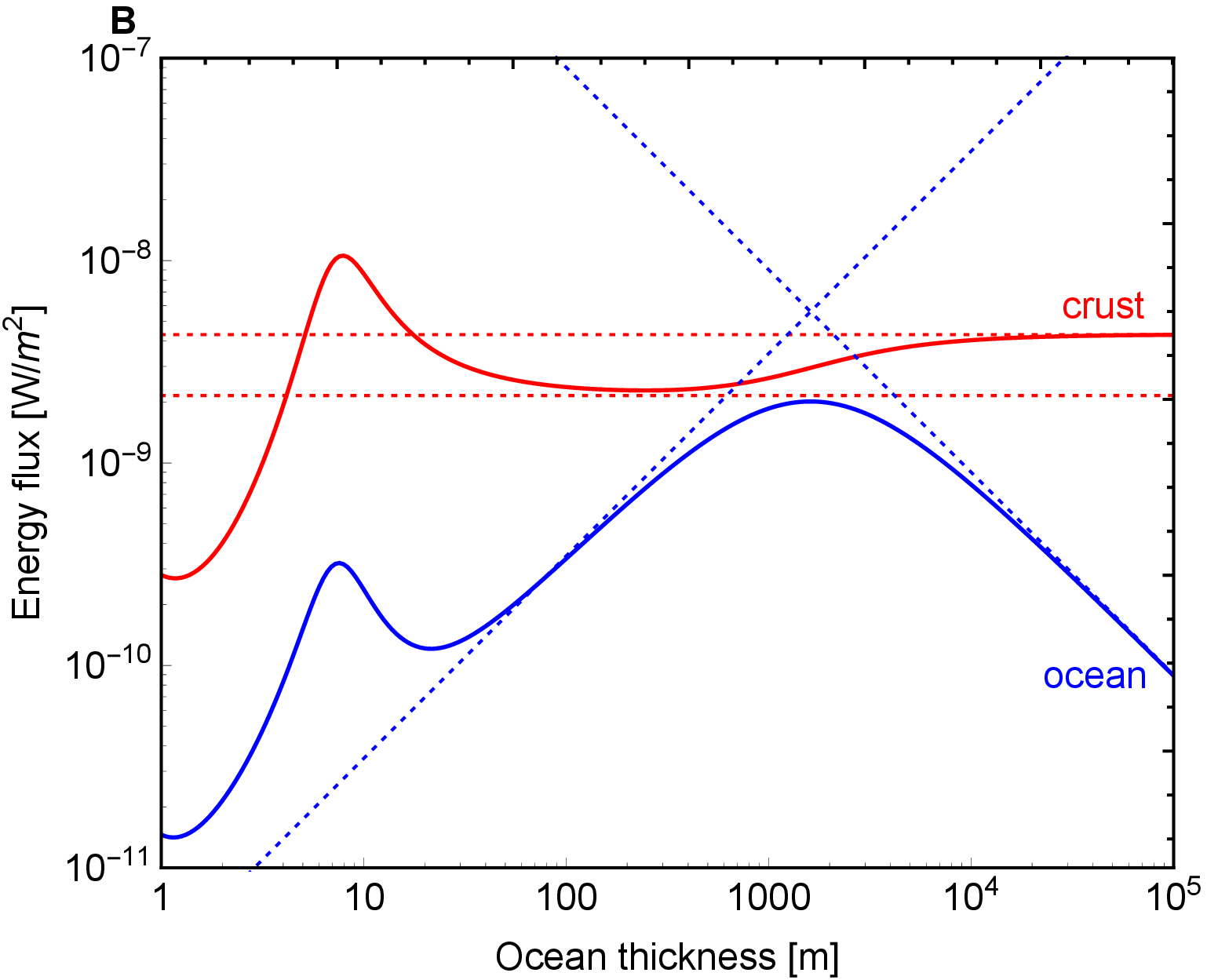}
   \caption[Dissipation in the crust and ocean due to obliquity tides]{
    Dissipation in the crust and ocean due to obliquity tides: (A) conductive crust, (B) convective crust.
    The contributions of crustal and oceanic dissipation to the surface energy flux are shown as separate curves.
    In each panel, the horizontal dotted lines show the lower ($f_Q=1/2$) and upper ($f_Q=1$) bounds of the deep-ocean limit for crustal dissipation (Eq.~(\ref{ECobli})).
    The oblique dotted lines show the shallow- and deep-ocean asymptotic limits for oceanic WOT dissipation (Eqs.~(\ref{EWOTshallowlim})-(\ref{EWOTdeeplim})).
    See Section~\ref{DissExamples} for details.
}
   \label{FigDissObliVisco}
   \end{minipage}
\end{figure}

In the deep-ocean limit, the radial displacement tends to the equilibrium tide (Eq.~(\ref{deepocean})) except maybe for the westward obliquity tide (Eq.~(\ref{deepoceanW})).
For eccentricity tides, the formula for crustal dissipation (Eq.~(\ref{EdissC3})) becomes
\begin{equation}
\dot E_{C}^{rigid} \approx \frac{3}{2} \, \xi_1 \, \frac{\Omega R}{G} \, Im(\Lambda_2) \, |Z_2|^2 \, U_{2}^{sq} \, ,
\label{ECecc1}
\end{equation}
where $U_2^{sq}$ is equal to $(\Omega R)^4(21/5)e^2$ (see Eq.~(\ref{Unsq})).
Combining Eqs.~(\ref{ImK2}) and (\ref{ECecc1}), one sees that the crustal dissipation rate for eccentricity tides tends to
\begin{equation}
\dot E_{C}^{rigid} \approx - \frac{21}{2} \, \frac{(\Omega R)^5}{G} \, e^2 \, Im(K_2) \, ,
\label{ECecc2}
\end{equation}
which is the classical formula for dissipation due to eccentricity tides in the static limit (e.g.\ Eq.~(42) of \citet{beuthe2013}).
This limit is illustrated in Fig.~\ref{FigDissEccVisco}.

For obliquity tides, the deep-ocean limit is not necessarily the same for eastward and westward tides (compare Eqs.~(\ref{deepocean}) and (\ref{deepoceanW})).
Thus,  the crustal dissipation rate for obliquity tides tends to
\begin{equation}
\dot E_{C}^{rigid} \approx - \frac{3}{2} \, f_Q \, \frac{(\Omega R)^5}{G} \, \sin^2 I \, Im(K_2) \, ,
\label{ECobli}
\end{equation}
where
\begin{equation}
f_Q = \frac{1}{2} \left( 1 + \frac{1}{1+|\epsilon_L Q_1/(20\beta_2)|^2} \right) .
\end{equation}
Eq.~(\ref{ECobli}) differs by the factor $1/2<f_Q<1$ from the classical formula for dissipation due to static obliquity tides (e.g.\ Eq.~(42) of \citet{beuthe2013}).
This limit is illustrated in Fig.~\ref{FigDissObliVisco}.

\subsubsection{Examples}
\label{DissExamples}

In this section, I compute dissipation using two crude models of crustal rheology.
More detailed models are described by \citet{hussmann2002}, \citet{moore2006}, and \citet{castillo2011}, among others.
Any linear (or linearized) rheological law could do, but I have chosen Maxwell rheology for its simplicity (see Appendix~\ref{AppendixMembrane}).
The crust is 25~km thick, as suggested by libration data \citep{thomas2016}.
Oceanic dissipation results from linear top and bottom drag with low viscosity ($Q_n=100$ for all $n$).

In the first model, heat is transported through the crust by conduction.
The temperature varies linearly through the crust, from the surface temperature $T_S=75\rm\,K$ to the melting temperature $T_M=273\rm\,K$ at the bottom (in a more accurate model, crustal dissipation and sphericity would introduce nonlinear corrections).
The viscosity of ice is related to temperature by an Arrhenius relation, $\zeta(T)=\zeta_0\exp(l(T_m/T-1))$, where $l\approx27$ and $\zeta_0$ is the viscosity of ice at the melting point.
The latter quantity depends on the deformation mechanism and on the grain size.
I assume here that $\zeta_0=\zeta_c$, where $\zeta_c=\mu_E/\Omega=6.6\times10^{13}\rm\,Pa.s$ is the Maxwell critical viscosity.
Given the viscosity profile and the elastic parameters of Table~\ref{TableParamInterior}, I can compute the effective viscoelastic parameters of the conductive model with the equations of Appendix~\ref{AppendixMembrane}:
\begin{equation}
\bar\mu=(3.44+0.13 i) {\rm\,GPa}
\hspace{10mm}
\bar\nu=0.333 - 0.004{i}
\hspace{10mm}
\beta_2=24.59+0.84{i}
\, .
\end{equation}

In the second model, the crust is divided into an upper conductive layer (10~km thick) and a lower convective layer (15~km thick).
The rheology of the conductive layer is described as in the first model while the lower layer is uniformly at critical viscosity.
As above, I can compute the effective parameters of the convective model:
\begin{equation}
\bar\mu=(2.43+1.10 i) {\rm\,GPa}
\hspace{10mm}
\bar\nu=0.386 - 0.031{i}
\hspace{10mm}
\beta_2=18.15+7.61{i}
\, .
\end{equation}
Comparing the values of $\beta_2$ for the conductive and convective models, we expect the latter to be more dissipative.

Fig.~\ref{FigDissEccVisco} shows dissipation in the crust and ocean, for eccentricity tides, as a function of ocean depth.
In both models, crustal dissipation dominates oceanic dissipation whatever the ocean depth.
This is not true of all models: oceanic dissipation could dominate crustal dissipation at small ocean depth.
If the ocean is deeper than $100\rm\,m$, crustal dissipation tends to the constant value predicted for static tides (Eq.~(\ref{ECecc2})), whereas oceanic dissipation tends to the asymptotic limit given by Eq.~(\ref{EOeccdeeplim}).
The viscoelastic crust damps the resonance peaks: the effect is particularly visible in the more dissipative convective model.

Fig.~\ref{FigDissObliVisco} shows the same thing as Fig.~\ref{FigDissEccVisco} but for obliquity tides.
The overall magnitude is much smaller than for eccentricity tides because the obliquity is very small.
In the first model, oceanic dissipation is dominant for a wide range of ocean depths ($50{\rm\,m}<D<20\rm\,km$).
If the ocean is deeper than $10\rm\,km$, crustal dissipation tends to the constant value predicted for static tides (Eq.~(\ref{ECobli}) with $f_Q=1$), whereas oceanic dissipation tends to the asymptotic limit given by Eq.~(\ref{EWOTdeeplim}).
In an intermediate range ($50{\rm\,m}<D<1\rm\,km$), crustal dissipation is close to the lower bound of Eq.~(\ref{ECobli}) because the westward radial tide nearly vanishes.
In the same range, oceanic dissipation stays close to the shallow-ocean asymptotic limit for WOT dissipation (Eq.~(\ref{EWOTshallowlim})).
As for eccentricity tides, the viscoelastic crust damps the resonance peaks, especially in the more dissipative convective model.
By contrast, the WOT peak (the peak in oceanic dissipation due to the westward obliquity tide) does not differ much between the two models: its magnitude indeed depends on $|\beta_2|+Im(\Lambda_2)$ (see Eq.~(\ref{Ewotmax})) which is nearly the same in the two models.


\section{Summary}

Global subsurface oceans transform icy moons into `membrane worlds', for which the tidal response mainly depends on the properties of the ice and water layer \citep{beuthe2015}.
Tidal frequencies are so low that dynamical effects are completely negligible in the solid crust, but ocean dynamics have a different nature.
One often assumes, however, that the fluid layer reacts to tidal forcing as if it were a solid layer with vanishing shear modulus.
For a thin fluid layer, this implies that the tidal flow has no toroidal component (zero vorticity).
Moreover, the flow does not dissipate energy.
By contrast, a few studies concerned with oceanic dissipation turned to the Laplace Tidal Equations (LTE) for more realistic dynamics but crustal effects were ignored.

In this paper, I develop the first-ever model for dissipative tides in a subsurface ocean.
This requires only a slight modification of the LTE, with a new forcing term provided by the membrane approach.
Thus, the standard numerical and analytical methods for solving LTE remain applicable.
I have used here a semi-analytical method (i.e.\ matrix inversion done numerically if the matrix is large).
Nonlinear dissipation requires a numerical solution on a grid \citep{sears1995,chen2014}, though the effective viscosity approach could also be helpful (see Section~\ref{nonlindiss}).
The LTE forcing term depends on the size and rheology of the crust and mantle, but mantle deformations are negligible for Enceladus.
The most important crustal parameter is $\Lambda_2\approx2\hat\mu(d/R)$, where $\hat\mu=\bar\mu/(\rho{}gR)$ is the effective shear modulus of the viscoelastic crust and $d/R$ is the relative crust thickness.
Crustal effects are generally substantial for Enceladus because $\hat\mu\approx100\gg1$, in contrast with larger satellites such as Europa or Titan.

If the crust is elastic, there is a simple scaling relation (Eq.~(\ref{DresApprox2}) and Table~\ref{TableEigenApprox}) between the surface and subsurface solutions, whether the resonances, the kinetic energy, or the dissipation rate.
This scaling rule depends on the membrane-bending spring constant of the crust ($\Lambda_2$) and on oceanic self-attraction ($\xi_2=3\rho/5\rho_b$) through the factor $\beta_2=1-\xi_2+\Lambda_2$ (mantle deformations are neglected here).
In brief, the resonances and the velocity potentials are given by the solutions for a surface ocean, shifted to smaller ocean depths by the factor $(1-\xi_2)/\beta_2$ (Fig.~\ref{FigEigenValCrust}).
Consequently, the energy spectrum and the dissipation rate are doubly scaled: they are shifted to smaller ocean depths and reduced in magnitude by the same factor $(1-\xi_2)/\beta_2$ (Figs.~\ref{EkinSubsurface} and \ref{FigDissElastic}).
For a deep ocean, this double scaling is equivalent to a dependence in $1/|\beta_2|^2$ of the dissipation rate (Eqs.~(\ref{EOeccdeeplim})-(\ref{EdotOwot})).
The scaling rule works particularly well for bodies (such as Enceladus) whose ocean-to-bulk density ratio is close to 0.6.

If the crust is viscoelastic, tidal resonances are damped (in addition to being damped by ocean viscosity) and energy is dissipated in the crust.
Resonances are shifted to smaller ocean depths by the factor $(1-\xi_2)/Re(\beta_2)$ (Eq.~(\ref{DresApprox3})).
For a shallow ocean, the dissipation rates in the crust and ocean exhibit the same resonances (Fig.~\ref{FigDissEccVisco}).
For a deep ocean, crustal dissipation tends to a constant value (static tides) whereas oceanic dissipation tends to zero.
The westward obliquity tide is a special case, because there is no radial tide unless the ocean is deep and the viscosity is high enough (Fig.~\ref{FigDissObliVisco}).
Therefore, crustal dissipation due to dynamical obliquity tides can differ by up to a factor of two from the standard prediction assuming static tides (Eq.~(\ref{ECobli})).
The same effect could play an important role for tectonics due to obliquity tides.

Crustal effects are thus severe for Enceladus' dynamical tides.
If the crust is $25\rm\,km$ thick, resonances occur for an ocean less than $20\rm\,m$ deep (Figs.~\ref{FigEigenValCrust} and \ref{FigDissEccVisco}), whereas the ocean depth is probably two or three orders of magnitude larger at the present time.
Therefore, oceanic dissipation is nowadays negligible and crustal dissipation can be computed in the static limit.
As mentioned in Section~\ref{LaplaceTidalEquations}, this conclusion can be avoided if the ocean is stratified, but it is doubtful that density stratification can be maintained if tidal heating occurs at the observed level.
As regards obliquity tides, they exhibit interesting dynamical effects in deep oceans, but are way too small to be relevant to Enceladus unless future observations show that the obliquity is much larger than the theoretical upper bound for the Cassini state \citep{baland2016}.
In the past, tidal resonances could have played a role in a forming or freezing ocean less than $100\rm\,m$ deep.
However, the displacement of resonances to very shallow depths compounds problems that already existed for a shallow surface ocean.
First, the resonant response is blocked as soon as the deforming crust comes into contact with the mantle (or as the water dries out somewhere in a surface ocean).
Second, the reliefs of the seafloor and `seaceiling' are comparable to the ocean depth and certainly alter the flow.

The LTE-membrane approach has been developed here with maximum generality, so that it can easily be applied to other icy moons.
Enceladus' crust is actually at the limit of validity of the membrane approach: five to ten percents of the surface radius, as indicated by librations \citep{thomas2016,vanhoolst2016}.
As shown in Section~\ref{MembraneWorld}, crustal effects further increase when finite crust thickness is properly taken into account (Figs.~\ref{FigLove} and \ref{FigResNR}).
However, no new physical effects are expected for ocean tides under a thick crust.
The big question mark is rather the shallow water assumption.
Is it a good approximation for subsurface oceans that are several tens (or even hundreds) of kilometers deep and which probably undergo mixing due to turbulent convection?
Tidal flows in thick rotating shells exhibit complex features which are far from being understood.
Ongoing research focuses on planetary fluid cores (under a very thick rigid shell) and on gaseous envelopes of giant planets and stars (with a free surface)\citep{ogilvie2014,lebars2015}.
By contrast, large-scale dynamics of deep oceans under an elastic crust have received scant attention.

I will end this paper on a historical note.
Four years before \citet{lewis1971} predicted liquid mantles in icy satellites,
Syd Barrett sang of the underground icy waters of giant planets' moons in Pink Floyd's `Astronomy Domine'.
We now have solid evidence that Europa, Enceladus and Titan have subsurface oceans and we suspect the same for Ganymede, Callisto, Dione, and maybe Mimas, but the moons of Uranus and Neptune are still shrouded in mystery.

\section*{Acknowledgments}
This work is financially supported by the Belgian Federal Science Policy Office through the Brain Pioneer contract BR/314/PI/LOTIDE.
I thank Antony Trinh, Attilio Rivoldini, and Tim Van Hoolst for discussions that inspired this project.

\begin{appendices}
\section{Spherical harmonics}
\label{AppendixSH}
\renewcommand{\theequation}{A.\arabic{equation}} 
\setcounter{equation}{0}  

The functions $Y_n^m(\theta,\varphi)$ are the spherical harmonics of degree $n$ and order $m$ \citep{arfken2013}:
\begin{equation}
Y_n^m(\theta,\varphi) = N_n^m \, P_n^m(\cos\theta) \, e^{im\varphi} \, .
\end{equation}
The associated Legendre functions $P_n^m(\cos\theta)$ and the normalization factors $N_n^m$ are real: the only effect of complex conjugation on spherical harmonics (as in Eqs.~(\ref{UnT}) and (\ref{FT})) is to transform $e^{im\varphi}$ into $e^{-im\varphi}$.
Following \citet{arfken2013} and several papers on ocean tides \citep{tyler2011,chen2014,matsuyama2014}, $P_n^m(\cos\theta)$ is defined with the Condon-Shortley phase $(-1)^m$, although this is not the usual practice in geophysics.
In any case this phase is irrelevant if eccentricity tides ($m=0,2$) and obliquity tides ($m=1$) are studied separately.
The Legendre functions of degree two are particularly important because they dominate the tidal potential and the tidal response:
\begin{equation}
\left( P_2^0 , P_2^1 , P_2^2 \right)(\cos\theta) = \left( \frac{1}{2}(3\cos^2\theta-1) , -3\sin\theta\cos\theta , 3\sin^2\theta \right) .
\end{equation}
If the spherical harmonics are normalized to one ($\int_S Y_n^m (Y_{n'}^{m'})^*dS=\delta_{nn'}\delta_{mm'}$), the normalization factor reads
\begin{equation}
N_n^m = \left( \frac{2n+1}{4\pi} \, \frac{(n-m)!}{(n+m)!} \right)^{1/2} \, ,
\label{Nnm}
\end{equation}
otherwise $N_n^m=1$ if spherical harmonics are not normalized.
Spherical harmonics are eigenvectors of the spherical Laplacian $\Delta$ with eigenvalues $-n(n+1)$.
The spherical Laplacian is the restriction of the 3D Laplacian to the surface of a unit sphere:
\begin{equation}
\Delta = \frac{\partial^2}{\partial \theta^2} + \cot\theta \, \frac{\partial}{\partial \theta} + \frac{1}{\sin^2\theta} \, \frac{\partial^2}{\partial\varphi^2} \, .
\label{defLaplacian}
\end{equation}

\section{Membrane and bending spring constants}
\label{AppendixMembrane}
\renewcommand{\theequation}{B.\arabic{equation}} 
\setcounter{equation}{0}  

The crust is modeled as a thin spherical shell of radius $R$, uniform thickness $d$ and density $\rho$, separated from the mantle by an ocean.
Its viscoelastic shear modulus $\mu_V$ and Poisson's ratio $\nu_V$ do not vary laterally, but they depend on depth if the crust is viscoelastic.
For Maxwell rheology (Appendix~C of \citet{beuthe2014}), the viscoelastic parameters are given by
\begin{equation}
\left( \mu_V, \nu_V \right) = \left( \frac{\mu_E}{1-i\delta} \, , \frac{3\,\nu_E-i \left(1+\nu_E \right) \delta }{3-2 i \left(1+\nu_E \right) \delta} \right) ,
\label{maxwell}
\end{equation}
where $\mu_E$ and $\nu_E$ are the shear modulus and Poisson's ratio of elastic ice, and $\delta=\mu_E/(\omega\zeta)$ in which $\zeta$ is the crustal viscosity.
The critical viscosity $\zeta_{crit}=\mu_E/\omega$ corresponds to $\delta=1$.

If there is no tangential load, the membrane equation relating the deflection $\eta^{top}$ to the normal load $q$ (positive outward) reads
\begin{equation}
2\bar\mu \left(1+\bar\nu \right) d \, \Delta'  \eta^{top} = R^2 \left( \Delta'-1-\bar\nu \right) q \, ,
\label{membeq}
\end{equation}
where $\Delta'=\Delta+2$ in which $\Delta$ is the spherical Laplacian (Eq.~(\ref{defLaplacian})).
Eq.~(\ref{membeq}) can be obtained by combining Eqs.~(15) and (16) of \citet{beuthe2014}.
The eigenvalues of $-\Delta'$ are denoted
\begin{equation}
x_n = (n-1)(n+2) \, .
\label{defxn}
\end{equation}
The parameters $\bar\mu$ and $\bar\nu$ are the effective shear modulus and effective Poisson's ratio of the crust, respectively:
\begin{eqnarray}
\bar\mu &=& \frac{1}{d} \int_d \mu_V \, dr \, ,
\nonumber \\
\bar\nu &=& \left( \int_d  \frac{\mu_V}{1-\nu_V} \, dr  \right)^{-1}  \, \int_d \frac{\mu_V}{1-\nu_V} \, \nu_V \, dr \, .
\label{munubar}
\end{eqnarray}
The membrane equation (\ref{membeq}) can be solved in spherical harmonics (Eq.~(27) of \citet{beuthe2014}).
The load deforming the shell at harmonic degree $n$ is proportional to the radial displacement of the surface (Hooke's law):
\begin{equation}
q_n = \rho g \, \Lambda^M_n \, \eta^{top}_{n} \, \, .
\label{hookeMAppendix}
\end{equation}
The membrane spring constant $\Lambda^M_n$ (nondimensional) characterizes crustal resistance to extension:
\begin{equation}
\Lambda^M_n = 2 x_n \, \frac{1+\bar\nu}{x_n+1+\bar\nu} \, \frac{\bar\mu}{\rho g R} \, \frac{d}{R} \, ,
\label{LambdaM}
\end{equation}
where $g$ is the surface gravity.

If the load is of short wavelength, bending moments become large so that the membrane approximation breaks down.
If the shell is homogeneous, the thin shell equation including the membrane and bending responses is given by
\begin{equation}
\left( D_V \, \Delta \, \Delta'+ R^2 E_V d \right)  \Delta'  \eta^{top} = R^4 \left( \Delta'-1-\nu_V \right) q \, .
\label{thinshelleq}
\end{equation}
You can obtain this equation from Eq.~(88) of \citet{beuthe2008} in which you set the tangential load to zero ($\Omega=0$), take the thin shell limit ($\zeta=1$ where $\zeta$ has the meaning given in \citet{beuthe2008}), and change the sign of the vertical load (positive outward instead of inward).
The parameters $E_V$ and $D_V$ are Young's modulus and the bending rigidity, respectively:
\begin{equation}
\left( E_V , D_V \right) = \left( 2(1+\nu_V)\mu_V ,  \frac{\mu_V \, d^3}{6(1-\nu_V)} \right) .
\end{equation}
Solving Eq.~(\ref{thinshelleq}) in the spherical harmonic domain yields a Hooke's law similar to Eq.~(\ref{hookeMAppendix}), except that the membrane spring constant is replaced by the thin-shell spring constant $\Lambda_n=\Lambda^M_n+\Lambda^B_n$.
The bending spring constant $\Lambda^B_n$ (nondimensional) characterizes crustal resistance to bending:
\begin{equation}
\Lambda^B_n = \left( x_n \right)^2 \, \frac{x_n+2}{x_n+1+\nu_V} \, \frac{D_V}{\rho{}gR^4} \, .
\label{LambdaB}
\end{equation}
This formula is only valid for a homogeneous crust: the viscoelastic parameters $(\mu_V,\nu_V)$ must be uniform.
It is possible to extend this formula to a viscoelastic crust with depth-dependent rheology but this will not be necessary in this paper because bending effects are negligible for tidal deformations.

\section{Tidal and load Love numbers}
\label{AppendixLoveTL}
\renewcommand{\theequation}{C.\arabic{equation}} 
\setcounter{equation}{0}  

Love numbers are nondimensional numbers characterizing the viscoelastic-gravitational response of a spherically symmetric body to some kind of forcing (indexed by $J$), the two main types of which are tidal perturbations ($J=T$) and surface mass loads ($J=L$) (see for example \citet{saito1974}).
The radial and gravitational Love numbers $h^J_n$ and $k^J_n$ relate the forcing potential $U^J_n$ to the radial displacement of the surface ($\eta^{top}$) and to the induced gravitational potential at the surface ${U^J_n}'$, respectively:
\begin{eqnarray}
(\eta^{top})_n &=& h^J_n \, \frac{U^J_n}{g} \, ,
\label{hnJ}  \\
{U^J_n}' &=& k^J_n \, U^J_n \, .
\label{knJ}
\end{eqnarray}
Whatever the type of forcing, it can always be expressed in the form of a gravitational potential.
For example, a mass load of surface density $\sigma_n$ directly produces the gravitational potential $U^L_n$ \citep[][Eq.~(2.1.25)]{kaula1968}:
\begin{equation}
U^L_n = \frac{3}{2n+1} \, \frac{g \sigma_n}{\rho_b} \, .
\label{UL}
\end{equation}
The mass load deforms the surface and this deformation generates in turn the induced gravitational potential ${U^L_n}'$.
The radial stress at the surface reads
\begin{equation}
\left( \sigma_{rr}\right)_n = -g \sigma_n = - \frac{2n+1}{3} \, \rho_b \, U^L_n \, .
\end{equation}

The standard formulation of the viscoelastic-gravitational problem involves six radial functions $y_i$ for which I follow the conventions of \citet{takeuchi1972} (in particular, $y_2$ corresponds to the radial stress).
If the forcing is tidal ($T$) or due to a mass load ($L$), the surface boundary conditions are \citep{saito1974}:
\begin{eqnarray}
\left( y_2^T \, , y_4^T \, , y_6^T \right)\Big|_{r=R} &=& \left( 0 \, , 0 \, , \frac{2n+1}{R} \right) ,
\label{bcT} \\
\left( y_2^L \, , y_4^L \, , y_6^L \right)\Big|_{r=R} &=& \left( - \frac{2n+1}{3}\, \rho_b \, , 0 \, , \frac{2n+1}{R} \right) .
\label{bcL}
\end{eqnarray}
The index $n$ is omitted on the functions $y^J_i$ so as to simplify the notation.
The Love numbers $(h^J_n,k^J_n)$ are related to the surface values of $(y^J_1, y^J_5)$:
\begin{eqnarray}
h^J_n &=& g y^J_1(R) \, ,
\nonumber \\
k^J_n &=& y^J_5(R) - 1  \hspace{5mm} \mbox{with} \hspace{2mm} J=(T,L) \, .
\label{kJ}
\end{eqnarray}
Tidal and load Love numbers are related by the Saito-Molodensky relation \citep{molodensky1977,saito1978}:
\begin{equation}
k^L_n=k^T_n-h^T_n \, .
\label{saitomolo}
\end{equation}
The diminishing factors (or tilt factors) are combinations of Love numbers defined by
\begin{equation}
\gamma^J_n = 1 +  k^J_n - h^J_n \hspace{5mm} \mbox{with} \hspace{2mm} J=(T,L) \, .
\label{gammaJ}
\end{equation}
The Saito-Molodensky relation yields
\begin{equation}
k^L_n = \gamma^T_n - 1 \, .
\label{kLgammaT}
\end{equation}

\section{Love numbers if uniform density}
\label{AppendixUniform}
\renewcommand{\theequation}{D.\arabic{equation}} 
\setcounter{equation}{0}  

Additional relations exist between Love numbers if the body is incompressible and of uniform density $\rho_b$.
The gravitational perturbation is then proportional to $r^n$ so that $(\partial/\partial{r}){y_5^J}(r)=(n/r)y_5^J(r)$.
Substituting this constraint into the fifth elastic-gravitational equation (Eq.~(17) of \citet{beuthe2015}) and applying the boundary condition $y_6^J(R)=(2n+1)/R$ with $J=T$ or $L$, I get the well-known relation
\begin{equation}
k_n^J = \frac{3}{2n+1} \, h_n^J \hspace{5mm} \mbox{with} \hspace{2mm} J=(T,L) \, ,
\label{LovekH}
\end{equation}
which can be combined with the Saito-Molodensky relation to give
\begin{equation}
\left( k^L_n, \, h^L_n \right) = \left( -\frac{2(n-1)}{2n+1} , \, -\frac{2(n-1)}{3} \right) h_n^T \, .
\label{LoveH}
\end{equation}
Furthermore, one can write $h_n^T$ as
\begin{equation}
h_n^T = \frac{2n+1}{2(n-1)} \, \frac{1}{1+ S_n \, \hat \mu_m } \, ,
\label{hnTincomp}
\end{equation}
where $\hat\mu_m$ is the nondimensional shear modulus of the uppermost layer of the mantle: $\hat\mu_m=\mu_m/(\rho_b{}gR)$.
$S_n$ is a nondimensional factor.
If the body is made of discrete layers, $S_n$ depends on the ratios of radii and shear moduli between successive layers (Appendix~F of \citet{beuthe2013}).
If the body is homogeneous, the condition of zero surface stress ($y_2^T(R)=y_4^T(R)=0$) yields
\begin{equation}
S_n = \frac{2n^2+4n+3}{n} \, .
\label{Sn}
\end{equation}
Eqs.~(\ref{LovekH})-(\ref{Sn}) are equivalent to Eqs.~(5.7.1) and (5.8.3) of \citet{munk1960}.
If there is either a liquid core or an infinitely rigid core of radius $R_C$, $S_n$ depends on $(n,x)$, with $x=R_C/R$ (see Appendix~G of \citet{beuthe2013} for the case $n=2$).

\section{Pressure Love numbers}
\label{AppendixLoveP}
\renewcommand{\theequation}{E.\arabic{equation}} 
\setcounter{equation}{0}  

Suppose that the body is submitted to a surface pressure $q_n$ (positive downwards) instead of a mass load \citep{molodensky1977,guo2004}.
Though this type of forcing does not directly produces a gravitational potential, it is convenient to associate a fictitious potential with the pressure, in analogy with Eq.~(\ref{UL}):
\begin{equation}
U^P_n = \frac{3}{2n+1} \, \frac{q_n}{\rho_b} \, .
\label{UP}
\end{equation}
Pressure (also called pressure-load) Love numbers $(k^P_n, h^P_n)$ are defined by Eqs.~(\ref{hnJ})-(\ref{knJ}) with $J=P$.
The radial stress at the surface reads
\begin{equation}
\left( \sigma_{rr}\right)_n = -q_n = - \frac{2n+1}{3} \, \rho_b \, U^P_n \, .
\end{equation}

The fact that the pressure forcing is potential-free, however, implies two differences with respect to tidal and load Love numbers.
First, the boundary conditions on $(y_2,y_4)$ are the same as for mass loading (Eq.~(\ref{bcL})), but $y_6(R)$ vanishes:
\begin{equation}
\left( y_2^P \, , y_4^P \, , y_6^P \right)\Big|_{r=R} = \left( - \frac{2n+1}{3}\, \rho_b \, , 0 \, , 0 \right) .
\label{bcP}
\end{equation}
Second, the Love number $h^P_n$ is related to $y_1^P$ as in Eq.~(\ref{kJ}), but the definition of $k^P_n$ differs from Eq.~(\ref{kJ}) because $y^P_5$ is only due to the induced gravitational potential $\Phi_n$:
\begin{equation}
k^P_n = y^P_5(R) \, .
\label{kP}
\end{equation}

As the viscoelastic-gravitational problem is linear, it is always possible to superpose two solutions $A$ and $B$ into a new one satisfying a superposition of the boundary conditions for $A$ and $B$.
By comparing the $T$, $L$, and $P$ boundary conditions (Eqs.~(\ref{bcT}), (\ref{bcL}), and (\ref{bcP})), one sees that the pressure Love numbers are linear combinations of tidal and load Love numbers,
\begin{eqnarray}
h^P_n &=& h^L_n - h^T_n \, ,
\nonumber \\
k^P_n &=& k^L_n - k^T_n \, = \, - h^T_n \, ,
\label{hPkP}
\end{eqnarray}
where the last equality results from the Saito-Molodensky relation (Eq.~(\ref{saitomolo})).

\section{Equilibrium tide for a 3-layer model}
\label{AppendixEquil}
\renewcommand{\theequation}{F.\arabic{equation}} 
\setcounter{equation}{0}  

Let us compute the degree-two equilibrium tide for a model made of three viscoelastic layers: mantle (not distinguished from the core), ocean and crust.
The crust is thin and has the same density $\rho$ as the upper layer of the ocean.
With these assumptions, the static tidal Love numbers read \citep{beuthe2014}
\begin{eqnarray}
k^T_2 + 1 &=& \frac{  k_2^\circ +1 }{ 1 + \xi_2 \left( k_2^\circ+1 \right) \frac{\Lambda_2}{1+\Lambda_2} } \, ,
\nonumber \\
h^T_2 &=& \frac{ h_2^\circ }{ 1 +  \left( 1 + \xi_2 \, h_2^\circ \right) \Lambda_2 } \, ,
\label{h3layers}
\end{eqnarray}
where $\xi_2$ and $\Lambda_2$ are defined by Eqs.~(\ref{xin}) and (\ref{LambdaMB}), respectively.
The fluid-crust radial Love number $h_2^\circ=k_2^\circ+1$ depends on the deep interior structure (the ocean and below).
If the ocean is homogeneous and the mantle is infinitely rigid, $h_2^\circ=1/(1-\xi_2)$ so that
\begin{equation}
h^T_{2,rigid} = \frac{1}{1-\xi_2+\Lambda_2} \, .
\label{h2r}
\end{equation}
Eq.~(\ref{h3layers}) is identical to Eqs.~(57)-(58) of \citet{beuthe2014}, except that $\Lambda_2$ includes here a contribution due to bending (which was not considered in \citet{beuthe2014} because it is negligible at degree two).

The displacement of the mantle-ocean boundary can be quantified by an internal Love number, 
\begin{eqnarray}
h_2^m &=& gy_1(R_m)
\nonumber \\
&=& \frac{k^T_2 + 1}{k_2^\circ+1} \, h_2^{\circ{}m} \, ,
\label{hTM}
\end{eqnarray}
where $h_2^{\circ m}=gy_1^\circ(R_m)$ is the corresponding fluid-crust tidal Love number.
The second line is obtained by gravity scaling (Eqs.~(141)-(142) of \citet{beuthe2014}).

Suppose now that the ocean is homogeneous and the core-mantle system below the ocean is approximated by a homogeneous and incompressible sphere with density $\rho_b$ and shear modulus $\mu_m$.
The fluid-crust tidal Love numbers of this model are given by  Eqs.~(118)-(121) of \citet{beuthe2014}.
In the limit of a thin crust and shallow ocean (i.e.\ the mantle radius tends to the surface radius), these equations become
\begin{eqnarray}
h_2^\circ &=& 5 \, \frac{5\left(1-\xi_1\right) + 19 \, \hat\mu_m}{10\left(1-\xi_1\right)+19\left(5-3\xi_1\right) \hat\mu_m} \, ,
\nonumber \\
h_2^{\circ{}m} &=& \frac{5 \left(1-\xi_1\right)}{5\left(1-\xi_1\right) + 19 \, \hat\mu_m} \, h_2^\circ \, ,
\end{eqnarray}
where $\xi_1=\rho/\rho_b$ and $\hat\mu_m=\mu_m/(\rho_b{}gR)$.
Substituting these expressions into Eqs.~(\ref{h3layers}) and (\ref{hTM}), I obtain the dependence of the degree-two equilibrium ocean tide on the interior structure (see Eq.~(\ref{Zn3layers})):
\begin{eqnarray}
Z_2 &=& h_2^{T} - h_2^m
\nonumber \\
&=& \frac{ 57\hat\mu_m + 5 \left( 5\xi_2 - 3 \right) \Lambda_2 }{ 2 \left( 3 - 5\xi_2 \right) + 57 \left( 1 - \xi_2 + \Lambda_2 \right) \hat\mu_m + \left( 6 + 5\xi_2 \left(1-5\xi_2\right) \right) \Lambda_2} \, .
\label{Zn3layersPMM}
\end{eqnarray}

\section{Mantle dissipation}
\label{AppendixMantleDissipation}
\renewcommand{\theequation}{G.\arabic{equation}} 
\setcounter{equation}{0}  

In the non-rotating model, the dissipation rate within the body can be expressed in terms of the radial deformation functions $(y_1,y_3,y_4)$ \citep{beuthe2013}.
For dynamical tides, the $y_i$ formalism is still valid in the mantle as long as its unperturbed state is spherically symmetric.
One can thus solve for the deformation functions within the mantle given the correct boundary conditions at the top of the mantle.

At the mantle-ocean boundary, the radial displacement must be equal to $\eta^{bot}$ (Eq.~(\ref{zetabot})) and the shear stress vanishes (free-slip condition).
If the mantle is of uniform density, these conditions are sufficient to determine $(y_1,y_3,y_4)$ because deformations can be solved independently of the gravitational potential (see Appendix~F of \citet{beuthe2013}).
For simplicity, assume that the mantle is incompressible so that analytical propagator matrix method is applicable.

In order to use the $y_i$ formalism, I write Eq.~(\ref{zetabot}) in terms of a fictitious potential $U_n^M$,
\begin{equation}
\eta_n^{bot} = \frac{h_n^T}{g} \, U_n^M \, ,
\end{equation}
where $U_n^M=U^T_n +  (h^L_n \, U^L_n + h^P_n \, U^P_n)/h^T_n$.
From Eqs.~(\ref{LoveH}) and (\ref{hPkP}), one sees that $U_n^M$ does not depend on the internal structure of the mantle (i.e.\ on $S_n\hat\mu_m$) so that it makes sense to consider $U_n^M$ as an external potential.
Supposing that $U_n^M$ is known, I write the boundary conditions as
\begin{equation}
\left( y_1(R) , y_4(R) \right) = \left( h_n^T/g , 0 \right) ,
\end{equation}
where the functions $y_i$ do not carry superscripts because they correspond to a superposition of tidal, mass load, and pressure problems.
The boundary conditions on $y_2$ and $y_6$ are not the same as Eq.~(\ref{bcT}) but they are not needed here.

The straightforward (and laborious) way to solve the problem consists in:
\begin{enumerate}
\item finding the deformation functions $(y_1,y_3,y_4)$ at degree $n$ (Eqs.~(99) and (107) of \citet{beuthe2013} extended to $n\neq2$),
\item computing the radial weights $(f_A , f_B , f_C)$ at degree $n$ (Eq.~(24) of \citet{beuthe2013} extended to $n\neq2$),
\item computing ${\cal I} = \int_0^R Im(\mu) H_\mu \, dr$ where $H_\mu=f_A+f_B+f_C$,
\item computing the global dissipation rate in the mantle as $\dot E_M = \sum_n 2\pi\Omega\,{\cal I}\,U_n^{M,sq}$ (Eqs.~(39) and (41) of \citet{beuthe2013}), where $U_n^{M,sq}$ refers to the spatial average of the squared norm of $U_n^M$ (Eq.~(\ref{Unsq})),
\item replacing $(h_n^T/g)U_n^M$ by $\eta_n^{bot}$.
\end{enumerate}
As a shortcut to steps 1-2-3 , I apply the fundamental micro-macro relation (Eq.~(40) of \citet{beuthe2013} extended to $n\neq2$):
\begin{equation}
\int_0^R  Im(\mu) \, H^{}_\mu \, dr = - \frac{(2n+1) R}{4 \pi G} \,  Im(k_n^T) \, .
\end{equation}
Substituting Eqs.~(\ref{LovekH}) and (\ref{hnTincomp}) into this equation, I directly obtain the quantity required in step~4:
\begin{equation}
{\cal I} = \frac{2(n-1)}{2n+1} \, R \, Im( \mu_m \, S_n ) \, \frac{|h_n^T|^2}{g^2} \, ,
\end{equation}
where $S_n$  is given by Eq.~(\ref{Sn}).
Finally, steps~4 and 5 yield the global dissipation rate for an incompressible mantle of uniform density:
\begin{equation}
\dot E_M = \Omega R \sum_{m=0}^2 \left( 1+\delta_{m0} \right) \sum_{n=m}^N \frac{n-1}{2n+1} Im \left( \mu_m \, S_n \right) \sum_{X=E,W} \left| (\eta^{bot})_{nX}^m \right|^2 .
\label{EMapp}
\end{equation}

\section{Whole body dissipation}
\label{AppendixTotalDissipation}
\renewcommand{\theequation}{H.\arabic{equation}} 
\setcounter{equation}{0}  

Let us transform the general formula for the whole body dissipation rate (Eq.~(\ref{Etot1})) in order to express it as a function of the ocean tide.
First, I evaluate the dissipation rate in terms of spherical harmonic coefficients:
\begin{equation}
\dot E = \frac{\Omega}{2} \, \frac{5R}{4\pi{}G} \sum_{m=0}^2 \left( 1+\delta_{m0} \right) \sum_{X=E,W} Im \left( U_{2X}^m \, {\Gamma_{2X}^m}^* \right) .
\label{Etot2}
\end{equation}
Using Eq.~(\ref{Gamman}), I can write the last factor of this equation as
\begin{equation}
Im \left( U_{2X}^m \, {\Gamma_{2X}^m}^* \right) =  -Im \left( k^T_2 \right) |U_{2X}^m|^2  + g \xi_2 \,  Im \left( \left( \gamma_2^T \, \eta_{2X}^m - h_2^T \Lambda_2 \, (\eta^{top})_{2X}^m \right) {U_{2X}^m}^*  \right) ,
\label{Etot3}
\end{equation}
where I used relations between Love numbers (Eqs.~(\ref{kLgammaT}) and (\ref{hPkP})).

Finally, I express $\eta^{top}$ in terms of $\eta$ with Eq.~(\ref{zetatop}), before expressing $\eta$ in terms of the velocity potential $\Phi$ with Eq.~(\ref{zetanm}).
The result can be written as the sum of three terms:
\begin{equation}
\dot E = \dot E_1 + \dot E_2 + \dot E_3 \, ,
\label{Ediss3}
\end{equation}
where
\begin{eqnarray}
\dot E_1 &=& - \frac{5\Omega R}{2G} \, U_2^{sq} \, Im(k_2^T) \, ,
\nonumber \\
\dot E_2 &=&  \frac{5\Omega R}{2G} \, U_2^{sq} \, Im \left( \frac{ \xi_2 \, (h^T_2)^2 \Lambda_2 }{1+\xi_2 \left( h^T_2 - h^L_2 \right) \Lambda_2} \right) ,
\nonumber \\
\dot E_3 &=& -3\rho D \sum_{m=0}^2 \left( 1+\delta_{m0} \right) \sum_{X=E,W} Im \left( \upsilon_2 \, \Phi_{2X}^m \, {U_{2X}^{m}}^* \right) .
\label{E1E2E3}
\end{eqnarray}
In the first two equations, $U_2^{sq}$ is given by Eq.~(\ref{Unsq}):
$U_2^{sq}=(\Omega R)^4(21/5)e^2$ for eccentricity tides whereas $U_2^{sq}=(\Omega R)^4(3/5)\sin^2 I$ for obliquity tides.

In general, the terms $\dot E_j$ cannot be identified with the dissipation rate of a specific layer (mantle, crust, or ocean).
For example, $\dot E_1$ (called `body tide' in \citet{zschau1978} and \citet{platzman1984}) is the dissipation rate of the body without a crust and an ocean, but this quantity is not observable.
The terms $\dot E_j$ vanish under the following conditions:
\begin{enumerate}
\item $\dot E_1=0$ if the mantle is elastic ($Im(k_2^T)=0$) or rigid ($k_2^T=0$).
\item $\dot E_2=0$ either if the mantle and the crust are both elastic ($Im(h_2^T)=Im(h_2^L)=Im(\Lambda_2)=0$), or if the mantle is rigid ($h_2^T=0$) and the crust is viscoelastic.
\item $\dot E_3=0$ if the mantle and the crust are both elastic ($Im(\upsilon_2)=0$) and the fluid is inviscid ($Im(\Phi_{2X}^m)=0$).
\end{enumerate}

\end{appendices}
\newpage

\bibliographystyle{agufull04}
\renewcommand{\baselinestretch}{0.5}

\end{document}